%% file: thesis.tex
\documentclass{buthesis}
\usepackage[usenames]{color} %used for font color
\usepackage{amssymb} %maths
\usepackage{amsmath} %maths
\usepackage[utf8]{inputenc} %useful to type directly diacritic characters
\usepackage{graphicx}

\thesistype{Ph.D.~}
\teztipi{Doktora}

\anahtarsoz{Haritalamayı Oku, Yakla\c{s}ık Dize E\c{s}le\c{s}tirme, Hizalamayı Oku, Levenshtein Mesafesi, String Algoritmaları, Mesafeyi Düzenle, hızlı {\"o}n hizalama filtresi, Alan programlanabilir kapı dizileri (FPGA)}

\title{ACCELERATING THE UNDERSTANDING OF LIFE'S CODE THROUGH BETTER ALGORITHMS AND HARDWARE DESIGN}

\baslik{YA\c{S}AMIN KODUNU ANLAMAYI DAHA \.{I}Y\.{I} ALGOR\.{I}TMALAR VE DONANIM TASARIMLARIYLA HIZLANDIRMAK}
\author{Mohammed H. K. Alser}
\dept{Computer Engineering}
\bolum{Bilgisayar M\"{u}hendisli\u{g}i}
\principaladvisor{Can Alkan}
\tezyoneticisi{Can Alkan}
\coadvisor{Onur Mutlu} % uncomment this line if there is a co-advisor
\ikincitezyoneticisi{Onur Mutlu} % uncomment this line if there is a co-advisor
\firstreader{Ergin Atalar}
\secondreader{Mehmet Koyut\"{u}rk}
\thirdreader{Nurcan Tun\c cba\u{g}}
\fourthreader{O\u{g}uz Ergin}
\director{Ezhan Kara\c san}
\submitdate{June 2018}
\tarih{Haziran 2018}

\begin{document}
\vspace*{-\baselineskip}
\vspace*{-\baselineskip}
\vspace*{-\baselineskip}
\vspace*{-\baselineskip}
  \centerline{ \textcolor{blue}{This thesis received the 2018 IEEE Turkey Doctoral Dissertation Award}}
\centerline{ \textcolor{blue}{**This copy was updated on October 2019**}}
\vspace*{\baselineskip}
\vspace*{\baselineskip}

%\titlepageMS %comment out if PhD thesis
\titlepagePhD %comment out if MS thesis
%\setcounter{page}{2}
%\signaturepageMS %comment out if PhD thesis
\signaturepagePhD %comment out if MS thesis

\thispagestyle{plain}
\begin{abstract}
Our understanding of human genomes today is affected by the ability of modern computing technology to quickly and accurately determine an individual's entire genome. Over the past decade, high throughput sequencing (HTS) technologies have opened the door to remarkable biomedical discoveries through its ability to generate hundreds of millions to billions of DNA segments per run along with a substantial reduction in time and cost. However, this flood of sequencing data continues to overwhelm the processing capacity of existing algorithms and hardware. To analyze a patient\rq{}s genome, each of these segments -called \textit{reads}- must be mapped to a reference genome based on the similarity between a read and “candidate" locations in that reference genome. The similarity measurement, called alignment, formulated as an approximate string matching problem, is the computational bottleneck because: (1) it is implemented using quadratic-time dynamic programming algorithms, and (2) the majority of candidate locations in the reference genome do not align with a given read due to high dissimilarity. Calculating the alignment of such incorrect candidate locations consumes an overwhelming majority of a modern read mapper’s execution time. Therefore, it is crucial to develop a fast and effective filter that can detect incorrect candidate locations and eliminate them before invoking computationally costly alignment algorithms.

In this thesis, we introduce four new algorithms that function as a pre-alignment step and aim to filter out most incorrect candidate locations. We call our algorithms GateKeeper, Shouji, MAGNET, and SneakySnake. The first key idea of our proposed pre-alignment filters is to provide high filtering accuracy by correctly detecting all similar segments shared between two sequences. \newpage The second key idea is to exploit the massively parallel architecture of modern FPGAs for accelerating our four proposed filtering algorithms. We also develop an efficient CPU implementation of the SneakySnake algorithm for commodity desktops and servers, which are largely available to bioinformaticians without the hassle of handling hardware complexity. We evaluate the benefits and downsides of our pre-alignment filtering approach in detail using 12 real datasets across different read length and edit distance thresholds. In our evaluation, we demonstrate that our hardware pre-alignment filters show two to three orders of magnitude speedup over their equivalent CPU implementations. We also demonstrate that integrating our hardware pre-alignment filters with the state-of-the-art read aligners reduces the aligner\rq{}s execution time by up to 21.5x. Finally, we show that efficient CPU implementation of pre-alignment filtering still provides significant benefits. We show that SneakySnake on average reduces the execution time of the best performing CPU-based read aligners Edlib and Parasail, by up to 43x and 57.9x, respectively. The key conclusion of this thesis is that developing a fast and efficient filtering heuristic, and developing a better understanding of its accuracy together leads to significant reduction in read alignment\rq{}s execution time, without sacrificing any of the aligner\rq{} capabilities. We hope and believe that our new architectures and algorithms catalyze their adoption in existing and future genome analysis pipelines.
\end{abstract}

\begin{ozet}
G\"{u}n\"{u}m\"{u}zde insan genomları konusundaki anlayı\c{s}ımız, modern bili\c{s}im teknolojisinin bir bireyin t\"{u}m genomunu hızlı ve do\u{g}ru bir \c{s}ekilde belirleyebilme yetene\u{g}inden etkilenmektedir. Ge\c{c}ti\u{g}imiz on yıl boyunca, y\"{u}ksek verimli dizileme (HTS) teknolojileri, zaman ve maliyette \"{o}nemli bir azalma ile birlikte, tek bir \c{c}alı\c{s}mada y\"{u}z milyonlardan milyarlarcaya kadar DNA par\c{c}ası \"{u}retme kabiliyeti sayesinde dikkat \c{c}ekici biyomedikal ke\c{s}iflere kapı a\c{c}mı\c{s}tır. Ancak, bu dizileme verisi bollu\u{g}u mevcut algoritmaların ve donanımların i\c{s}lem kapasitelerinin sınırlarını zorlamaya devam etmektedir. Bir hastanın genomunu analiz etmek i\c{c}in, "okuma" adı verilen bu par\c{c}aların her biri referans genomundaki aday b\"{o}lgelerle olan benzerliklerine bakılarak, referans genomu \"{u}zerine yerle\c{s}tirilir. Yakla\c{s}ık karakter dizgisi e\c{s}le\c{s}tirme problemi \c{s}eklinde form\"{u}le edilen ve hizalama olarak adlandırılan benzerlik hesaplaması, i\c{s}lemsel bir darbo\u{g}azdır \c{c}\"{u}nk\"{u}: (1) ikinci dereceden devingen programlama algoritmaları kullanılarak hesaplanır ve (2) referans genomundaki aday b\"{o}lgelerin b\"{u}y\"{u}k bir b\"{o}l\"{u}m\"{u} ile verilen okuma par\c{c}ası birbirlerinden y\"{u}ksek d\"{u}zeyde farklılık g\"{o}sterdiklerinden dolayı hizalanamaz. Bu \c{s}ekilde yanlı\c{s} belirlenen aday b\"{o}lgelerin hizalanabilirli\u{g}in hesaplanması, g\"{u}n\"{u}m\"{u}zdeki okuma haritalandırıcı algoritmaların \c{c}alı\c{s}ma s\"{u}relerinin b\"{u}y\"{u}k b\"{o}l\"{u}m\"{u}n\"{u} olu\c{s}turmaktadır. Bu nedenle, hesaplama olarak maliyetli bu hizalama algoritmalarını \c{c}alı\c{s}tırmadan \"{o}nce, do\u{g}ru olmayan aday b\"{o}lgeleri tespit edebilen ve bu b\"{o}lgeleri aday b\"{o}lge olmaktan \c{c}ıkaran, hızlı ve etkili bir filtre geli\c{s}tirmek \c{c}ok \"{o}nemlidir.
 
Bu tezde, \"{o}n hizalama a\c{s}aması olarak i\c{s}lev g\"{o}ren ve yanlı\c{s} aday konumlarının \c{c}o\u{g}unu filtrelemeyi hedefleyen d\"{o}rt yeni algoritma sunuyoruz. Algoritmalarımızı GateKeeper, Shouji, MAGNET ve SneakySnake olarak adlandırıyoruz. \newpage \"{O}nerilen \"{o}n hizalama filtrelerinin ilk temel fikri, iki dizi arasında payla\c{s}ılan t\"{u}m benzer segmentleri do\u{g}ru bir \c{s}ekilde tespit ederek y\"{u}ksek filtreleme do\u{g}rulu\u{g}u sa\u{g}lamaktır. İkinci temel fikir, \"{o}nerilen d\"{o}rt filtreleme algoritmamızın hızlandırılması i\c{c}in modern FPGA'ların \c{c}ok b\"{u}y\"{u}k \"{o}l\c{c}ekte paralel mimarisini kullanmaktır. SneakySnake’i esas olarak biyoinformatisyenlerin mevcut olan, donanım karma\c{s}ıklı\u{g}ı ile u\u{g}ra\c{s}mak zorunda olmadıkları emtia masa\"{u}st\"{u} ve sunucularında kullanabilmeleri i\c{c}in geli\c{s}tirdik. \"{O}n okuma filtreleme yakla\c{s}ımımızın avantaj ve dezavantajlarını 12 ger\c{c}ek veri setini, farklı okuma uzunlukları ve mesafe e\c{s}ikleri kullanarak ayrıntılı olarak de\u{g}erlendirdik. De\u{g}erlendirmemizde, donanım \"{o}n hizalama filtrelerimizin e\c{s}de\u{g}er CPU uygulamalarına g\"{o}re iki ila \"{u}\c{c} derece hızlı olduklarını g\"{o}steriyoruz. Donanım \"{o}n hizalama filtrelerimizi son teknoloji okuma hizalayıcılarıyla entegre etmenin hizalayıcının \c{c}alı\c{s}ma s\"{u}resini d\"{u}zenleme mesafesi e\c{s}i\u{g}ine ba\u{g}lı olarak 21.5x. Son olarak, \"{o}n hizalama filtrelerinin etkin CPU uygulamasının hala \"{o}nemli faydalar sa\u{g}ladı\u{g}ını g\"{o}steriyoruz. SneakySnake'in en iyi performansa sahip CPU tabanlı okuma ayarlayıcıları Edlib ve Parasail'in y\"{u}r\"{u}tme s\"{u}relerini sırasıyla 43x ve 57,9x'e kadar azalttı\u{g}ını g\"{o}steriyoruz. Bu tezin ana sonucu, hızlı ve verimli bir filtreleme mekanizması geli\c{s}tirilmesi ve bu mekanizmanın do\u{g}rulu\u{g}unun daha iyi anla\c{s}ılması, hizalayıcıların yeteneklerinden hi\c{c}bir \c{s}ey \"{o}d\"{u}n vermeden, okuma hizalamasının \c{c}alı\c{s}ma s\"{u}resinde \"{o}nemli bir azalmaya yol a\c{c}maktadır. Yeni mimarilerimizin ve algoritmalarımızın, mevcut ve gelecekteki genom analiz planlarında benimsenmelerini katalize etti\u{g}imizi umuyor ve buna inanıyoruz.
\end{ozet}

\begin{ack}
The last nearly four years at Bilkent University have been most exciting and fruitful time of my life, not only in the academic arena, but also on a personal level. For that, I would like to seize this opportunity to acknowledge all people who have supported and helped me to become who I am today. First and foremost, the greatest of all my appreciation goes to the almighty Allah for granting me countless blessings and helping me stay strong and focused to accomplish this work.

I would like to extend my sincere thanks to my advisor Can Alkan, who introduced me to the realm of bioinformatics. I feel very privileged to be part of his lab and I am proud that I am going to be his first PhD graduate. Can generously provided the resources and the opportunities that enabled me to publish and collaborate with researchers from other institutions. I appreciate all his support and guidance at key moments in my studies while allowing me to work independently with my own ideas. I truly enjoyed the memorable time we spent together in Ankara, Izmir, Istanbul, Vienna, and Los Angeles. Thanks to him, I get addicted to the dark chocolate with almonds and sea salt.

I am also very thankful to my co-advisor Onur Mutlu for pushing my academic writing to a new level. It is an honor that he also gave me the opportunity to join his lab at ETH Z\"{u}rich as a postdoctoral researcher. His dedication to research is contagious and inspirational. 

I am grateful to the members of my thesis committee: Ergin Atalar, Mehmet Koyut\"{u}rk, Nurcan Tun\c cba\u{g}, O\u{g}uz Ergin, and Ozcan Ozt\"{u}rk for their valuable comments and discussions.

I would also like to thank Eleazar Eskin and Jason Cong for giving me the opportunity to join UCLA during the summer of 2017 as a staff research associate. I would like to extend my thanks to Serghei Mangul, David Koslicki, Farhad Hormozdiari, and Nathan LaPierre who provided the guidance to make my visit fruitful and enjoyable. I am also thankful to Akash Kumar for giving me the chance to join the Chair for Processor Design during the winter of 2017-2018 as a visiting researcher at TU Dresden.
\newpage
I would like to acknowledge Nordin Zakaria and Izzatdin B A Aziz for providing me the opportunity to continue carrying out PETRONAS projects while i am in Turkey. I am grateful to Nor Hisham Hamid for believing in me, hiring me as a consultant for one of his government-sponsored projects, and inviting me to Malaysia. It was my great fortune to work with them.

I want to thank all my co-authors and colleagues: Eleazar Eskin, Erman Ayday, Hongyi Xin, Hasan Hassan, Serghei Mangul, O\u{g}uz Ergin, Akash Kumar, Nour Almadhoun, Jeremie Kim, and Azita Nouri.

I would like to acknowledge all past and current members of our research group for being both great friends and colleagues to me. Special thanks to Fatma Kahveci for being a great labmate who taught me her native language and tolerated me for many years. Thanks to G\"{u}lfem Demir for all her valuable support and encouragement. Thanks to Shatlyk Ashyralyyev for his great help when I prepared for my qualifying examination and for many valuable discussions on research and life. Thanks to Handan Kulan for her friendly nature and priceless support. Thanks to Can Firtina and Damla \c{S}enol Cali (CMU) for all the help in translating the abstract of this thesis to Turkish. I am grateful to other members for their companionship: Marzieh Eslami Rasekh, Azita Nouri, Elif Dal, Zülal Bingöl, Ezgi Ebren, Arda Söylev, Halil İbrahim Özercan, Fatih Karaoğlanoğlu, Tuğba Doğan, and Balanur İçen.

I would like to express my deepest gratitude to Ahmed Nemer Almadhoun, Khaldoun Elbatsh (and his wonderful family), Hamzeh Ahangari, Soe Thane, Mohamed Meselhy Eltoukhy (and his kind family), Abdel-Haleem Abdel-Aty, Heba Kadry, and Ibrahima Faye for all the good times and endless generous support in the ups and downs of life. I feel very grateful to all my inspiring friends who made Bilkent a happy home for me, especially Tamer Alloh, Basil Heriz, Kadir Akbudak, Amirah Ahmed, Zahra Ghanem, Muaz Draz, Nabeel Abu Baker, Maha Sharei, Salman Dar, Elif Do\u{g}an Dar, Ahmed Ouf, Shady Zahed, Mohammed Tareq, Abdelrahman Teskyeh, Obada and many others.

My graduate study is an enjoyable journey with my lovely wife, Nour Almadhoun, and my sons, Hasan and Adam. Their love and never ending support always give me the internal strength to move on whenever I feel like giving up or whenever things become too hard. \linebreak Pursuing PhD together with Nour in the same department is our most precious memory in lifetime. We could not have been completed any of our PhD work without continuously supporting and encouraging each other. 

My family has been a pillar of support, encouragement and comfort all throughout my journey. Thanks to my parents, Hasan Alser and Itaf Abdelhadi, who raised me with a love of science and supported me in all my pursuits. Thanks to my lovely sister Deema and my brothers Ayman (and his lovely family), Hani, Sohaib, Osaid, Moath, and Ayham for always believing in me and being by my side. Thanks to my parents-in-law, Mohammed Almadhoun and Nariman Almadhoun, my brotherss-in-law, Ahmed Nemer, Ezz Eldeen, Moatasem, Mahmoud, sisters-in-law, Narmeen, Rania, and Eman, and their families for all their understanding, great support, and love.

Finally, I gratefully acknowledge the funding sources that made my PhD work possible. I was honored to be a T\"{U}B\.{I}TAK Fellow for 4 years offered by the Scientific and Technological Research Council of Turkey under 2215 program. In 2017, I was also honored to receive the HiPEAC Collaboration Grant. I am grateful to Bilkent University for providing generous financial support and funding my academic visits. This thesis was supported by NIH Grant (No. HG006004) to Onur Mutlu and Can Alkan and a Marie Curie Career Integration Grant (No. PCIG-2011-303772) to Can Alkan under the Seventh Framework Programme.\\

\textit{Mohammed H. K. Alser}

\textit{June 2018, Ankara, Turkey}
\end{ack}

\tableofcontents
\listoffigures
\listoftables
\newpage
\newpage
\newpage 
\pagestyle{headings}
\makeatother

\setcounter{secnumdepth}{5} %% depth of subsections
\setcounter{tocdepth}{5} %% depth of Table of Contents

\pagestyle{plain}
\pagenumbering{arabic}
\setcounter{page}{1}
\setcounter{section}{1}
\parskip 0.5cm
\parindent 0.5cm

\input{chapter1}
\input{chapter2}
\input{chapter3}
\input{chapter4}

\input{chapter5} 
\input{chapter6}
\input{chapter7}
\input{chapter8} 
\input{chapter9} 
\input{chapter10} 
\input{chapter11}

\bibliography{whole}
\bibliographystyle{ieeetr}
\input{appendix}

\end{document}

%% file: chapter1.tex
\chapter{Introduction} 

Genome is the \textit{code of life} that includes set of instructions for making everything from humans to elephants, bananas, and yeast. Analyzing the life's code helps, for example, to determine differences in genomes from human to human that are passed from one generation to the next and may cause diseases or different traits \cite{lane2017genome, dobigny2017chromosomal, sudmant2015integrated, korte2013advantages, visscher2012five, antonacci2009characterization}. One benefit of knowing the genetic variations is better understanding and diagnosis diseases such as cancer and autism \cite{han2017taxonomy, visscher201710, chin2011cancer, ding2010analysis} and the development of efficient drugs \cite{pritchard2017strategies, esplin2014personalized, metzker2010sequencing, hamburg2010path, ginsburg2001personalized}. The first step in genome analysis is to reveal the entire content of the subject genome – a process known as DNA sequencing \cite{reuter2015high}. Until today, it remains challenging to sequence the entire DNA molecule as a whole. As a workaround, high throughput DNA sequencing (HTS) technologies are used to sequence random fragments of copies of the original molecule. These fragments are called short reads and are 75-300 basepairs (bp) long. The resulting reads lack information about their order and origin (i.e., which part of the subject genome they are originated from). Hence the main challenge in genome analysis is to construct the donor’s complete genome with respect to a reference genome. 
\newpage
During a process, called read mapping, each read is mapped onto one or more possible locations in the reference genome based on the similarity between the read and the reference sequence segment at that location (like solving a jigsaw puzzle). The similarity measurement is referred to as optimal read alignment (i.e., verification) and could be calculated using the Smith-Waterman local alignment algorithm \cite{smith1981textordfeminineidentification}. It calculates the alignment that is an ordered list of characters representing possible edit operations and matches required to change one of the two given sequences into the other. Commonly allowed edit operations include deletion, insertion and substitution of characters in one or both sequences. As any two sequences can have several different arrangements of the edit operations and matches (and hence different alignments), the alignment algorithm usually involves a backtracking step. This step finds the alignment that has the highest alignment score (called optimal alignment). The alignment score is the sum of the scores of all edits and matches along the alignment implied by a user-defined scoring function. However, this approach is infeasible as it requires \textit{O(mn)} running time, where \textit{m} is the read length (few hundreds of bp) and \textit{n} is the reference length ($\sim$ 3.2 billion bp for human genome), for each read in the data set (hundreds of millions to billions).

To accelerate the read mapping process and reduce the search space, state-of-the-art mappers employ a strategy called \textit{seed-and-extend}. In this strategy, a mapper applies heuristics to first find candidate map locations (seed locations) of subsequences of the reads using hash tables (BitMapper \cite{cheng2015bitmapper}, mrFAST with FastHASH \cite{xin2013accelerating}, mrsFAST \cite{hach2010mrsfast}) or BWT-FM indices (BWA-MEM \cite{li2013aligning}, Bowtie 2 \cite{langmead2012fast}, SOAP3-dp \cite{luo2013soap3}). It then aligns the read in full \textit{only} to those seed locations. Although the strategies for finding seed locations vary among different read mapping algorithms, seed location identification is typically followed by alignment step. The general goal of this step is to compare the read to the reference segment at the seed location to check if the read aligns to that location in the genome with fewer differences (called \textit{edits}) than a threshold \cite{navarro2001guided}.

\section{Research Problem}
\textbf{The alignment step is the performance bottleneck of today's read mapper taking over 70\% to 90\% of the total running time \cite{kim2018grim, cheng2015bitmapper, xin2013accelerating}}. We pinpoint three specific problems that cause, affect, or exacerbate the long alignment's execution time. 

\begin{figure}
\centering
\includegraphics[width=\linewidth]{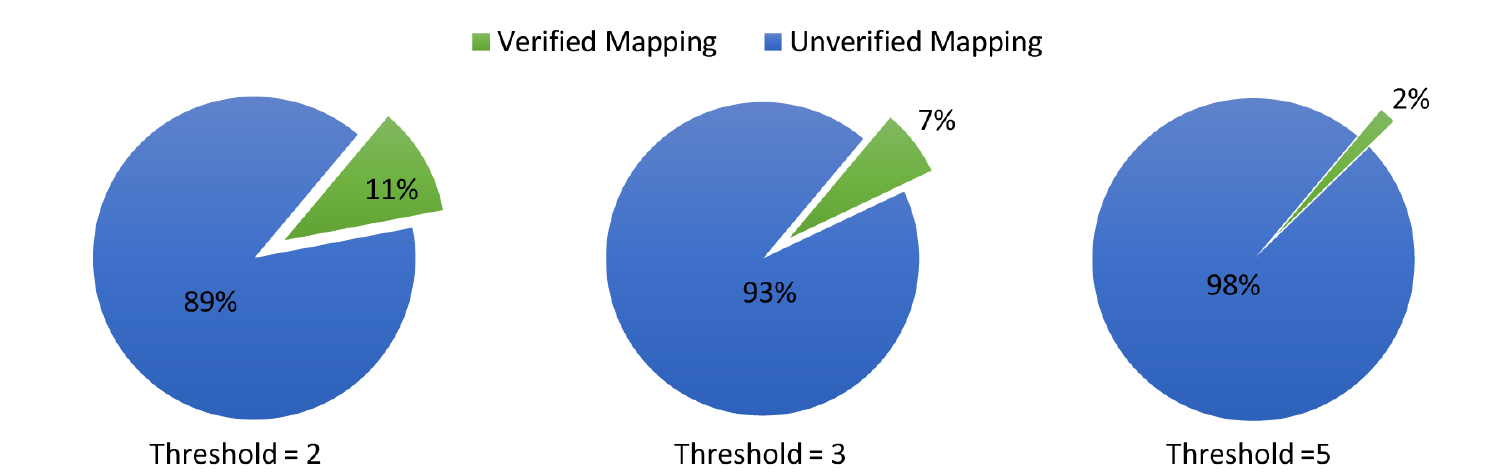}
\caption{Rate of verified and unverified read-reference pairs that are generated by mrFAST mapper and are fed into its read alignment algorithm. We set the threshold to 2, 3, and 5 edits.}
\label{fig:figure1incorrectmappings}
\end{figure}

(1) We find that across our data set (see \textbf{Chapter 9}), an overwhelming majority (more than 90\% as we present in Figure \ref{fig:figure1incorrectmappings}) of the seed locations, that are generated by a state-of-the-art read mapper, mrFAST with FastHASH \cite{xin2013accelerating}, exhibit more edits than the allowed threshold. These particular seed locations impose a large computational burden as they waste 90\% of the alignment's execution time in verifying these incorrect mappings. This observation is also in line with similar results for other read mappers \cite{xin2013accelerating, kim2018grim, cheng2015bitmapper, marco2012gem}. 

(2) Typical read alignment algorithm also needs to tolerate sequencing \textit{errors} \cite{firtina2019apollo, fox2014accuracy} as well as genetic variations \cite{mckernan2009sequence}. The read alignment approach is non-additive measure \cite{calude2002additive}. This means that if we divide the sequence pair into two consecutive subsequence pairs, the edit distance of the entire sequence pair is not necessarily equivalent to the sum of the edit distances of the shorter pairs. Instead, we need to examine all possible prefixes of the two input sequences and keep track of the pairs of prefixes that provide an optimal solution. Therefore, the read alignment step is implemented using dynamic programming algorithms to avoid re-examining the same prefixes many times. This includes Levenshtein distance \cite{levenshtein1966binary}, Smith-Waterman \cite{smith1981textordfeminineidentification}, Needleman-Wunsch \cite{needleman1970general} and their improved implementations. These algorithms are inefficient as they run in a quadratic-time complexity in the read length, \textit{m}, (i.e., \textit{O(m\textsuperscript{2})}). 

(3) This computational burden is further aggravated by the unprecedented flood of sequencing data which continues to overwhelm the processing capacity of existing algorithms and compute infrastructures \cite{canzar2017short}. While today’s HTS machines (e.g., Illumina HiSeq4000) can generate more than 300 million bases per minute, state-of-the-art read mapper can only map 1\% of these bases per minute \cite{escalona2016comparison}. The situation gets even worse when one tries to to understand a complex disease (e.g., autism and cancer) \cite{visscher201710, chin2011cancer, eberle2017reference, zheng2016haplotyping, griffith2015optimizing, iossifov2014contribution, meyerson2010advances} or profile a metagenomics sample \cite{lapierre2019micop, qin2012metagenome, segata2012metagenomic, afshinnekoo2015geospatial}, which requires sequencing hundreds of thousands of genomes. The long execution time of modern-day read alignment can severely hinder such studies.
There is also an urgent need for rapidly incorporating clinical sequencing into clinical practice for diagnosis of genetic disorders in critically ill infants \cite{delaney2016toward, perez2016urgent, kingsmore2015emergency, willig2015whole}. While early diagnosis in such infants shortens the clinical course and enables optimal outcomes \cite{sweeney2018case, farnaes2018rapid, berg2013processes}, it is still challenging to deliver efficient clinical sequencing for tens to hundreds of thousands of hospitalized infants each year \cite{ferreira2017medical}.

Tackling these challenges and bridging the widening gap between the execution time of read alignment and the increasing amount of sequencing data necessitate the development of fundamentally new, fast, and efficient read alignment algorithms. In the next sections, we provide the motivation behind our proposed work to considerably boost the performance of read alignment. We also provide further background information and literature study in Chapter 2.

\section{Motivation}
\begin{figure}
\centering
\includegraphics[width=\linewidth]{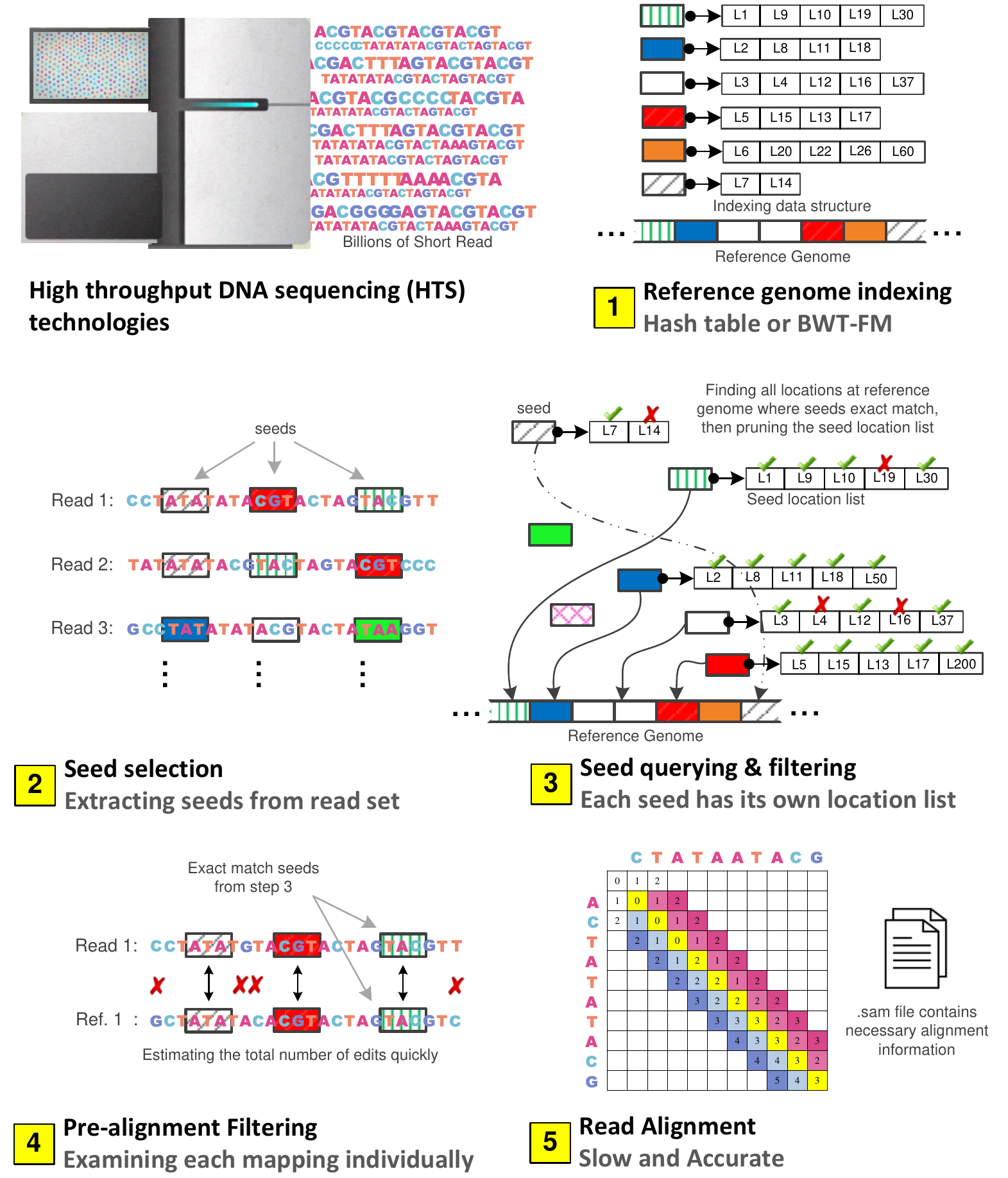}
\caption{The flow chart of seed-and-extend based mappers that includes five main steps. (1) Typical mapper starts with indexing the reference genome using a scheme based on hash table or BWT-FM. (2) It extracts number of seeds from each read that are produced using sequencing machine. (3) It obtains a list of possible locations within the reference genome that could result in a match with the extracted seeds. (4) It applies fast heuristics to examine the alignment of the entire read with its corresponding reference segment of the same length. (5) It uses expensive and accurate alignment algorithm to determine the similarity between the read and the reference sequences.}
\label{fig:figure1seed-and-extend}
\end{figure}

We present in Figure \ref{fig:figure1seed-and-extend} the flow chart of a typical seed-and-extend based mapper during the mapping stage. The mapper follows five main steps to map a read set to the reference genome sequence. (1) In step 1, typically a mapper first constructs fast indexing data structure (e.g., large hash table) for short segments (called seeds or k-mers or q-maps) of the reference sequence. (2) In step 2, the mapper extracts short subsequences from a read and uses them to query the hash table. (3) In step 3, The hash table returns all the occurrence hits of each seed in the reference genome. Modern mappers employ seed filtering mechanism to reduce the false seed locations that leads to incorrect mappings. (4) In step 4, for each possible location in the list, the mapper retrieves the corresponding reference segment from the reference genome based on the seed location. The mapper can then examine the alignment of the entire read with the reference segment using fast filtering heuristics that reduce the need for the dynamic programming algorithms. It rejects the mappings if the read and the reference are obviously dissimilar. 

Otherwise, the mapper proceeds to the next step. (5) In step 5, the mapper calculates the optimal alignment between the read sequence and the reference sequence using a computationally costly sequence alignment (i.e., verification) algorithm to determine the similarity between the read sequence and the reference sequence. Many attempts were made to tackle the computationally very expensive alignment problem. Most existing works tend to follow one of three key directions: (1) accelerating the dynamic programming algorithms \cite{banerjee2018asap, fei2018fpgasw, daily2016parasail, houtgast2015fpga, georganas2015meraligner, liu2015gswabe, waidyasooriya2014fpga, luo2013soap3, liu2013cudasw++, arram2013hardware}, (2) developing seed filters that aim to reduce the false seed locations \cite{cheng2015bitmapper, xin2013accelerating, marco2012gem, weese2012razers, ahmadi2011hobbes, rizk2010gassst, weese2009razers}, and (3) developing pre-alignment filtering heuristics \cite{xin2015shifted}.

The first direction takes advantage of parallelism capabilities of high performance computing platforms such as central processing units (CPUs) \cite{daily2016parasail, georganas2015meraligner}, graphics processing units (GPUs) \cite{liu2015gswabe, luo2013soap3, liu2013cudasw++}, and field-programmable gate arrays (FPGAs) \cite{banerjee2018asap, fei2018fpgasw, nishimura2017accelerating, chen2014accelerating, houtgast2015fpga, waidyasooriya2014fpga, arram2013hardware}. Among these computing platforms, FPGA accelerators seem to yield the highest performance gain \cite{fei2018fpgasw, waidyasooriya2016hardware}. However, many of these efforts either \textit{simplify} the scoring function, or only take into account accelerating the computation of the dynamic programming matrix \textit{without} providing the optimal alignment (i.e., backtracking) as in \cite{nishimura2017accelerating, chen2014accelerating}. Different scoring functions are typically needed to better quantify the similarity between the read and the reference sequence segment \cite{edgar2004comparison}. The backtracking step required for optimal alignment computation involves unpredictable and irregular memory access patterns, which poses difficult challenge for efficient hardware implementation.

The second direction to accelerate read alignment is to use filtering heuristics to reduce the size of the seed location list. This is the basic principle of nearly all mappers that employ seed-and-extend approach. Seed filter applies heuristics to reduce the output location list. The location list stores all the occurrence locations of each seed in the reference genome. The returned location list can be tremendously large as a mapper searches for an exact matches of short segment (typically 10 bp -13 bp for hash-based mappers) between two very long homologous genomes \cite{xin2015optimal}. Filters in this category suffer from low filtering accuracy as they can only look for exact matches with the help of hash table. Thus, they query a few number of seeds per read (e.g., in Bowtie 2 \cite{langmead2012fast}, it is 3 fixed length seeds at fixed locations) to maintain edit tolerance. mrFAST \cite{xin2013accelerating} uses another approach to increase the seed filtering accuracy by querying the seed and its shifted copies. This idea is based on the observation that indels cause the trailing characters to be shifted to one direction. If one of the shifted copies of the seed, generated from the read sequence, or the seed itself matches the corresponding seed from the reference, then this seed has zero edits. Otherwise, this approach calculates the number of edits in this seed as a single edit (that can be a single indel or a single substitution). Therefore, this approach fails to detect the correct number of edits for these case, for example, more than one substitutions in the same seed, substitutions and indel in the same seed, or more than one indels in the last seed). Seed filtering successes to eliminate some of the incorrect locations but it is still unable to eliminate sufficiently enough large portion of the false seed locations, as we present in Figure \ref{fig:figure1incorrectmappings}.
\newpage
The third direction to accelerate read alignment is to minimize the number of incorrect mappings on which alignment is performed by incorporating filtering heuristics. This is the last line of defense before invoking computationally expensive read alignment. Such filters come into play before read alignment (i.e., hence called pre-alignment filter), discarding incorrect mappings that alignment would deem a poor match. Though the seed filtering and the pre-alignment filtering have the same goal, they are fundamentally different problems. In pre-alignment filtering approach, a filter needs to examine the entire mapping. They calculate a best guess estimate for the alignment score between a read sequence and a reference segment. If the lower bound exceeds a certain number of edits, indicating that the read and the reference segment do not align, the mapping is eliminated such that no alignment is performed. Unfortunately, the best performing existing pre-alignment filter, such as shifted Hamming distance (SHD), is slow and its mechanism introduces inaccuracy in its filtering unnecessarily as we show in our study in Chapter 3 and in our experimental evaluation, Chapter 9. 

\begin{figure}
\centering
\includegraphics[width=\linewidth]{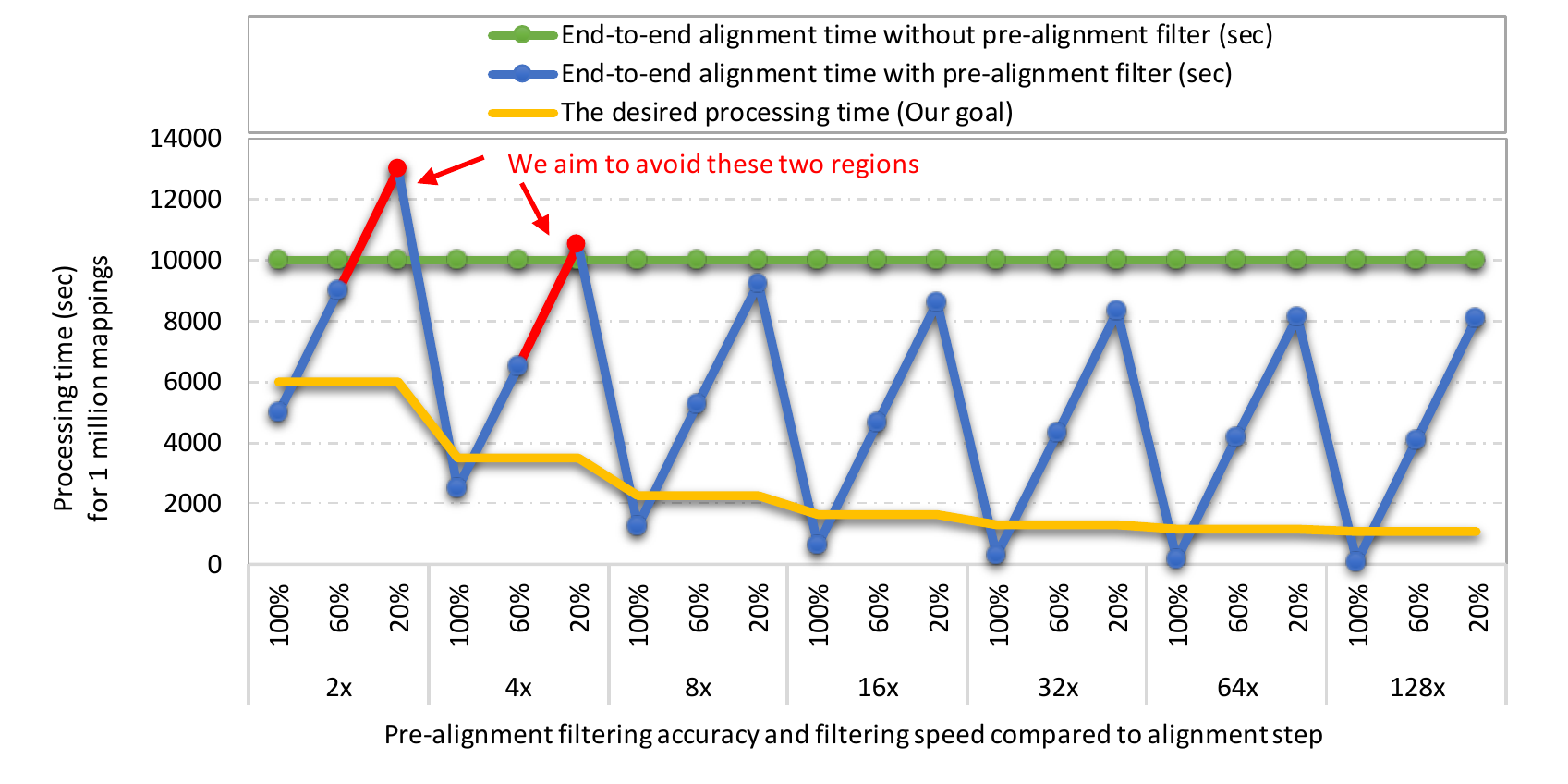}
\caption{End-to-end execution time (in seconds) for read alignment, with (blue plot) and without (green plot) pre-alignment filter. Our goal is to significantly reduce the alignment time spent on verifying incorrect mappings (highlighted in yellow). We sweep the percentage of rejected mappings and the filtering speed compared to alignment algorithm in the x-axis.}
\label{fig:figure2effect}
\end{figure}

\newpage
\textbf{Pre-alignment filter enables the acceleration of read alignment and meanwhile offers the ability to make the best use of existing read alignment algorithms}. 

These benefits come \textit{without} sacrificing any capabilities of these algorithms, as pre-alignment filter does \textit{not} modify or replace the alignment step. This motivates us to focus our improvement and acceleration efforts on pre-alignment filtering. We analyze in Figure \ref{fig:figure2effect} the effect of adding pre-alignment filtering step before calculating the optimal alignment and after generating the seed locations. We make two key observations. (1) The reduction in the end-to-end processing time of the alignment step largely depends on the accuracy and the speed of the pre-alignment filter. (2) Pre-alignment filtering can provide unsatisfactory performance (as highlighted in red) if it can not reject more than about 30\% of the potential mappings while it's only 2x-4x faster than read alignment step. 

We conclude that it is important to understand well what makes pre-alignment filter inefficient, such that we can devise new filtering technique that is much faster than read alignment and yet maintains high filtering accuracy.

\section{Thesis Statement}
Our goal in this thesis is to significantly reduce the time spent on calculating the optimal alignment in genome analysis from hours to mere seconds, given limited computational resources (i.e., personal computer or small hardware). This would make it feasible to analyze DNA routinely in the clinic for personalized health applications. Towards this end, we analyze the mappings that are provided to read alignment algorithm, and explore the causes of filtering inaccuracy. Our thesis statement is:

\textit{\textbf{read alignment can be substantially accelerated using computationally inexpensive and accurate pre-alignment filtering algorithms designed for specialized hardware.}}

Accurate filter designed on a specialized hardware platform can drastically expedite read alignment by reducing the number of locations that must be verified via dynamic programming. To this end, we (1) develop four hardware-acceleration-friendly filtering algorithms and highly-parallel hardware accelerator designs which greatly reduce the need for alignment verification in DNA read mapping, (2) introduce fast and accurate pre-alignment filter for general purpose processors, and (3) develop a better understanding of filtering inaccuracy and explore speed/accuracy trade-offs.

\section{Contributions}
The overarching contribution of this thesis is the new algorithms and architectures that reduce read alignment's execution time in read mapping. More specifically, this thesis makes the following main contributions:
\begin{enumerate}
\item We provide a detailed investigation and analysis of four potential causes of filtering inaccuracy in the state-of-the-art alignment filter, SHD \cite{xin2015shifted}. We also provide our recommendations on eliminating these causes and improving the overall filtering accuracy.

\item \textbf{GateKeeper.} We introduce the first hardware pre-alignment filtering, GateKeeper, which substantially reduces the need for alignment verification in DNA read mapping. GateKeeper is highly parallel and heavily relies on bitwise operations such as look-up table, shift, XOR, and AND. GateKeeper can examine up to 16 mappings in parallel, on a single FPGA chip with a logic utilization of less than 1\% for a single filtering unit. It provides two orders of magnitude speedup over the state-of-the-art pre-alignment filter, SHD. It also provides up to 13.9x speedup to the state-of-the-art aligners. GateKeeper is published in Bioinformatics \cite{alser2017gatekeeper} and also available in arXiv \cite{alser2016gatekeeper}.

\item \textbf{Shouji.} We introduce Shouji, a highly accurate and parallel pre-alignment filter which uses a sliding window approach to quickly identify dissimilar sequences without the need for computationally expensive alignment algorithms. Shouji can examine up to 16 mappings in parallel, on a single FPGA chip with a logic utilization of up to 2\% for a single filtering unit. It provides, on average, 1.2x to 1.4x more speedup than what GateKeeper provides to the state-of-the-art read aligners due to its high accuracy. Shouji is 2.9x to 155x more accurate than GateKeeper. Shouji is published in Bioinformatics \cite{alser2019shouji} and also available in arXiv \cite{alser2018slider}.

\item \textbf{MAGNET.} We introduce MAGNET, a highly accurate pre-alignment filter which employs greedy divide-and-conquer approach for identifying all non-overlapping long matches between two sequences. MAGNET can examine 2 or 8 mappings in parallel depending on the edit distance threshold, on a single FPGA chip with a logic utilization of up to 37.8\% for a single filtering unit. MAGNET is, on average, two to four orders of magnitude more accurate than both Shouji and GateKeeper. This comes at the expense of its filtering speed as it becomes up to 8x slower than Shouji and GateKeeper. It still provides up to 16.6x speedup to the state-of-the-art read aligners. MAGNET is published in IPSI \cite{alser2017magnet1}, available in arXiv \cite{alser2017magnet}, and presented in AACBB2018 \cite {alser2018exploring}.
\newpage
\item \textbf{SneakySnake.} We introduce SneakySnake, the fastest and the most accurate pre-alignment filter. SneakySnake reduces the problem of finding the optimal alignment to finding a snake's optimal path (with the least number of obstacles) in linear time complexity in read length. We provide a cost-effective CPU implementation for our SneakySnake algorithm that accelerates the state-of-the-art read aligners, Edlib \cite{vsovsic2017edlib} and Parasail \cite{daily2016parasail}, by up to 43x and 57.9x, respectively, without the need for hardware accelerators. We also provide a scalable hardware architecture and hardware design optimization for the SneakySnake algorithm in order to further boost its speed. The hardware implementation of SneakySnake accelerates the existing state-of-the-art aligners by up to 21.5x when it is combined with the aligner. SneakySnake is up to one order, four orders, and five orders of magnitude more accurate compared to MAGNET, Shouji, and GateKeeper, while preserving all correct mappings. SneakySnake also reduces the memory footprint of Edlib aligner by 50\%. This work is yet to be published.

\item We provide a comprehensive analysis of the asymptotic run time and space complexity of our four pre-alignment filtering algorithms. We perform a detailed experimental evaluation of our proposed algorithms using 12 real datasets across three different read lengths (100 bp, 150 bp, and 250 bp) an edit distance threshold of 0\% to 10\% of the read length. We explore different implementations for the edit distance problem in order to compare the performance of SneakySnake that calculate approximate edit distance with that of the efficient implementation of exact edit distance. This also helps us to develop a deep understanding of the trade-off between the accuracy and speed of pre-alignment filtering.

\end{enumerate}

Overall, we show in this thesis that developing a hardware-based alignment filtering algorithm and architecture together is both feasible and effective by building our hardware accelerator on a modern FPGA system. We also demonstrate that our pre-alignment filters are more effective in boosting the overall performance of the alignment step than only accelerating the dynamic programming algorithms by one to two orders of magnitude. \newpage This thesis provides a foundation in developing fast and accurate pre-alignment filters for accelerating existing and future read mappers.

\section{Outline}
This thesis is organized into 11 chapters. Chapter 2 describes the necessary background on read mappers and related prior works on accelerating their computations. Chapter 3 explores the potential causes of filtering inaccuracy and provides recommendations on tackling them. Chapter 4 presents the architecture and implementation details of our hardware accelerator that we use for boosting the speed of our proposed pre-alignment filters. Chapter 5 presents GateKeeper algorithm and architecture. Chapter 6 presents Shouji algorithm and its hardware architecture. Chapter 7 presents MAGNET algorithm and its hardware architecture. Chapter 8 presents SneakySnake algorithm. Chapter 9 presents the detailed experimental evaluation for all our proposed pre-alignment filters along with comprehensive comparison with the state-of-the-art existing works. Chapter 10 presents conclusions and future research
directions that are enabled by this thesis. Finally, Appendix A extends Chapter 9 with more detailed information and additional experimental data/results. 

%% file: chapter2.tex
\chapter{Background}
In this chapter, we provide the necessary background on two key read mapping methods. We highlight the strengths and weaknesses of each method. We then provide an extensive literature review on the prior, existing, and recent approaches for accelerating the operations of read mappers. We devote the provided background materials only to the reduction of read mapper's execution time.

\section{Overview of Read Mapper}
With the presence of a reference genome, read mappers maintain a large index database ($\sim$ 3 GB to 20 GB for human genome) for the reference genome.
This facilitates querying the whole reference sequence quickly and efficiently. Read mappers can use one of the following indexing techniques: suffix trees \cite{weiner1973linear}, suffix arrays \cite{ozsu2011principles}, Burrows-Wheeler transformation \cite{burrows1994block} followed by Ferragina-Manzini index \cite{ferragina2000opportunistic} (BWT-FM), and hash tables \cite{xin2013accelerating, rumble2009shrimp, ahmadi2011hobbes}. The choice of the index affects the query size, querying speed, and memory footprint of the read mapper, and even access patterns. \newpage Unlike hash tables, suffix-array or suffix tree can answer queries of variable length sequences. Based on the indexing technique used, short read mappers typically fall into one of two main categories \cite{canzar2017short}: (1) Burrows-Wheeler Transformation \cite{burrows1994block} and Ferragina-Manzini index \cite{ferragina2000opportunistic} (BWT-FM)-based methods and (2) Seed-and-extend based methods. Both types have different strengths and weaknesses. The first approach (implemented by BWA \cite{li2009fast}, BWT-SW \cite{li2010fast}, Bowtie \cite{langmead2009ultrafast}, SOAP2 \cite{li2009soap2}, and SOAP3 \cite{liu2012soap3}) uses aggressive algorithms to optimize the candidate location pools to find closest matches, and therefore may not find many potentially-correct mappings \cite{firtina2016genomic}. Their performance degrades as either the sequencing error rate increases or the genetic differences between the subject and the reference genome are more likely to occur \cite{medina2016highly, li2009fast}. To allow mismatches, BWT-FM mapper exhaustively traverses the data structure and match the seed to each possible path. Thus, Bowtie \cite{langmead2009ultrafast}, for example, performs a depth-first search (DFS) algorithm on the prefix trie and stops when the first hit (within a threshold of less than 4) is found. Next, we explain SOAP2, SOAP3, and Bowtie as examples of this category.

SOAP2 \cite{li2009soap2} improves the execution time and the memory utilization of SOAP \cite{li2008soap} by replacing its hash index technique with the BWT index. SOAP2 divides the read into non-overlapping (i.e., consecutive) seeds based on the number of allowed edits (default five). To tolerate two edits, SOAP2 splits a read into three consecutive seeds to search for at least one exact match seed that allows for up to two mismatches. SOAP3 \cite{liu2012soap3} is the first read mapper that leverage graphics processing unit (GPU) to facilitate parallel calculations, as the authors claim in \cite{liu2012soap3}. It speeds up the mapping process of SOAP2 \cite{li2009soap2} using a reference sequence that is indexed by the combination of the BWT index and the hash table. The purpose of this combination is to address the issue of random memory access while searching the BWT index, which is challenging for a GPU implementation. Both SOAP2 \cite{li2009soap2} and SOAP3 \cite{liu2012soap3} can support alignment with an edit distance threshold of up to four bp.
\newpage
Bowtie \cite{langmead2009ultrafast} follows the same concept of SOAP2. However, it also provides a backtracking step that favors high-quality alignments. It also uses a 'double BWT indexing' approach to avoid excessive backtracking by indexing the reference genome and its reversed version. Bowtie fails to align reads to a reference for an edit distance threshold of more than three bp.

The second category uses a hash table to index short seeds presented in either the read set (as in SHRiMP \cite{rumble2009shrimp}, Maq \cite{li2008mapping}, RMAP \cite{smith2008using}, and ZOOM \cite{lin2008zoom}) or the reference (as in most of the other modern mappers in this category). The idea of the hash table indexing can be tracked back to BLAST \cite{altschul1990basic}. Examples of this category include BFAST \cite{homer2009bfast}, BitMapper \cite{cheng2015bitmapper}, mrFAST with FastHASH \cite{xin2013accelerating}, mrsFAST \cite{hach2010mrsfast}, SHRiMP \cite{rumble2009shrimp}, SHRiMP2 \cite{david2011shrimp2}, RazerS \cite{weese2009razers}, Maq \cite{li2008mapping}, Hobbes \cite{ahmadi2011hobbes}, drFAST \cite{hormozdiari2011sensitive}, MOSAIK \cite{lee2014mosaik}, SOAP \cite{li2008soap}, Saruman \cite{blom2011exact} (GPU), ZOOM \cite{lin2008zoom}, and RMAP \cite{smith2008using}. Hash-based mappers build a very comprehensive but overly large candidate location pool and rely on seed filters and local alignment techniques to remove incorrect mappings from consideration in the verification step. Mappers in this category are able to find all correct mappings of a read, but waste computational resources for identifying and rejecting incorrect mappings. As a result, they are slower than BWT-FM-based mappers. Next, we explain mrFAST mapper as an example of this category.

mrFAST ($>$ version 2.5) \cite{xin2013accelerating} first builds a hash table to index fixed-length seeds (typically 10-13 bp) from the reference genome . It then applies a seed location filtering mechanism, called Adjacency Filter, on the hash table to reduce the false seed locations. It divides each query read into smaller fixed-length seeds to query the hash table for their associated seed locations. Given an edit distance threshold, Adjacency Filter requires \textit{N-E} seeds to exactly match adjacent locations, where \textit{N} is the number of the seeds and \textit{E} is the edit distance threshold. Finally, mrFAST tries to extend the read at each of the seed locations by aligning the read to the reference fragment at the seed location via Levenshtein edit distance \cite{levenshtein1966binary} with Ukkonen’s banded algorithm \cite{ukkonen1985algorithms}. One drawback of this seed filtering is that the presence of one or more substitutions in any seed is counted by the Adjacency Filter as a single mismatch. The effectiveness of the Adjacency Filter for substitutions and indels diminishes when \textit{E} becomes larger than 3 edits. 

A recent work in \cite{yorukoglu2016compressive} shows that by removing redundancies in the reference genome and also across the reads, seed-and-extend mappers can be faster than BWT-FM-based mappers. This space-efficient approach uses a similar idea presented in \cite{lin2008zoom}. A hybrid method that incorporates the advantages of each approach can be also utilized, such as BWA-MEM \cite{li2013aligning}.

\section{Acceleration of Read Mappers}
\begin{figure}
\includegraphics[width=\linewidth]{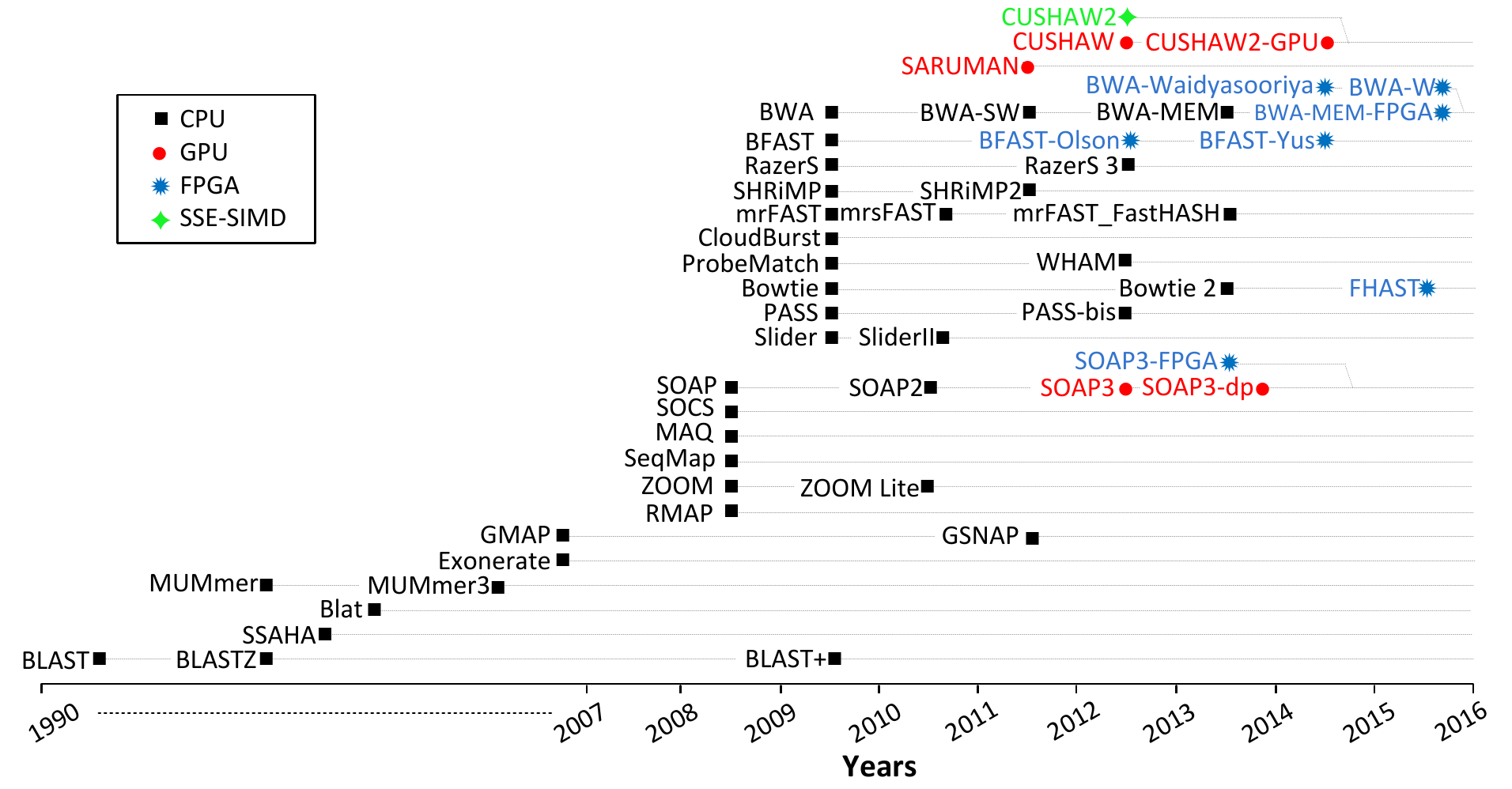}
\caption{Timeline of read mappers. CPU-based mappers are plotted in black, GPU accelerated mappers in red, FPGA accelerated mappers in blue and SSE-based mappers in green. Grey dotted lines connect related mappers (extensions or new versions). The names in the timeline are exactly as they appear in publications, except: SOAP3-FPGA \cite{arram2013reconfigurable}, BWA-MEM-FPGA \cite{houtgast2015fpga}, BFAST-Olson \cite{olson2012hardware}, BFAST-Yus \cite{sogabe2014fpga}, BWA-Waidyasooriya \cite{waidyasooriya2014fpga}, and BWA-W \cite{waidyasooriya2016hardware}.}
\label{fig:figure3timeline}
\end{figure}

A majority of read mappers are developed for machines equipped with the general-purpose central processing units (CPUs). As long as the gap between the CPU computing speed and the very large amount of sequencing data widens, CPU-based mappers become less favorable due to their limitations in accessing data \cite{banerjee2018asap, fei2018fpgasw, daily2016parasail, houtgast2015fpga, georganas2015meraligner, liu2015gswabe, waidyasooriya2014fpga, luo2013soap3, liu2013cudasw++, arram2013hardware}. To tackle this challenge, many attempts were made to accelerate the operations of read mapping. We survey in Figure \ref{fig:figure3timeline} the existing read mappers implemented in various acceleration platforms. FPGA-based read mappers often demonstrate one to two orders of magnitude speedups against their GPU-based counterparts \cite{fei2018fpgasw, waidyasooriya2016hardware}. Most of the existing works used hardware platforms to only accelerate the dynamic programming algorithms (e.g., Smith-Waterman algorithm \cite{smith1981textordfeminineidentification}), as these algorithms contributed significantly to the overall running time of read mappers. Most existing works can be divided into three main approaches: (1) Developing seed filtering mechanism to reduce the seed location list, (2) Accelerating the computationally expensive alignment algorithms using algorithmic development or hardware accelerators, and (3) Developing pre-alignment filtering heuristics to reduce the number of incorrect mappings before being examined by read alignment. We describe next each of these three acceleration efforts in detail.

\subsection{Seed Filtering}
The first approach to accelerate today’s read mapper is to filter the seed location list before performing read alignment. \newpage This is the basic principle of nearly all seed-and-extend mappers. Seed filtering is based on the observation that if two sequences are potentially similar, then they share a certain number of seeds. Seeds (sometimes called q-grams or k-mers) are short subsequences that are used as indices into the reference genome to reduce the search space and speed up the mapping process. Modern mappers extract short subsequences from each read and use them as a key to query the previously built large reference index database. The database returns the location lists for each seed. The location list stores all the occurrence locations of each seed in the reference genome. The mapper then examines the optimal alignment between the read and the reference segment at each of these seed locations. The performance and accuracy of seed-and-extend mappers depend on how the seeds are selected in the first stage. Mappers should select a large number of non-overlapping seeds while keeping each seed as infrequent as possible for full sensitivity \cite{xin2015optimal, kielbasa2011adaptive, xin2013accelerating}. There is also a significant advantage to selecting seeds with unequal lengths, as possible seeds of equal lengths can have drastically different levels of frequencies. Finding the optimal set of seeds from read sequences is challenging and complex, primarily because the associated search space is large and it grows exponentially as the number of seeds increases. There are other variants of seed filtering based on the pigeonhole principle \cite{cheng2015bitmapper, weese2012razers}, non-overlapping seeds \cite{xin2013accelerating}, gapped seeds \cite{egidi2013better, rizk2010gassst}, variable-length seeds \cite{xin2015optimal}, random permutation of subsequences \cite{lederman2013random}, or full permutation of all possible subsequences \cite{kim2018grim, kim2017grim, kim2017genome}.

\subsection{Accelerating Read Alignment}
The second approach to boost the performance of read mappers is to accelerate read alignment step. One of the most fundamental computational steps in most bioinformatics analyses is the detection of the differences/similarities between two genomic sequences. Edit distance and pairwise alignment are two approaches to achieve this step, formulated as approximate string matching \cite{navarro2001guided}. Edit distance approach is a measure of how much the sequences differ. It calculates the minimum number of edits needed to convert one sequence into the other. \newpage The higher the distance, the more different the sequences from one another. Commonly allowed edit operations include deletion, insertion, and substitution of characters in one or both sequences. Pairwise alignment is a way to identify regions of high similarity between sequences. Each application employs a different edit model (called scoring function), which is then used to generate an alignment score. The latter is a measure of how much the sequences are alike. Any two sequences have a single edit distance value but they can have several different alignments (i.e., ordered lists of possible edit operations and matches) with different alignment scores. Thus, alignment algorithms usually involve a backtracking step for providing the optimal alignment (i.e., the best arrangement of the possible edit operations and matches) that has the highest alignment score. Depending on the demand, pairwise alignment can be performed as global alignment, where two sequences of the same length are aligned end-to-end, or local alignment, where subsequences of the two given sequences are aligned. It can also be performed as semi-global alignment (called glocal), where the entirety of one sequence is aligned towards one of the ends of the other sequence. 

The edit distance and pairwise alignment approaches are non-additive measures \cite{calude2002additive}. This means that if we divide the sequence pair into two consecutive subsequence pairs, the edit distance of the entire sequence pair is not necessarily equivalent to the sum of the edit distances of the shorter pairs. Instead, we need to examine all possible prefixes of the two input sequences and keep track of the pairs of prefixes that provide an optimal solution. Enumerating all possible prefixes is necessary for tolerating edits that result from both sequencing errors \cite{fox2014accuracy} and genetic variations \cite{mckernan2009sequence}. Therefore, they are typically implemented as dynamic programming algorithms to avoid re-computing the edit distance of the prefixes many times. These implementations, such as Levenshtein distance \cite{levenshtein1966binary}, Smith-Waterman \cite{smith1981textordfeminineidentification}, and Needleman-Wunsch \cite{needleman1970general}, are still inefficient as they have quadratic time and space complexity (i.e., O(\textit{m}\textsuperscript{2}) for a sequence length of \textit{m}). Many attempts were made to boost the performance of existing sequence aligners. Despite more than three decades of attempts, the fastest known edit distance algorithm \cite{masek1980faster} has a running time of O(\textit{m}\textsuperscript{2}/$\log_2 m$) for sequences of length \textit{m}, which is still nearly quadratic \cite{backurs2017edit}. 

Therefore, more recent works tend to follow one of two key new directions to boost the performance of sequence alignment and edit distance implementations: (1) Accelerating the dynamic programming algorithms using hardware accelerators. (2) Developing filtering heuristics that reduce the need for the dynamic programming algorithms, given an edit distance threshold. \textbf{Hardware accelerators are becoming increasingly popular for speeding up the computationally-expensive alignment and edit distance algorithms \cite{al2017survey, aluru2014review, ng2017reconfigurable, sandes2016parallel}}. Hardware accelerators include multi-core and SIMD (single instruction multiple data) capable central processing units (CPUs), graphics processing units (GPUs), and field-programmable gate arrays (FPGAs). The classical dynamic programming algorithms are typically accelerated by computing only the necessary regions (i.e., diagonal vectors) of the dynamic programming matrix rather than the entire matrix, as proposed in Ukkonen’s banded algorithm \cite{ukkonen1985algorithms}. The number of the diagonal bands required for computing the dynamic programming matrix is 2\textit{E}+1, where \textit{E} is a user-defined edit distance threshold. The banded algorithm is still beneficial even with its recent sequential implementations as in Edlib \cite{vsovsic2017edlib}. The Edlib algorithm is implemented in C for standard CPUs and it calculates the banded Levenshtein distance. Parasail \cite{daily2016parasail} exploits both Ukkonen’s banded algorithm and SIMD-capable CPUs to compute a banded alignment for a sequence pair with user-defined scoring matrix and affine gap penalty. SIMD instructions offer significant parallelism to the matrix computation by executing the same vector operation on multiple operands at once. 

Multi-core architecture of CPUs and GPUs provides the ability to compute alignments of many sequence pairs independently and concurrently \cite{georganas2015meraligner, liu2015gswabe}. GSWABE \cite{liu2015gswabe} exploits GPUs (Tesla K40) for a highly-parallel computation of global alignment with affine gap penalty. CUDASW++ 3.0 \cite{liu2013cudasw++} exploits the SIMD capability of both CPUs and GPUs (GTX690) to accelerate the computation of the Smith-Waterman algorithm with affine gap penalty. CUDASW++ 3.0 provides only the optimal score, not the optimal alignment (i.e., no backtracking step). \newpage Other designs, for instance FPGASW \cite{fei2018fpgasw}, exploit the very large number of hardware execution units in FPGAs (Xilinx VC707) to form a linear systolic array \cite{kung1982systolic}. Each execution unit in the systolic array is responsible for computing the value of a single entry of the dynamic programming matrix. The systolic array computes a single vector of the matrix at a time. The data dependencies between the entries restrict the systolic array to computing the vectors sequentially (e.g., top-to-bottom, left-to-right, or in an anti-diagonal manner). FPGA acceleration platform can also provide more speedup to big-data computing frameworks -such as Apache Spark- for accelerating BWA-MEM \cite{li2013aligning}. By this integration, Chen et al. \cite{chen2016apache} achieve 2.6x speedup over the same cloud-based implementation but without FPGA acceleration \cite{chen2015cs}. FPGA accelerators seem to yield the highest performance gain compared to the other hardware accelerators \cite{banerjee2018asap, fei2018fpgasw, waidyasooriya2016hardware, waidyasooriya2014fpga}. However, many of these efforts either simplify the scoring function, or only take into account accelerating the computation of the dynamic programming matrix without providing the optimal alignment as in \cite{nishimura2017accelerating, chen2014accelerating, liu2013cudasw++}. Different scoring functions are typically needed to better quantify the similarity between two sequences \cite{wang2011comparison, henikoff1992amino}. The backtracking step required for the optimal alignment computation involves unpredictable and irregular memory access patterns, which poses a difficult challenge for efficient hardware implementation. Comprehensive surveys on hardware acceleration for computational genomics appeared in \cite{al2017survey, aluru2014review, ng2017reconfigurable, sandes2016parallel, canzar2017short}

\subsection{False Mapping Filtering}
The third approach to accelerate read mapping is to incorporate a pre-alignment filtering technique within the read mapper, before read alignment step. This filter is responsible for quickly excluding incorrect mappings in an early stage (i.e., as a pre-alignment step) to reduce the number of false mappings (i.e., mappings that have more edits than the user-defined threshold) that must be verified via dynamic programming. Existing filtering techniques include the so-called shifted Hamming distance (SHD) \cite{xin2015shifted}, which we explain next.

\subsubsection{Shifted Hamming Distance (SHD)}
SHD enables pre-alignment filtering with the existence of indels and substitutions. Instead of building a single bit-vector using a pairwise comparison as Hamming distance does, SHD builds 2\textit{E}+1 bit-vectors, where \textit{E} is the user-defined edit distance threshold. 
This is similar to the Ukkonen’s banded algorithm \cite{ukkonen1985algorithms}. Each bit-vector is built by gradually shifting the read sequence and then performing a pairwise comparison. The shifting process is inevitable in order to skip the deleted (or inserted) character and examine the subsequent matches. SHD merges all masks using bitwise AND operation. Due to the use of AND operation, a zero (i.e., pairwise match) at any position in the 2\textit{E}+1 masks leads to a ‘0’ in the resulting output of the AND operation at the same position. The last step is to count the positions that have a value other than ‘0’. SHD decides if the mapping is correct based on whether the number of the mismatches exceeds the edit distance threshold or not. SHD heavily relies on bitwise operations such as shift, XOR, and AND. This makes SHD suitable for bitwise hardware implementations (e.g., FPGAs and SIMD-enabled CPUs). 

Our crucial observation is that SHD examines each mapping, throughout the filtering process, by performing expensive computations unnecessarily; as SHD uses the same amount of computation regardless the type of edit. SHD is also implemented using Intel SSE, which limits the supported read length up to only 128 bp (due to SIMD register size). The filtering mechanism of SHD also introduces inaccuracy in its filtering decision as we investigate and demonstrate in Chapter 3 and in our experimental evaluation, Chapter 9.

\section{Summary} 
We survey in this chapter the existing key directions that aim at accelerating all or part of the operations of modern read mappers. We analyze these attempts and provide the pros and cons of each direction. \newpage We present three main acceleration approaches, including (1) seed filtering, (2) accelerating the dynamic programming algorithm, (3) pre-alignment filtering. In Figure \ref{fig:figure1incorrectmappings}, we illustrate that the state-of-the-art mapper mrFAST with FastHASH \cite{xin2013accelerating} generates more than 90\% of the potential mappings as incorrect ones, although it implements a seed filtering mechanism (Adjacency Filter) and SIMD-accelerated banded Levenshtein edit distance algorithm. This demonstrates that the development of a fundamentally new, fast, and efficient pre-alignment filter is the utmost necessity. Note that there is still no work, to best of our knowledge, on specialized hardware acceleration of pre-alignment filtering techniques.

%% file: chapter3.tex
\chapter{Understanding and Improving Pre-alignment Filtering Accuracy}
In this chapter, we firstly provide performance metrics used to evaluate pre-alignment filtering techniques. The essential performance metrics are filtering speed and filtering accuracy. We secondly study the causes of filtering inaccuracy of the state-of-the-art pre-alignment filter, SHD \cite{xin2015shifted}, aiming at eliminating them. We find four key causes and provide a detailed investigation along with examples on these inaccuracy sources. This is the first work to comprehensively assess the filtering inaccuracy of the SHD algorithm \cite{xin2015shifted} and provide recommendations for desirable improvements.

\section{Pre-alignment Filter Performance Metrics}
An ideal pre-alignment filter should be both fast and accurate in rejecting the incorrect mappings. Meanwhile, it should also preserve all correct mappings. Incorrect mapping is defined as a sequence pair that differs by more than the edit distance threshold. Correct mapping is defined as a sequence pair that has edits less than or equal to the edit distance threshold. \newpage Next, we describe the performance metrics that are necessary to evaluate the speed and accuracy of existing and future pre-alignment filtering algorithms.

\subsection{Filtering Speed}
The filtering speed is defined as the time spent by the pre-alignment filter in examining all the incoming mappings. We always want to increase the speed of the pre-alignment filter to compensate the computation overhead introduced by its filtering technique. 

\subsection{Filtering Accuracy}
\subsubsection{False Accept Rate} 
The false accept rate (or false positive rate) is the ratio between the incorrect mappings that are falsely accepted by the filter and the incorrect mappings that are rejected by optimal read alignment algorithm. Similarly, a mapping is considered as a false positive if read alignment accepts it but pre-alignment filter rejects it. We always want to minimize the false accept rate.

\subsubsection{True Accept Rate}
The true accept rate (or true positive rate) is the ratio between the correct mappings that are accepted by the filter and the correct mappings that are accepted by optimal read alignment algorithm. The true accept rate should always equal to 1. 

\subsubsection{False Reject Rate}
The false reject rate (or false negative rate) is the ratio between the correct mappings that are rejected by the filter and the correct mappings that are accepted by optimal read alignment algorithm. The false reject rate should always equal to 0. 

\subsubsection{True Reject Rate}
The true reject rate (or true negative rate) is the ratio between the incorrect mappings that are rejected by the filter and the incorrect mappings that are rejected by optimal read alignment algorithm. We always want to maximize the true reject rate. In fact, the true reject rate is inversely proportional to the false accept rate. However, they can be equivalent in ratio in case of all mappings are correct and accepted by the filter.

\subsection{End-to-End Alignment Speed}
Can very fast filter with high false accept rate be better than more accurate filter at the cost of its speed? The answer to this question is not trivial because both speed and accuracy contribute to the overall speed of read alignment. The only way to answer this question is evaluate the effect of such filter on the overall speed of read alignment step. Thus, we need to evaluate the end-to-end alignment speed. This includes the integration of pre-alignment filter with read alignment step and evaluate the acceleration rate. Another crucial observation is that pre-alignment filter applies heuristic approach, which can be optimal for some alignment cases while it fails in other cases. Thus, filter that performs best for specific read set and edit distance threshold may not perform well for other read sets and edit distance thresholds. The user-defined edit distance threshold, \textit{E}, is usually less than 5\% of the read length \cite{cheng2015bitmapper, xin2015shifted, hatem2013benchmarking, ahmadi2011hobbes}. 

\section{On the False Filtering of SHD algorithm}
In this section, we investigate the potential causes of filtering inaccuracy that are introduced by the state-of-the-art filter, SHD \cite{xin2015shifted} (we describe the algorithm in Chapter 2). We also provide examples that illustrate each of these causes. Adding an additional fast filtering heuristic before the verification step in a read mapper can be beneficial. But, such a filter can be easily worthless if it allows a high false accept rate. Even though the incorrect mappings that pass SHD are discarded later by the read alignment step (as it has zero false accept rate and zero false reject rate), they can dramatically increase the execution time of read mapper by causing a mapping to be examined twice unnecessarily by both the filtering step as well as read alignment step. Below, we describe four major sources of false positives that are introduced by the filtering strategy of SHD.

\subsection{Random Zeros}
The first source of false accept rate of SHD \cite{xin2015shifted} is the random zeros that appear in the individual shifted Hamming mask. Although they result from a pairwise comparison between a shifted read and a reference segment, we refer to them as random zeros because they are sometimes meaningless and are not part of the correct alignment. SHD ANDs all shifted Hamming masks together with the idea that all ‘0’s in the individual Hamming masks propagate to the final bit-vector, thereby preserving the information of individual matching subsequences. Due to the use of AND operation, a zero at any position in the 2\textit{E}+1 Hamming masks leads to a ‘0’ in the resulting final bit-vector at the same position. Hence, even if some Hamming masks show a mismatch at that position, a zero in some other masks leads to a match (‘0’) at the same position. This tends to underestimate the actual number of edits and eventually causes some incorrect mappings to pass. To fix this issue, SHD proposes the so-called \textit{speculative removal of short-matches (SRS)} before ANDing the masks, which flips short streaks of ‘0’s in each mask into ‘1’s such that they do not mask out ‘1’s in other Hamming masks. 

\begin{figure}
\includegraphics[width=\linewidth]{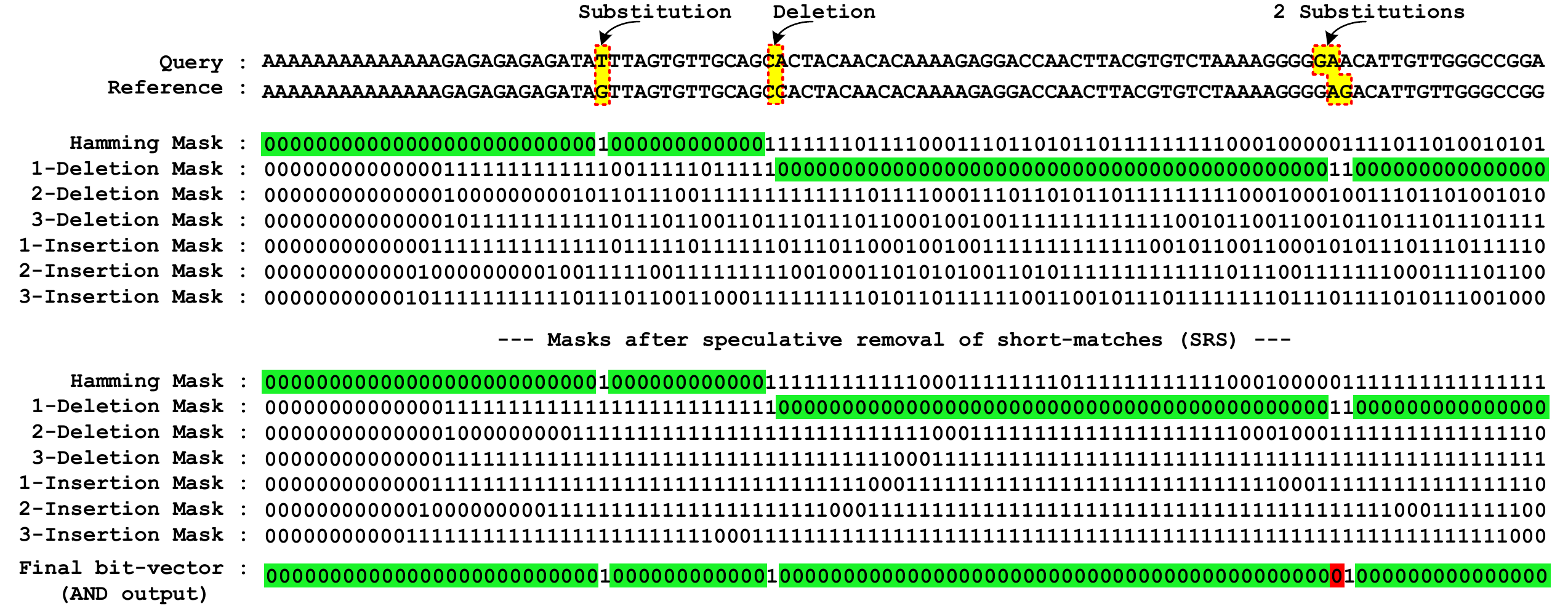}
\caption{An example of a mapping with all its generated masks, where the edit distance threshold (\textit{E}) is set to 3. The green highlighted subsequences are part of the correct alignment. The red highlighted bit in the final bit-vector is a wrong alignment provided by SHD. The correct alignment (highlighted in yellow) shows that there are three substitutions and a single deletion, while SHD detects only two substitutions and a single deletion.}
\label{fig:figure4SHD}
\end{figure}

\begin{figure}
\includegraphics[width=\linewidth]{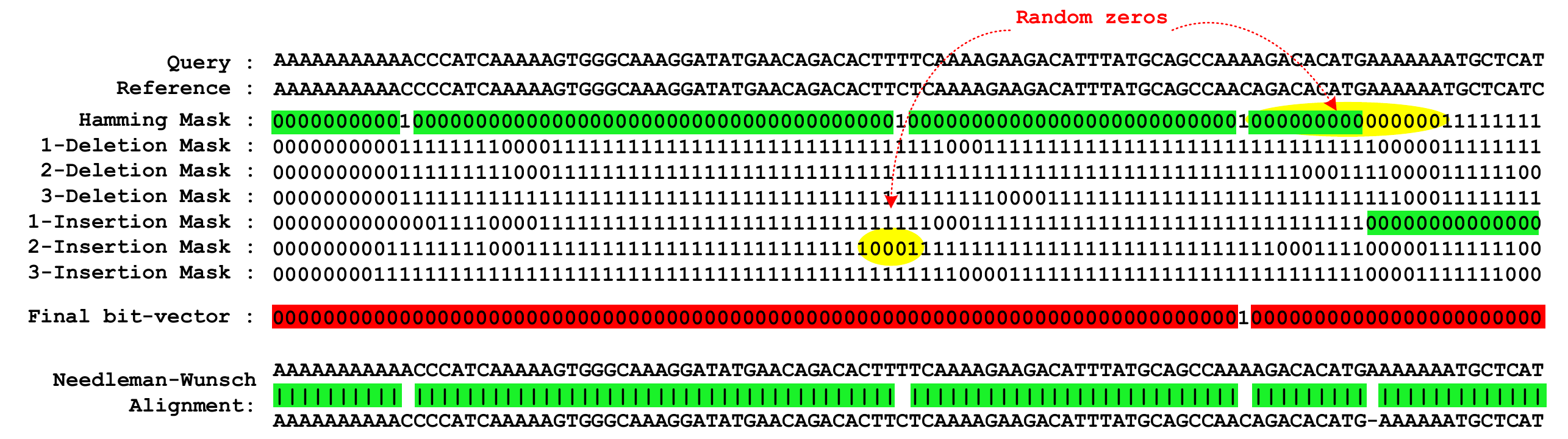}
\caption{Examples of an incorrect mapping that passes the SHD filter due to the random zeros. While the edit distance threshold is 3, a mapping of 4 edits (as examined at the end of the figure by Needleman-Wunsch algorithm) passes as a falsely accepted mapping.}
\label{fig:figure5SHD}
\end{figure}

We illustrate this method in Figure \ref{fig:figure4SHD}. The number of zeros to be amended (SRS threshold) is set by default to two. That is, bit streams such as 101, 1001 are replaced with 111 and 1111, respectively. The 2\textit{E}+1 masks contain other zeros that are part of the correct alignment. For example, Figure \ref{fig:figure4SHD} shown a segment of consecutive matches in one-step right-shifted mask. This segment indicates that there is a single deletion that occurred in the read sequence. Unlike these meaningful zeros, random amended zeros can be anywhere in the masks except the two ends of each mask. However, the length and the position of these zeros are unpredictable. They can have any length that makes the SRS method ineffective at handling these random zeros. There is no clear theory behind the exact SRS threshold to be used to eliminate such zeros. SRS successfully reduce some of the falsely accepted mappings, but it also introduces its own source of falsely accepted mappings. Choosing a small SRS threshold helps, but does not provide any guarantee, to get rid of some of these random zeros. Choosing a larger SRS threshold can be risky, since, with such a large threshold, SHD might no longer be able to distinguish whether any streak of consecutive zeros is generated by random chance or it is part of the correct alignment. This results in SHD ignoring most of the exact matching subsequences and causes an all-‘1’ final bit-vector. \newpage In Figure \ref{fig:figure5SHD}, we provide an example where random zeros dominate and lead to a zero in the final bit-vector at their corresponding locations. SRS can address the inaccuracy caused by the random 3-bit zeros, which are highlighted by the left arrow, using an SRS threshold of 3. However, SRS is still unable to solve the inaccuracy caused by the 15-bit zeros that are highlighted by the right arrow. This is due to the fact that the 15-bit zeros are part of the correct alignment and hence amending them to ones can introduce more falsely accepted mappings.

\subsection{Conservative Counting}
\begin{figure}
\includegraphics[width=\linewidth]{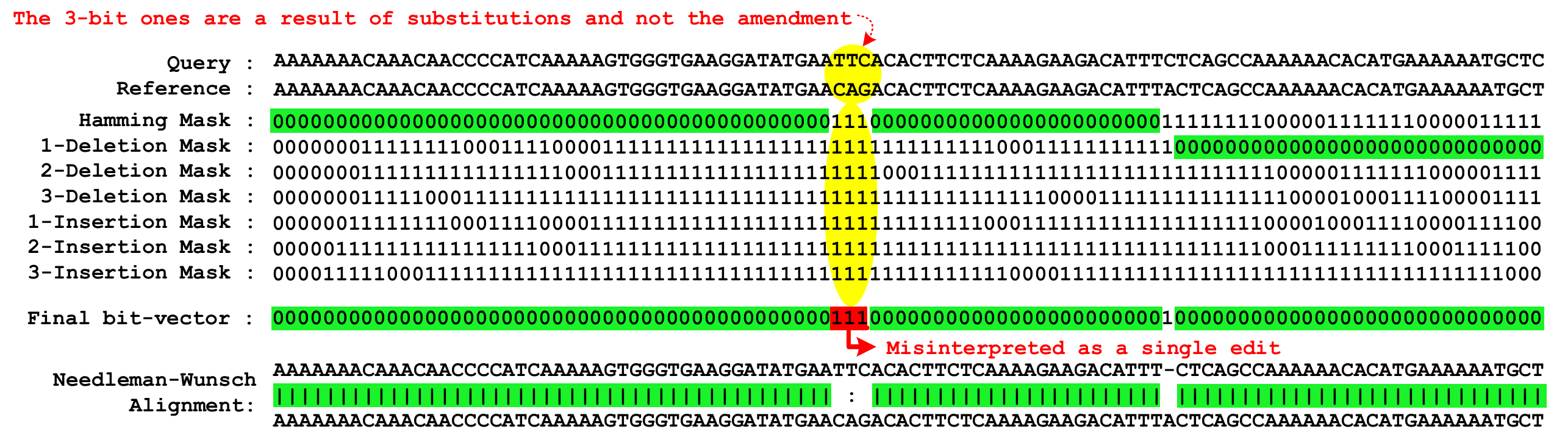}
\caption{An example of an incorrect mapping that passes the SHD filter due to conservative counting of the short streak of ‘1’s in the final bit-vector.}
\label{fig:figure6SHD}
\end{figure}

The second source of high false accept rate of SHD \cite{xin2015shifted} is related to the way in which SHD counts the edits in the final bit-vector. Amending short streaks of ‘0’s to ‘1’s could cause correct mappings to be mistakenly filtered out, as it may produce multiple ones in the final bit-vector. To ensure that it does not overcount the number of edits, SHD always assumes the streaks of ‘1’s in the final bit-vector as a side effect of the SRS amendment, and counts only the minimum number of edits that potentially generate such a streak of ‘1’s. The total number of edits reported by SHD can be much smaller than the actual number of edits. \newpage For instance, as illustrated in Figure \ref{fig:figure6SHD}, three consecutive substitutions render a streak of three ‘1’s in the final bit-vector. But since SHD always assumes the middle ‘1’ is the result of an amended ‘0’ by SRS, SHD will only consider the streak of three ‘1’s as a single edit and let it pass, even if the edit distance threshold is less than three. 

\subsection{Leading and Trailing Zeros}
\begin{figure}
\includegraphics[width=\linewidth]{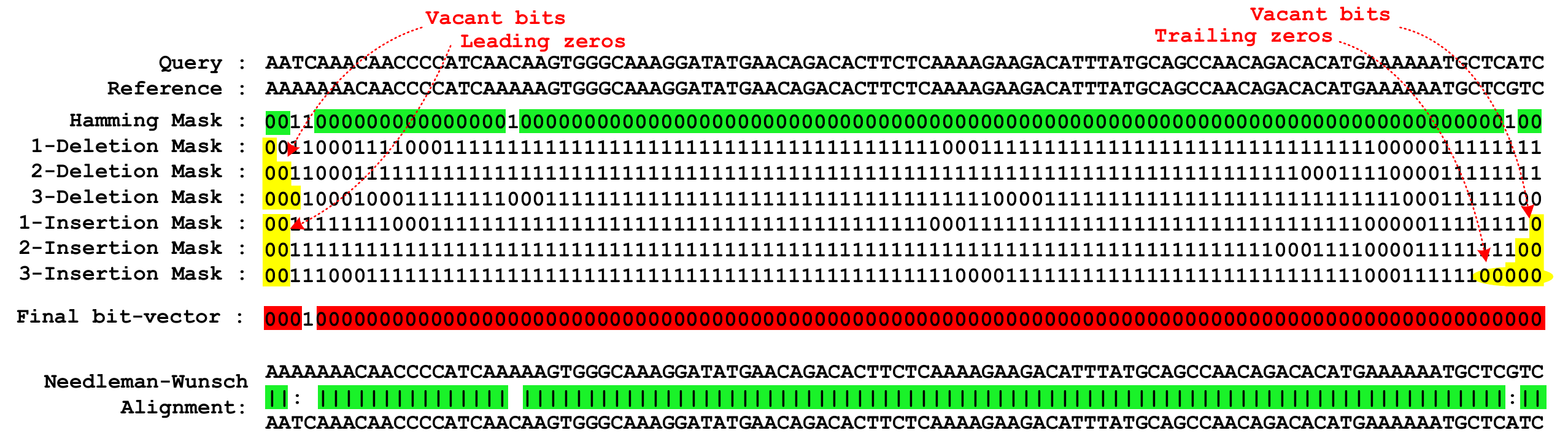}
\caption{Examples of an invalid mapping that passes the SHD filter due to the leading and trailing zeros. We use an edit distance threshold of 3 and an SRS threshold of 2. While the regions that are highlighted in green are part of the correct alignment, the wrong alignment provided by SHD is highlighted in red. The yellow highlighted bits indicate a source of falsely accepted mapping.}
\label{fig:figure7SHD}
\end{figure}

The third source of high false accept rate of SHD \cite{xin2015shifted} is the streaks of zeros that are located at any of the two ends of each mask. Hence we refer to them as leading and trailing zeros. These streaks of zeros can be in two forms: (1) the vacant bits that are caused by shifting the read against the reference segment and (2) the streaks of zeros that are not vacant bits. SHD generates 2\textit{E}+1 masks using arithmetic left-shift and arithmetic right-shift operations. For both the left and right directions, the right-most and the left-most vacant bits, respectively, are filled with ‘0’s. The number of vacant zeros depends on the number of shifted steps for each mask, which is at most equal to the edit distance threshold. The second form of the leading and trailing zeros is the zeros that are located at the two ends of the Hamming masks and are not vacant zeros. These streaks of zeros result from the pairwise comparison (i.e., bitwise XOR). They differ from the vacant bits in that their length is independent of the edit distance threshold. \newpage The main issue with both forms of leading and trailing zeros is that they always dominate, even if some Hamming masks show a mismatch at that position (due to the use of the AND operation). This gives the false impression that the read and the reference have a smaller edit distance, even when they differ significantly, as explained in Figure \ref{fig:figure7SHD}. SRS does not address the inaccuracy caused by the leading and trailing zeros by amending such zeros to ones, due to two reasons: (1) the number of these consecutive zeros is not fixed and thus they can be longer than the SRS threshold, (2) these consecutive zeros are not surrounded by ones and hence even if SRS threshold is greater than two bits, they are not eligible to be amended.

\subsection{Lack of Backtracking}
The last source of false accept rate of in SHD \cite{xin2015shifted} is the inability of SHD to backtrack (after generating the final bit-vector) the location of each long identical subsequence (i.e., the mask that originates the identical subsequence), which is part of the correct alignment. The source of each subsequence provides a key insight into the actual number of edits between each two subsequences. That is, if a subsequence is located in a 2-step right shifted mask, it should indicate that there are two deletions before this subsequence.

\begin{figure}
\includegraphics[width=\linewidth]{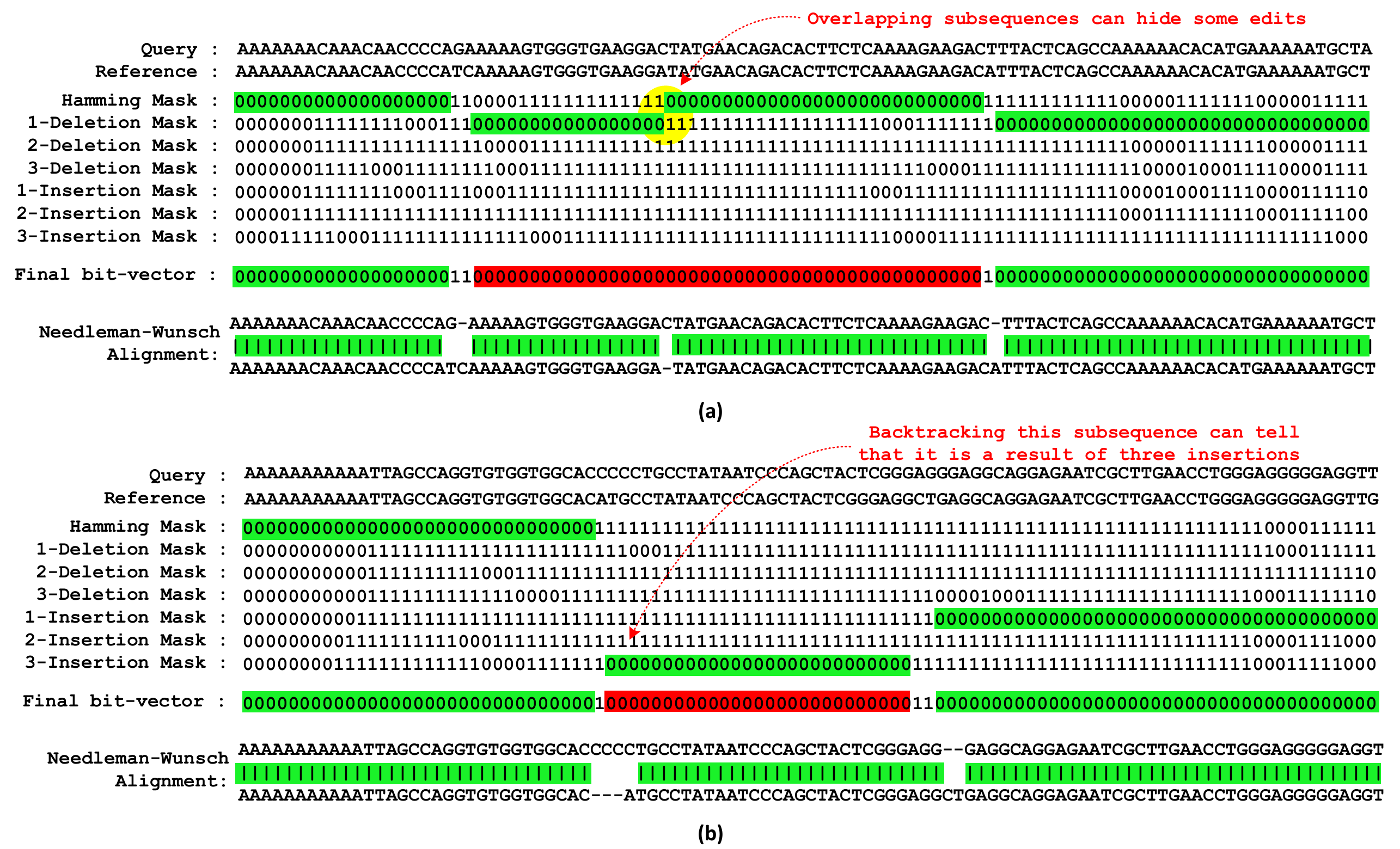}
\caption{Examples of incorrect mappings that pass the SHD filter due to (a) overlapping identical subsequences, and (b) lack of backtracking.}
\label{fig:figure8SHD}
\end{figure}

 SHD does not relate this important fact to the number of edits in the final bit-vector. The lack of backtracking causes two types of falsely accepted mapping: (1) the first type appears clearly when two of the identical subsequences, in the individual Hamming masks, are overlapped or nearly overlapped, (2) the second type happens when the identical subsequences come from different Hamming masks. The issue with the first type (i.e., overlapping subsequences) is the fact that they appear as a single identical subsequence in the final bit-vector, due to the use of AND operation. An example of this scenario is given in Figure \ref{fig:figure8SHD} (a). This tends to hide some of the edits and eventually causes some invalid mappings to pass. The second type of false positives caused by the lack of backtracking happens, for example, when an identical subsequence comes from the first Hamming mask (i.e., with no shift) and the next identical subsequence comes from the 3-step left shifted mask. This scenario reveals that the number of edits between the two subsequences should not be less than three insertions. \newpage However, SHD inaccurately reports it as a single edit (due to ANDing all Hamming masks without backtracking the source of each streak of zeros), as illustrated in Figure \ref{fig:figure8SHD} (b). Keeping track of the source mask of each identical subsequence prevents such false positives and helps to reveal the correct number of edits.

\section{Discussion on Improving the Filtering Accuracy}
In this section, we provide our own observations and recommendations based on our comprehensive accuracy analysis of SHD filter \cite{xin2015shifted}. We make two crucial observations. (1) The \textbf{first observation} is that handling the short streaks of ‘0’s (i.e., using the SRS method that we discuss above) is indeed inefficient. These “noisy” streaks do not have determined properties, as their length and number are unpredictable (random-like). They introduce their own sources of falsely accepted mappings and do not contribute any useful information. Therefore, future filtering strategies should avoid processing such short streaks of ‘0’s. (2) The \textbf{second observation} is that the correct (desired) alignment always contains all the longest non-overlapping identical subsequences. This turns our attention to focusing on the long matches (that are highlighted in green in all previous figures, i.e., Figure \ref{fig:figure4SHD} to Figure \ref{fig:figure8SHD}) in each Hamming mask. We find that the long non-overlapping subsequences of consecutive zeros have two interesting properties. (1) There is an upper bound on their quantity. With the existence of \textit{E} edits, there are at most \textit{E}+1 non-overlapping identical subsequences shared between a pair of sequences. The total length of these non-overlapping subsequences is equal to \textit{m-E}, where \textit{m} is the read length. (2) The source mask of each long subsequence provides an insight into the number of edits between this subsequence and its preceding one. These two observations motivate us to incorporate long-match-awareness into the design of our filtering strategy and ignore processing noisy short matches. 

\section{Summary}
We identify four causes that introduce the filtering inaccuracy of the SHD \cite{xin2015shifted} algorithm, namely, the random zeros, conservative counting, leading and trailing zeros, and lack of backtracking. Based on these four sources of falsely accepted mapping, we observe that there are still opportunities for further improvements on the accuracy of the state-of-the-art filter, SHD, which we discuss next.

%% file: chapter4.tex
\chapter{The First Hardware Accelerator for Pre-Alignment Filtering}

In this chapter, we introduce a new FPGA-based accelerator architecture for hardware-aware pre-alignment filtering algorithms. To our knowledge, this is the first work that exploit reconfigurable hardware platforms to accelerate pre-alignment filtering. A fast filter designed on a specialized hardware platform can drastically expedite alignment by reducing the number of locations that must be verified via dynamic programming. This eliminates many unnecessary expensive computations, thereby greatly improving overall run time. 

\section{FPGA as Acceleration Platform}
We select FPGA as an acceleration platform for our proposed pre-alignment filtering algorithms, as its architecture offers large amounts of parallelism \cite{aluru2014review, herbordt2007achieving, trimberger2015three}. The use of FPGA as an acceleration platform can yield significant performance improvements, especially for massively parallel algorithms. \newpage FPGAs are the most commonly used form of reconfigurable hardware engines today in bioinformatics \cite{mcvicar2018fpga, alachiotis2017versatile, ng2017reconfigurable}, and their computational capabilities are greatly increasing every generation due to increased number of transistors on the FPGA chip. An FPGA chip can be programmed (i.e., configured) to include a very large number of hardware execution units that are custom-tailored to the problem at hand.

\section{Overview of Our Accelerator Architecture}
\begin{figure}
\includegraphics[width=\linewidth]{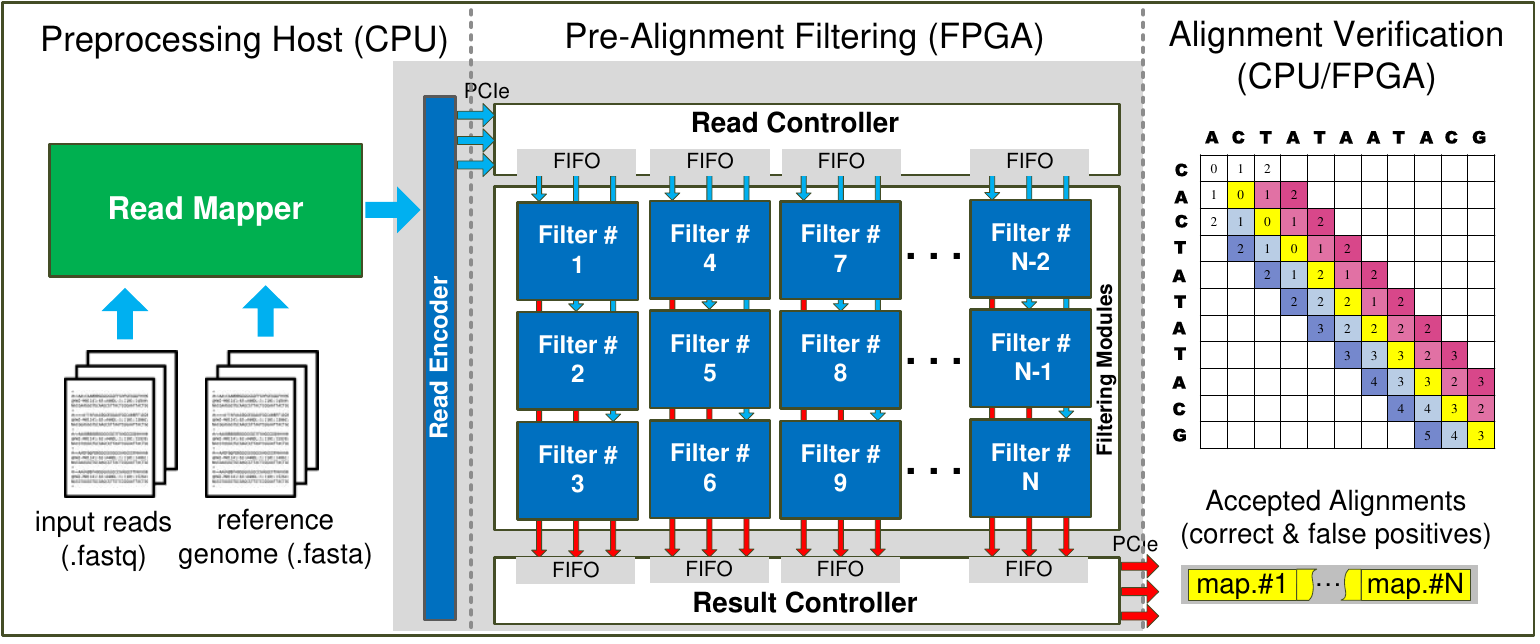}
\caption{Overview of our accelerator architecture.}
\label{fig:figure9GateKeeper}
\end{figure}

One of our aims is to accelerate our new pre-alignment filtering algorithms (that we describe in the next three chapters) by leveraging the capabilities and parallelism of FPGAs. To this end, we build our own hardware accelerator that consists of an FPGA engine as an essential component and a CPU. We present in Figure \ref{fig:figure9GateKeeper} the overall architecture of our FPGA-based accelerator. The CPU is responsible for acquiring and encoding the short reads and transferring the data to and from the FPGA. The FPGA engine is equipped with PCIe transceivers, Read Controller, Result Controller, and a set of filtering units that are responsible for examining the read alignment. The workflow of the accelerator starts with reading a repository of short reads and seed locations. All reads are then converted into their binary representation that can be understood by the FPGA engine. Encoding the reads is a preprocessing step and accomplished through a Read Encoder at the host before transmitting the reads to the FPGA chip. Next, the encoded reads are transmitted and processed in a streaming fashion through the fastest communication medium available on the FPGA board (i.e., PCIe). We design our system to perform alignment filtering in a streaming fashion: the accelerator receives a continual stream of short reads, examines each alignment in parallel with others, and returns the decision (i.e., a single bit of value ‘1’ for an accepted sequences and ‘0’ for a rejected sequences) back to the CPU instantaneously upon processing.

\subsection{Read Controller}
The Read Controller on the FPGA side is responsible for two main tasks. First, it permanently assigns the first data chunk as a reference sequence for all processing cores. Second, it manages the subsequent data chunks and distributes them to the processing cores. The first processing core receives the first read sequence and the second core receives the second sequence and so on, up to the last core. It iterates the data chunk management task until no more reads are left in the repository. 

\subsection{Result Controller}
Following similar principles as the Read Controller, the Result Controller gathers the output results of the filtering units. Both the Read Controller and the Result Controller preserve the original order of reads as in the repository (i.e., at the host). This is critical to ensure that each read will receive its own alignment filtering result. The results are transmitted back to the CPU side in a streaming fashion and then saved in the repository.

\section{Parallelization}
We design our hardware accelerator to exploit the large amounts of parallelism offered by FPGA architectures \cite{aluru2014review, herbordt2007achieving, trimberger2015three, assaad2011fpga,alser2018accelerating,alser2013wide, assaad2012design, alser2012design, alser2011design}. We take advantage of the fact that alignment filtering of one read is inherently independent of filtering of another read. We therefore can examine many reads in a parallel fashion. In particular, instead of handling each read in a sequential manner, as CPU-based filters (e.g., SHD) do, we can process a large number of reads at the same time by integrating as many hardware filtering units as possible (constrained by chip area) in the FPGA chip. Each filtering unit is a complete alignment filter and can handle a single read at a time. Our hardware accelerator contains large number of filtering units that their number can be configured by the user. Each filtering unit provides pre-alignment filtering individually from all other units. We use the term “filtering unit” in this work to refer to the entire operation of the filtering process involved. Filtering units are part of our architecture and are unrelated to the term “CPU core” or “thread”.

\begin{figure}
\includegraphics[width=\linewidth]{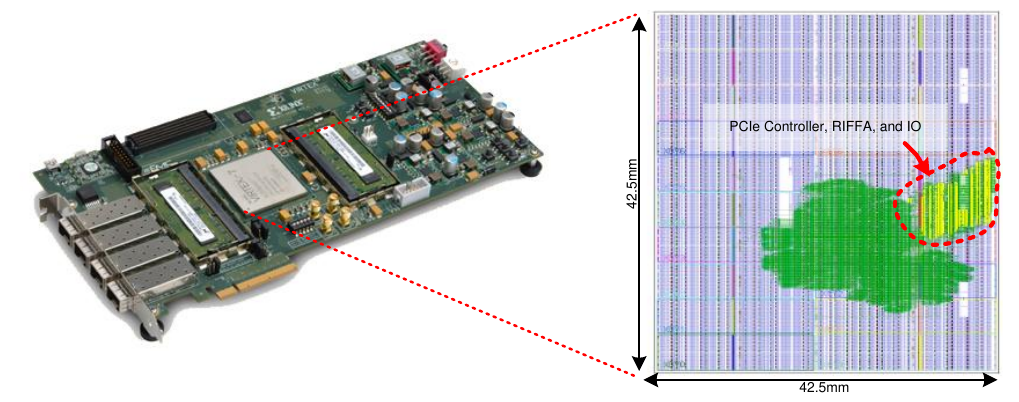}
\caption{Xilinx Virtex-7 FPGA VC709 Connectivity Kit and chip layout of our implemented accelerator.}
\label{fig:figure14GateKeeper}
\end{figure}

\section{Hardware Implementation}
Our hardware implementation of our accelerator is independent from specific FPGA-platform as it does not rely on any vendor-specific computing elements (e.g., intellectual property cores). However, each FPGA board has different features and hardware capabilities that can directly or indirectly affect the performance and the data throughput of the design. In fact, the number of filtering units is determined by the maximum data throughput and the available FPGA resources. We use a Xilinx Virtex 7 VC709 board \cite{virtex7fpga} to implement our accelerator architecture. We build the FPGA design with Vivado 2015.4 in synthesizable Verilog. We preset the chip layout of out hardware accelerator in Figure \ref{fig:figure14GateKeeper}. The maximum operating frequency of our accelerator and the VC709 board is 250 MHz. At this frequency, we observe a data throughput of nearly 3.3 GB/s, which corresponds to ~13.3 billion bases per second. This nearly reaches the peak throughput of 3.64 GB/s provided by the RIFFA \cite{jacobsen2015riffa} communication channel that feeds data into the FPGA using Gen3 4-lane PCIe.

\section{Summary}
We introduce in this chapter a new hardware accelerator architecture that exploit the large amounts of parallelism offered by FPGA architectures to boost the performance of our pre-alignment filters. Our hardware accelerator processes the pre-alignment filtering for each sequence pair independently from each another. We therefore can examine many reads in a parallel fashion. We build the hardware architecture of our hardware accelerator using many hardware filtering units, where each filtering unit is a complete pre-alignment filter and can handle a single read at a time. To take full advantage of the capabilities and parallelism of our FPGA accelerator, each pre-alignment filtering unit needs to be designed and implemented using FPGA-supported operations such as bitwise operations, bit shifts, and bit count. Next, we discuss our proposed pre-alignment filters that can be included in our FPGA accelerator as a filtering unit.

%% file: chapter5.tex
\chapter{GateKeeper: Fast Hardware Pre-Alignment Filter}
In this chapter, we introduce a new FPGA-based fast alignment filtering technique (called GateKeeper) that acts as a pre-alignment step in read mapping. Our filtering technique improves and accelerates the state-of-the-art SHD filtering algorithm \cite{xin2015shifted} using new mechanisms and FPGAs. 

\section{Overview}
Our new filtering algorithm has two properties that make it suitable for an FPGA-based implementation: (1) it is highly parallel, (2) it heavily relies on bitwise operations such as shift, XOR, and AND. Our architecture discards the incorrect mappings from the candidate mapping pool in a streaming fashion – data is processed as it is transferred from the host system. Filtering the mappings in a streaming fashion gives the ability to integrate our filter with any mapper that performs alignment, such as Bowtie 2 \cite{langmead2012fast} and BWA-MEM \cite{li2013aligning}. Our current filter implementation relies on several optimization methods to create a robust and efficient filtering approach. \newpage At both the design and implementation stages, we satisfy several requirements: (1) Ensuring a lossless filtering algorithm by preserving all correct mappings. (2) Supporting both Hamming distance and edit distance. (3) Examining the alignment between a read and a reference segment in a fast and efficient way (in terms of execution time and required resources). 

\section{Methods}
\begin{figure}
\centering
\includegraphics[width=10cm,keepaspectratio]{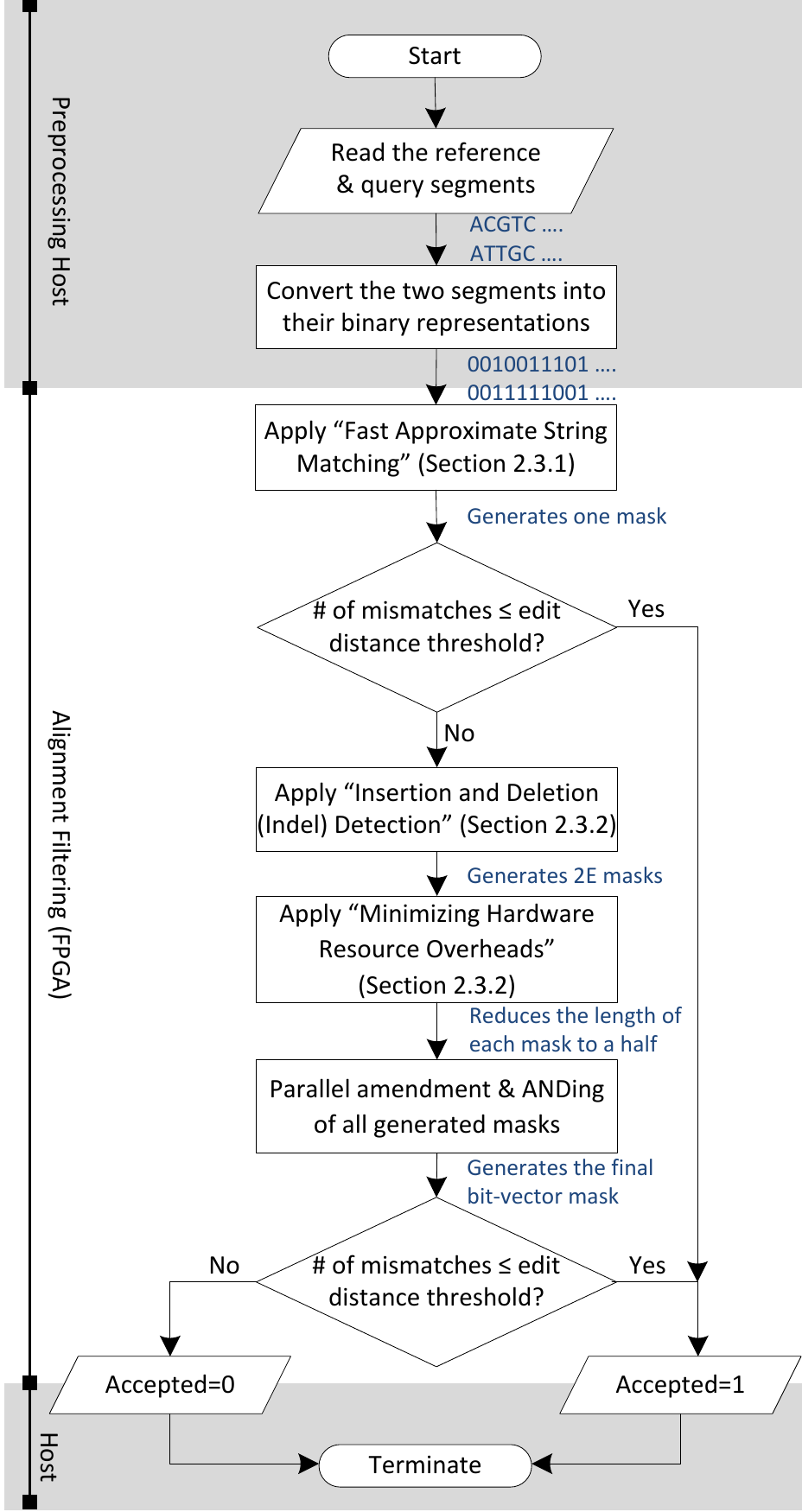}
\caption{A flowchart representation of the GateKeeper algorithm.}
\label{fig:figure10GateKeeper}
\end{figure}

Our primary purpose is to enhance the state-of-the-art SHD alignment filter such that we can greatly accelerate pre-alignment by taking advantage of the capabilities and parallelism of FPGAs. To achieve our goal, we design an algorithm inspired by SHD to reduce both the utilized resources and the execution time. These optimizations enable us to integrate more filtering units within the FPGA chip and hence examine many mappings at the same time. We present three new methods that we use in each GateKeeper filtering unit to improve execution time. Our first method introduces a new algorithmic method for performing alignment very rapidly compared to the original SHD. This method provides: (1) fast detection for exact matching alignment and (2) handling of one or more base-substitutions. Our second method supports calculating the edit distance with a new, very efficient hardware design. Our third method addresses the problem of hardware resource overheads introduced due to the use of FPGA as an acceleration platform. All methods are implemented within the hardware filtering unit of our accelerator (see Chapter 4) and thus are performed highly efficiently. We present a flowchart representation of all steps involved in our algorithm in Figure \ref{fig:figure10GateKeeper}. Next, we describe the three new methods.

\subsection{Method 1: Fast Approximate String Matching}
We first discuss how to examine a mapping with a given Hamming distance threshold, and later extend our solution to support edit distance. \newpage Our first method aims to quickly detect the obviously-correct alignments that contain no edits or only few substitutions (i.e., less than the user-defined threshold). If the first method detects a correct alignment, then we can skip the other two methods but we still need the optimal alignment algorithms. A read is mappable if the Hamming distance between the read and its seed location does not exceed the given Hamming distance threshold. Hence, the first step is to identify all bp matches by calculating what we call a Hamming mask. The Hamming mask is a bit-vector of ‘0’s and ‘1’s representing the comparison of the read and the reference, where a ‘0’ represents a bp match and a ‘1’ represents a bp mismatch. We need to count only occurrences of ‘1’ in the Hamming mask and examine whether their total number is equal to or less than the user-defined Hamming distance threshold. If so, the mapping is considered to be valid and the read passes the filter. Similarly, if the total number of ‘1’ is greater than the Hamming distance threshold then we cannot be certain whether this is because of the high number of substitutions, or there exist insertions and/or deletions; hence, we need to follow the rest of our algorithm. Our filter can detect not only substitutions but also insertions and deletions in an efficient way, as we discuss next.

\subsection{Method 2: Insertion and Deletion (Indel) Detection}
\begin{figure}
\centering
\includegraphics[width=7cm,keepaspectratio]{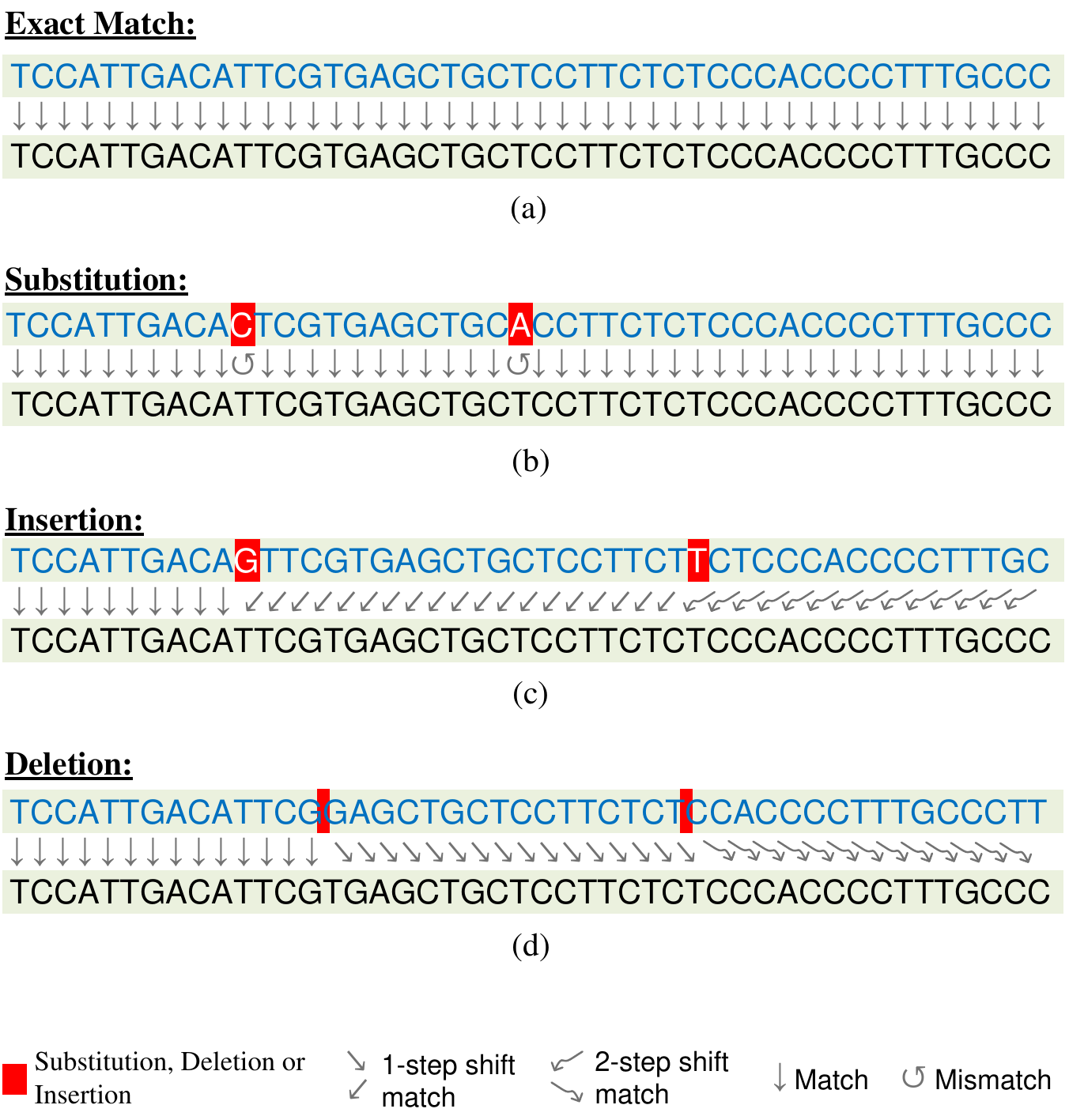}
\caption{An example showing how various types of edits affect the alignment of two reads. In (a) the upper read exactly matches the lower read and thus each base exactly matches the corresponding base in the target read. (b) shows base-substitutions that only affect the alignment at their positions. (c) and (d) demonstrate insertions and deletions, respectively. Each edit has an influence on the alignment of all the subsequent bases.}
\label{fig:figure11GateKeeper}
\end{figure}

Our indel detection algorithm is inspired by the original SHD algorithm presented in \cite{xin2015shifted}. If the substitution detection rejects an alignment, then GateKeeper checks if an insertion or deletion causes the violation (i.e., high number of edits). Figure \ref{fig:figure11GateKeeper} illustrates the effect of occurrence of edits on the alignment process. If there are one or more base-substitutions or the alignment is exactly matching, the matching and mismatching regions can be accurately determined using Hamming distance. It also helps detect the matches that are located before the first indel. However, this mask is already generated as part of the first method of the algorithm (i.e., Fast Approximate String Matching). On the other hand, each insertion and deletion can shift multiple trailing bases and create multiple edits in the Hamming mask. Thus, our indel detection method identifies whether the alignment locations of a read are valid, by shifting individual bases. 

We need to perform \textit{E} incremental shifts to the right (or left) direction to detect any read that has E deletions (or insertions), where \textit{E} is the edit distance threshold. The shift process guarantees to cancel the effect of indel. As we do not have prior knowledge about whether there exist substitutions, or indels, or combination of both, we need to test for every possible case in our algorithm. Thus, GateKeeper generates 2\textit{E}+1 Hamming masks regardless the source of the edit. The last step is to merge all the 2\textit{E}+1 Hamming masks using a bitwise AND operation. This step tells us where the relevant matching and mismatching regions reside in the presence of edits in the read compared to the reference segment. Similarly to SHD, we apply amending process to the 2\textit{E}+1 masks before performing AND operation. In SHD, the amending process is accomplished using a 4-bit packed shuffle (SIMD parallel table-lookup instruction), shift, and OR operations. The number of computations needed is 4 packed shuffle, 4\textit{m} bitwise OR, and three shift operations for each Hamming mask, which is (7+4\textit{m})(2\textit{E}+1) operations, where \textit{m} is the read length. We find that this is very inefficient for FPGA implementation.

\begin{figure}
\centering
\includegraphics[width=10cm,keepaspectratio]{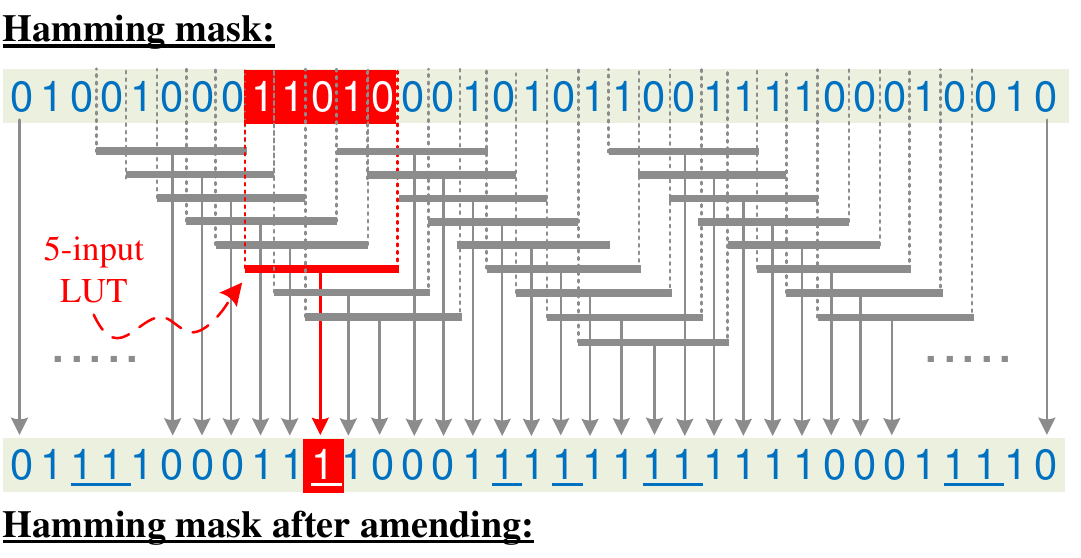}
\caption{Workflow of the proposed architecture for the parallel amendment operations.}
\label{fig:figure12GateKeeper}
\end{figure}

To reduce the number of operations, we build a new hardware-based amending process. We propose using dedicated hardware components in FPGA slices. More precisely, rather than shifting the read and then performing packed shuffle to replace patterns of 101 or 1001 to 111 or 1111 respectively, we perform only packed shuffle independently and concurrently for each bit of each Hamming mask. We present the proposed architecture for amendment operation in Figure \ref{fig:figure12GateKeeper}. In order to replace all patterns of 101 or 1001 to 111 or 1111 respectively, we use a single 5-input look-up table (LUT) for each bit of the Hamming mask. The first LUT copies the bit value of the first input regardless of its value; even if it is zero, it will not be amended as it is not contributing to the 101 or 1001 pattern. 

Likewise for the last LUT. Thus, the total number of LUTs needed is equal to the length of the short read in bases minus 2 for the first and last bases. In each LUT, we consider a single bit of the Hamming mask and two of its right neighboring bits and two of its left neighboring bits. If the input that corresponds to the output has a bit value of one, then the output copies the value of that input bit (as we only amend zeros). Otherwise, using the previous two bits and the following two bits with respect to the input bit, we can replace any zero of the “101” or “1001” patterns independently from other output bits. All bits of the amended masks are generated at the same time, as the propagation delay through an FPGA look-up table is independent of the implemented function \cite{guide7series}. \newpage Thus we can process all masks in a parallel fashion without affecting the correctness of the filtering decision. Using this dedicated architecture, we are able to get rid of the four shifting operations and perform the amending process concurrently for all bits of any Hamming mask. Thus, the required number of operations is only (2\textit{E}+1) instead of (7+4\textit{m})(2\textit{E}+1) for a total of (2\textit{E}+1) Hamming masks. This saves a considerable amount of the filtering time, reducing it by two orders of magnitude for a read that is 100bp long.

\subsection{Method 3: Minimizing Hardware Resource Overheads}
The short reads are composed of a string of nucleotides from the DNA alphabet \{A, C, G, T\}. Since the reads are processed in an FPGA platform, the symbols have to be encoded into a unique binary representation. We need 2 bits to encode each symbol. Hence, encoding a read sequence of length m results in a 2\textit{m}-bit word. Encoding the reads into a binary representation introduces overhead to accommodate not only the encoded reads but also the Hamming masks as their lengths also double (i.e., 2\textit{m}). The issue introduced by encoding the read can be even worse when we apply certain operations on these Hamming masks. For example, the number of LUTs required for performing the amending process on the Hamming masks will be doubled, mainly due to encoding the read. 

To reduce the complexity of the subsequent operations on the Hamming masks and save about half of the required amount of FPGA resources, we propose a new solution. We observe that comparing a pair of DNA nucleotides is similar to comparing their binary representations (e.g., comparing A to T is similar to comparing ‘00’ to ‘11’). Hence, comparing each two bits from the binary representation of the read with their corresponding bits of the reference segment generates a single bit that represents one of two meanings; either match or mismatch between two bases. 

\begin{figure}
\centering
\includegraphics[width=8.5cm]{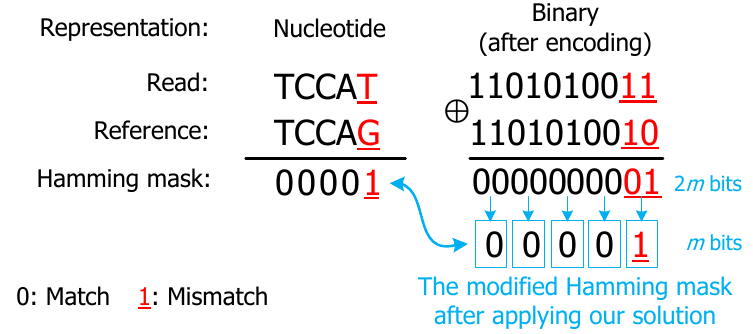}
\caption{An example of applying our solution for reducing the number of bits of each Hamming mask by half. We use a modified Hamming mask to store the result of applying the bitwise OR operation to each two bits of the Hamming mask. The modified mask maintains the same meaning of the original Hamming mask.}
\label{fig:figure13GateKeeper}
\end{figure}

\newpage This is performed by encoding each two bits of the result of the pairwise comparison (i.e., bitwise XOR) into a single bit of ‘0’ or ‘1’ using OR operations in a parallel fashion, as explained in Figure \ref{fig:figure13GateKeeper}. This makes the length of each Hamming mask equivalent to the length of the original read, without affecting the meaning of each bit of the mask. The modified Hamming masks are then merged together in 2\textit{E} bitwise AND operations. Finally, we count the number of ones (i.e., edits) in the final bit-vector mask; if the count is less than the edit distance threshold, the filter accepts the mapping.

\section{Analysis of GateKeeper Algorithm}
In this section, we provide the pseudocode of our GateKeeper algorithm. Algorithm 5.1 presents the main functions. Algorithms 5.2 presents the details of the amending process using in Algorithm 5.1.

\begin{center}
\includegraphics[width=\textwidth]{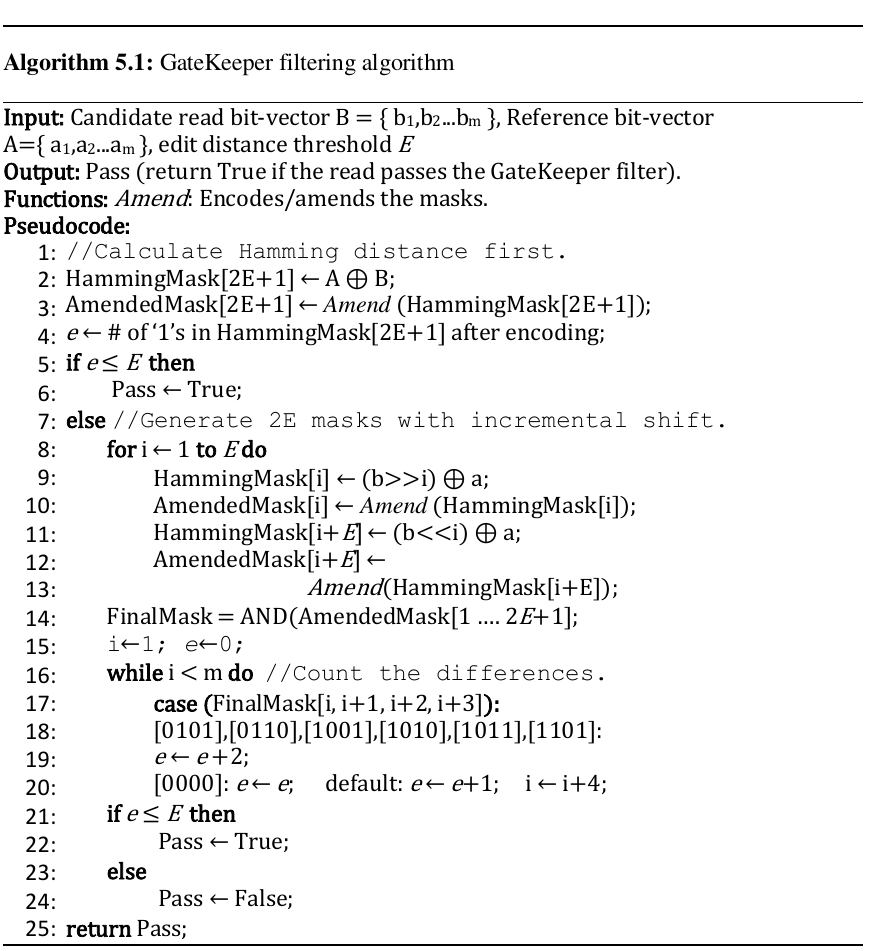}
\end{center}

\begin{center}
\includegraphics[width=\textwidth]{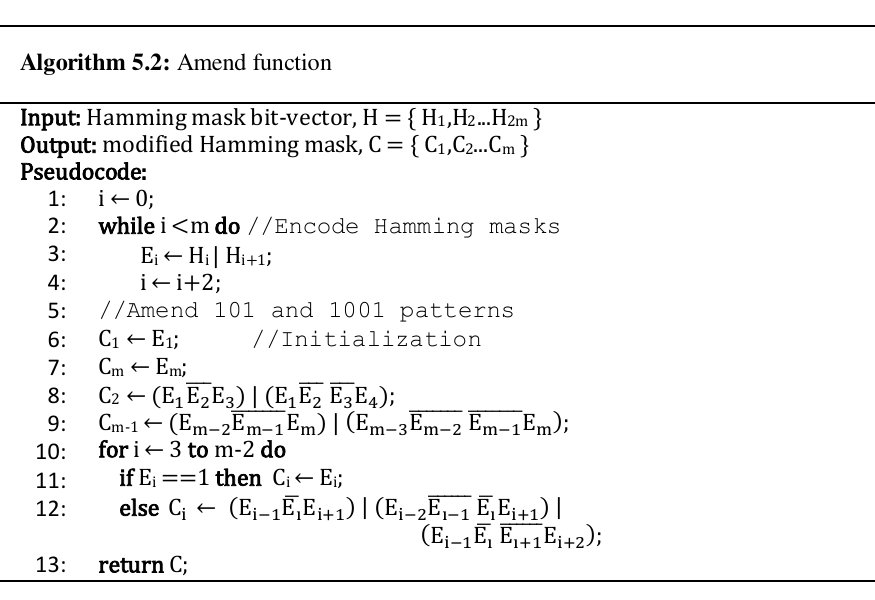}
\end{center}

\section{Discussion and Novelty}
GateKeeper is the only read mapping filter that takes advantage of the parallelism offered by FPGA architectures in order to expedite the alignment filtering process. GateKeeper supports both Hamming distance and edit distance in a fast and efficient way. Each GateKeeper filtering unit performs all operations defined in the GateKeeper algorithm. Table \ref{table:table1gatekeepernovelty} summarizes the relative benefits gained by each of the aforementioned optimization methods over the best previous filter, SHD (\textit{E} is the user-defined edit distance threshold and \textit{m} is the read length). When a read matches the reference exactly, or with few substitutions, GateKeeper requires only 2\textit{m} bitwise XOR operations, providing substantial speedup compared to SHD, which performs a much greater number of operations. However, this is not the only benefit we gain from our first proposed method (i.e., Fast Approximate String Matching). \newpage As this method provides an accurate examination for alignments with only substitutions (i.e., no deletions or insertions), we can directly skip calculating their optimal alignment using the computationally expensive alignment algorithms. For more general cases such as deletions and insertions, GateKeeper still requires far fewer operations (as shown in Table \ref{table:table1gatekeepernovelty}) than the original SHD filter, due to the optimization methods outlined above. Our improvements over SHD help drastically reduce the execution time of the filtering process. The rejected alignments by our GateKeeper filter are not further examined by read alignment.

\begin{table}
\centering
\caption{Overall benefits of GateKeeper over SHD in terms of number of operations performed.}
\label{table:table1gatekeepernovelty}
\includegraphics[width=\textwidth]{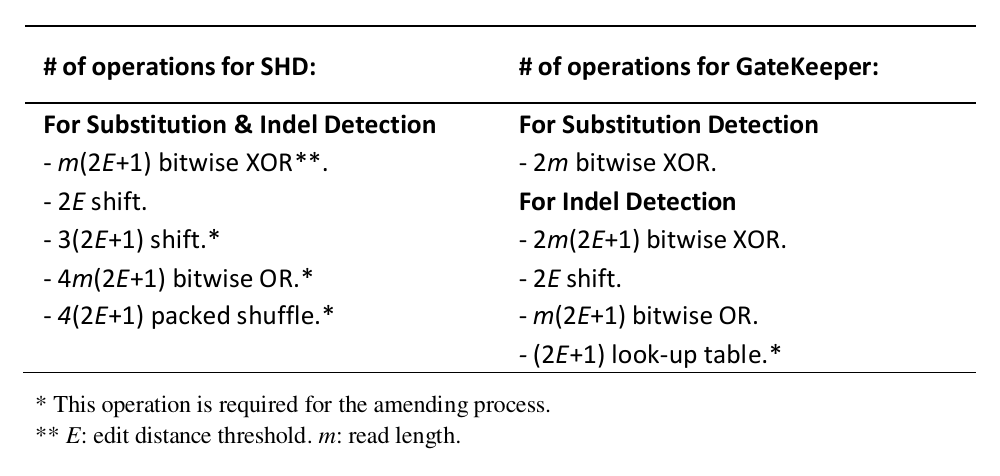}
\end{table}

\section{Summary}
We introduce the first hardware acceleration system for alignment filtering, called GateKeeper. We develop both a hardware-acceleration-friendly filtering algorithm and a highly-parallel hardware accelerator design. GateKeeper is a standalone filter and can be integrated with any existing reference-based mapper. GateKeeper does not replace read alignment. GateKeeper should be followed by read alignment step, which precisely verifies the mappings that pass our filter and eliminates the falsely accepted ones.

%% file: chapter6.tex
\chapter{Shouji: Fast and Accurate Hardware Pre-Alignment Filter}
Our primary purpose is to reject incorrect mappings accurately and quickly such that we reduce the need for the computationally expensive alignment step. In this chapter, we propose the Shouji algorithm to achieve highly accurate filtering. We then accelerate Shouji by taking advantage of the capabilities and parallelism of FPGAs to achieve fast filtering operations. Shouji's filtering strategy is inspired by our analytical study of SHD's filtering accuracy (see Chapter 3). We discuss the details of the Shouji algorithm next.

\section{Overview}
\begin{figure}
\centering
\includegraphics[width=10cm,keepaspectratio]{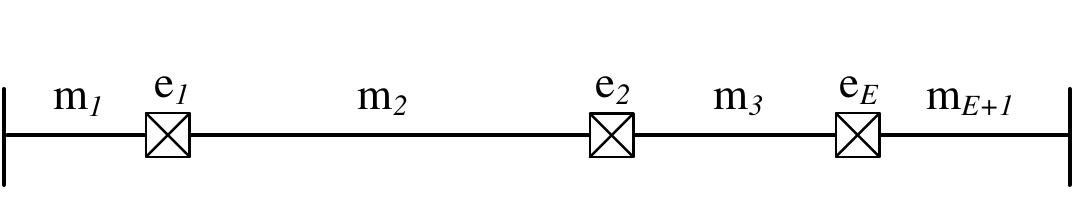}
\caption{Random edit distribution in a read sequence. The edits (\textit{e$_{1}$, e$_{2}$, …, e$_{E}$}) act as dividers resulting in several identical subsequences (\textit{m$_{1}$, m$_{2}$, …, m$_{E+1}$}) between the read and the reference.}
\label{fig:figure15SLIDER}
\end{figure}

The key filtering strategy of Shouji is inspired by the \textbf{\emph{pigeonhole principle}}, which states that if \textit{E} items are distributed into \textit{E}+1 boxes, then one or more boxes would remain empty. In the context of pre-alignment filtering, this principle provides the following key observation: if two sequences differ by \textit{E} edits, then the two sequences should share \emph{at least} a single common subsequence (i.e., free of edits) and at most \textit{E}+1 non-overlapping common subsequences, where \textit{E} is the edit distance threshold. With the existence of at most \textit{E} edits, the total length of these non-overlapping common subsequences should not be less than \textit{m-E}, where \textit{m} is the sequence length, as illustrated in Figure \ref{fig:figure15SLIDER}. Shouji employs the pigeonhole principle to decide whether or not two sequences are potentially similar. Shouji finds all the non-overlapping subsequences that exist in both sequences. If the total length of these common subsequences is less than \textit{m-E}, then there exist more edits than the allowed edit distance threshold, and hence Shouji rejects the two given sequences. Otherwise, Shouji accepts the two sequences. Next, we discuss the details of Shouji.

\section{Methods}
Shouji identifies the dissimilar sequences, without calculating the optimal alignment, in three main steps. (1) The first step is to construct what we call a \emph{neighborhood map} that visualizes the pairwise matches and mismatches between two sequences given an edit distance threshold of \textit{E} characters. (2) The second step is to find all the non-overlapping common subsequences in the neighborhood map using a sliding search window approach. (3) The last step is to accept or reject the given sequence pairs based on the length of the found matches. If the length of the found matches is small, then Shouji rejects the input sequence pair.

\subsection{Method 1: Building the Neighborhood Map}
\begin{figure}
\centering
\includegraphics[width=8cm,keepaspectratio]{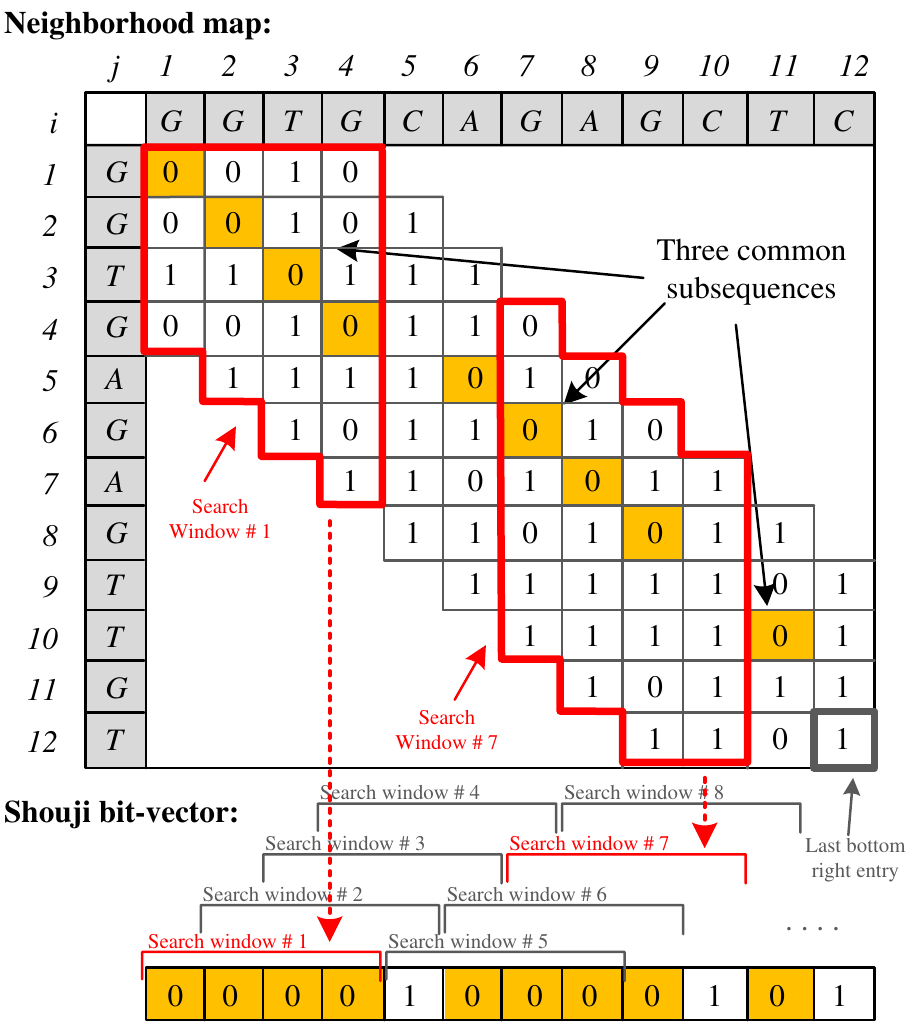}
\caption{Neighborhood map (\emph{N}) and the Shouji bit-vector, for text \emph{T} = GGTGCAGAGCTC, and pattern \emph{P} = GGTGAGAGTTGT for \emph{E}=3. The three common subsequences (i.e., GGTG, AGAG, and T) are highlighted in yellow. We use a search window of size 4 columns (two examples of which are high-lighted in red) with a step size of a single column. Shouji searches diagonally within each search window for the 4-bit vector that has the largest number of zeros. Once found, Shouji examines if the found 4-bit vector maximizes the number of zeros at the corresponding location of the 4-bit vector in the Shouji bit-vector. If so, then Shouji stores this 4-bit vector in the Shouji bit-vector at its corresponding location.}
\label{fig:figure16SLIDER}
\end{figure}

The neighborhood map, \textit{N}, is a binary \textit{m} by \textit{m} matrix, where \textit{m} is the read length. Given a text sequence \textit{T}[1…\textit{m}], a pattern sequence \textit{P}[1…\textit{m}], and an edit distance threshold \textit{E}, the neighborhood map represents the comparison result of the  $i^{th}$ character of \emph{P} with the  $j^{th}$ character of \emph{T}, where \textit{i} and \textit{j} satisfy 1$\leq$ \textit{i} $\leq$ \textit{m} and \textit{i-E} $\leq$ \textit{j} $\leq$ \textit{i+E}, the entry \textit{N}[\textit{i, j}] of the neighborhood map can be calculated as follows:
\begin{equation} \textit{N}[\textit{i,j}] =\begin{cases}0 & P[i]=T[j]\\1 & P[i]\neq T[j]\end{cases} \end{equation}
We present in Figure \ref{fig:figure16SLIDER} an example of a neighborhood map for two sequences, where sequence \textit{B} differs from sequence \textit{A} by three edits. 
The entry \textit{N}[\textit{i,j}] is set to zero if the $i^{th}$ character of the pattern matches the $j^{th}$ character of the text. Otherwise, it is set to one. The way we build our neighborhood map ensures that computing each of its entries is independent of every other, and thus the entire map can be computed all at once in a parallel fashion. Hence, our neighborhood map is well suited for highly-parallel computing platforms \cite{alser2017gatekeeper, seshadri2017ambit}. Note that in sequence alignment algorithms, computing each entry of the dynamic programming matrix depends on the values of the immediate left, upper left, and upper entries of its own. Different from\textit{\lq\lq{}dot plot\rq\rq{}} or \textit{\lq\lq{}dot matrix\rq\rq{}} (visual representation of the similarities between two closely similar genomic sequences) that is used in FASTA/FASTP \cite{lipman1985rapid}, our neighborhood map computes \textit{only} necessary diagonals near the main diagonal of the matrix (e.g., for \textit{E}=3, Shouji computes only 2\textit{E}+1 diagonal vectors of the neighborhood map).

\subsection{Method 2: Identifying the Diagonally-Consecutive Matches}
\begin{figure}
\centering
\includegraphics[width=10cm,height=7cm]{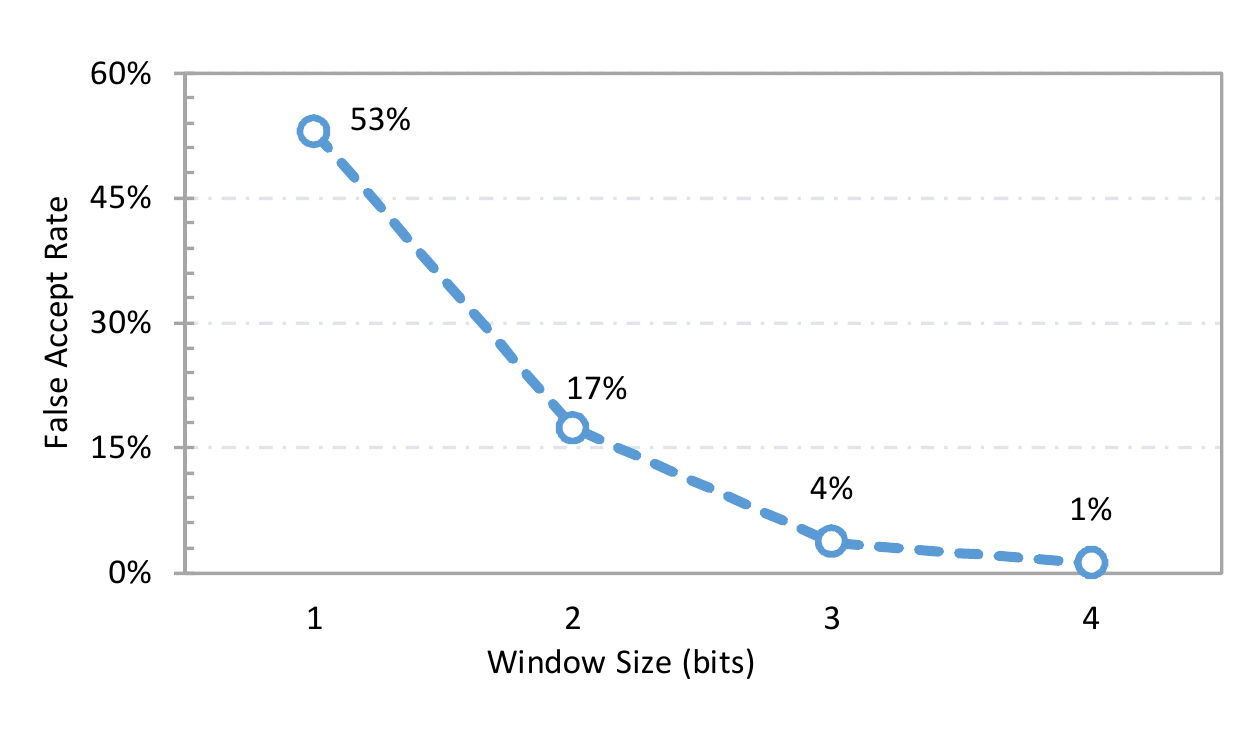}
\caption{The effect of the window size on the rate of the falsely accepted sequences (i.e., dissimilar sequences that are considered as similar ones by Shouji filter). We observe that a window width of 4 columns provides the highest accuracy. We also observe that as window size increases beyond 4 columns, more similar sequences are rejected by Shouji, which should be avoided.}
\label{fig:figure17SLIDER}
\end{figure}

The key goal of this step is to accurately find all the non-overlapping common subsequences shared between a pair of sequences. The accuracy of finding these subsequences is crucial for the overall filtering accuracy, as the filtering decision is made solely based on total subsequence length. With the existence of \emph{E} edits, there are at most \emph{E}+1 non-overlapping common subsequences (based on the pigeonhole principle) shared between a pair of sequences. Each non-overlapping common subsequence is represented as a streak of diagonally-consecutive zeros in the neighborhood map (as highlighted in yellow in Figure \ref{fig:figure16SLIDER}). These streaks of diagonally-consecutive zeros are distributed along the diagonals of the neighborhood map without any prior information about their length or number. One way of finding these common subsequences is to use a brute-force approach, which examines all the streaks of diagonally-consecutive zeros that start at the first column and selects the streak that has the largest number of zeros as the first common subsequences. It then iterates over the remaining part of the neighborhood map to find the other common subsequences. However, this brute-force approach is infeasible for highly-optimized hardware implementation as the search space is unknown at design time. Shouji overcomes this issue by dividing the neighborhood map into equal-size parts. We call each part a search window. Limiting the size of the search space from the entire neighborhood map to a search window has three key benefits. (1) It helps to provide a scalable architecture that can be implemented for any sequence length and edit distance threshold. (2) Downsizing the search space into a reasonably small sub-matrix with a known dimension at design time limits the number of all possible permutations of each bit-vector to 2\emph{n}, where \emph{n} is the search window size. This reduces the size of the look-up tables (LUTs) required for an FPGA implementation and simplifies the overall design. (3) Each search window is considered as a smaller sub-problem that can be solved independently and rapidly with high parallelism. Shouji uses a search window of 4 columns wide, as we illustrate in Figure \ref{fig:figure16SLIDER}. We need \emph{m} search windows for pro-cessing two sequences, each of which is of length \emph{m} characters. Each search window overlaps with its next neighboring search window by 3 columns. This ensures covering the entire neighborhood map and finding all the common subsequences regardless of their starting location. We select the width of each search window to be 4 columns to guarantee finding the shortest possible common subsequence, which is a single match located between two mismatches (i.e., ‘101’). However, we observe that the bit pattern ‘101’ is not always necessarily a part of the correct alignment (or the common subsequences). For example, the bit pattern ‘101’ exists once as a part of the correct alignment in Figure \ref{fig:figure16SLIDER}, but it also appears five times in other different locations that are not included in the correct alignment. To improve the accuracy of finding the diagonally-consecutive matches, we increase the length of the diagonal vector to be examined to four bits. We also experimentally evaluate different window sizes in Figure \ref{fig:figure17SLIDER}. We find that a window size of 4 columns provides the highest filtering accuracy without falsely rejecting similar sequences. This is because individual matches (i.e., single zeros) are usually useless and they are not necessarily part of the common subsequences. As we increase the search window size, we are ignoring these individual matches and instead we only look for longer streaks of consecutive zeros.

Shouji finds the diagonally-consecutive matches that are part of the common subsequences in the neighborhood map in two main steps. Step 1: For each search window, Shouji finds a 4-bit diagonal vector that has the largest number of zeros. Shouji greedily considers this vector as a part of the common subsequence as it has the least possible number of edits (i.e., 1’s). Finding always the maximum number of matches is necessary to avoid overestimating the actual number of edits and eventually preserving all similar sequences. Shouji achieves this step by comparing the 4 bits of each of the 2\emph{E}+1 diagonal vectors within a search window and selects the 4-bit vector that has the largest number of zeros. In the case where two 4-bit subsequences have the same number of zeros, Shouji breaks the ties by selecting the first one that has a leading zero. Then, Shouji slides the search window by a single column (i.e., step size = 1 column) towards the last bottom right entry of the neighborhood map and repeats the previous computations. Thus, Shouji performs “Step 1” \emph{m} times using \emph{m} search windows, where \emph{m} is the sequence length. Step 2: The last step is to gather the results found for each search window (i.e., 4-bit vector that has the largest number of zeros) and construct back all the diagonally-consecutive matches. For this purpose, Shouji maintains a Shouji bit-vector of length \emph{m} that stores all the zeros found in the neighborhood map as we illustrate in Figure \ref{fig:figure16SLIDER}. For each sliding search window, Shouji examines if the selected 4-bit vector maximizes the number of zeros in the Shouji bit-vector at the same corresponding location. If so, Shouji stores the selected 4-bit vector in the Shouji bit-vector at the same corresponding location. This is necessary to avoid overestimating the number of edits between two given sequences. The common subsequences are represented as streaks of consecutive zeros in the Shouji bit-vector.

\begin{figure}
\includegraphics[width=\linewidth]{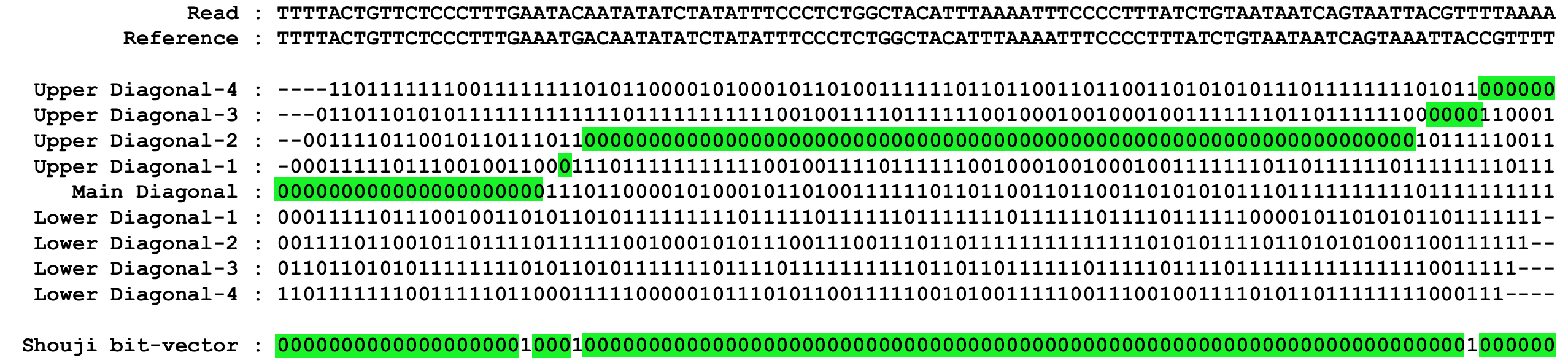}
\caption{An example of applying Shouji filtering algorithm to a sequence pair, where the edit distance threshold is set to 4. We present the content of the neighborhood map along with the Shouji bit-vector. We apply the Shouji algorithm starting from the leftmost column towards the rightmost column.}
\label{fig:figure18SLIDER}
\end{figure}

\subsection{Method 3: Filtering Out Dissimilar Sequences}
The last step of Shouji is to calculate the total number of edits (i.e., ones) in the Shouji bit-vector. Shouji examines if the total number of ones in the Shouji bit-vector is greater than \emph{E}. If so, Shouji excludes the two sequences from the optimal alignment calculation. Otherwise, Shouji considers the two sequences similar within the allowed edit distance threshold and allows their optimal alignment to be computed using optimal alignment algorithms. The Shouji bit-vector represents the differences between two sequences along the entire length of the sequence, \emph{m}. However, Shouji is not limited to end-to-end edit distance calculation. Shouji is also able to provide edit distance calculation in local and glocal (semi-global) fashion. For example, achieving local edit distance calculation requires ignoring the ones that are located at the two ends of the Shouji bit-vector. Achieving glocal edit distance calculation requires excluding the ones that are located at one of the two ends of the Shouji bit-vector from the total count of the ones in the Shouji bit-vector. This is important for correct pre-alignment filtering for global, local, and glocal alignment algorithms. We present an example of applying Shouji filtering algorithm in Figure \ref{fig:figure18SLIDER}.

\section{Analysis of Shouji algorithm}
Shouji filter does not filter out similar sequences; hence, it provides zero false reject rate. The reason behind that is the way we find the identical subsequences. We always look for the subsequences that has the largest number of zeros, such that we maximize the number of matches and minimize number of edits that cause the division of one long identical sequence into shorter subsequences. This also allows for a very small portion of dissimilar sequences to pass. Next, we analyze the computational complexity (i.e., asymptotic run time and space complexity) of our Shouji filter. Shouji filter divides the problem of finding the identical subsequences into at most \textit{m} subproblems, as described in Algorithm 6.1. Each subproblem examines each of the 2\textit{E}+1 bit-vectors and finds the 4-bit subsequence that has the largest number of zeros within the sliding window. Once found, Shouji filter also compares the found subsequence with its corresponding subsequence of the Shouji bit-vector. Now, let \textit{c} be a constant representing the run time of examining each 4 bits of each bit-vector. Then the time complexity of the Shouji algorithm is as follows:
\begin{equation} \textit{T$_{Shouji}$(m)} = c\,.\,m\,.\,(2E+1) \end{equation}
This demonstrates that the Shouji algorithm runs in linear time with respect to the sequence length and edit distance threshold. Shouji algorithm maintains 2\textit{E}+1 diagonal bit-vectors and an additional auxiliary bit-vector (i.e., Shouji bit-vector) for each two given sequences. The space complexity of the Shouji algorithm is as follows:
\begin{equation} \textit{D$_{Shouji}$(m)} = m\,.\,(2E+2) \end{equation}
Hence, the Shouji algorithm requires linear space with respect to the read length and edit distance threshold. Next, we outline the hardware implementation details of Shouji filter.

\begin{center}
\includegraphics[width=\textwidth]{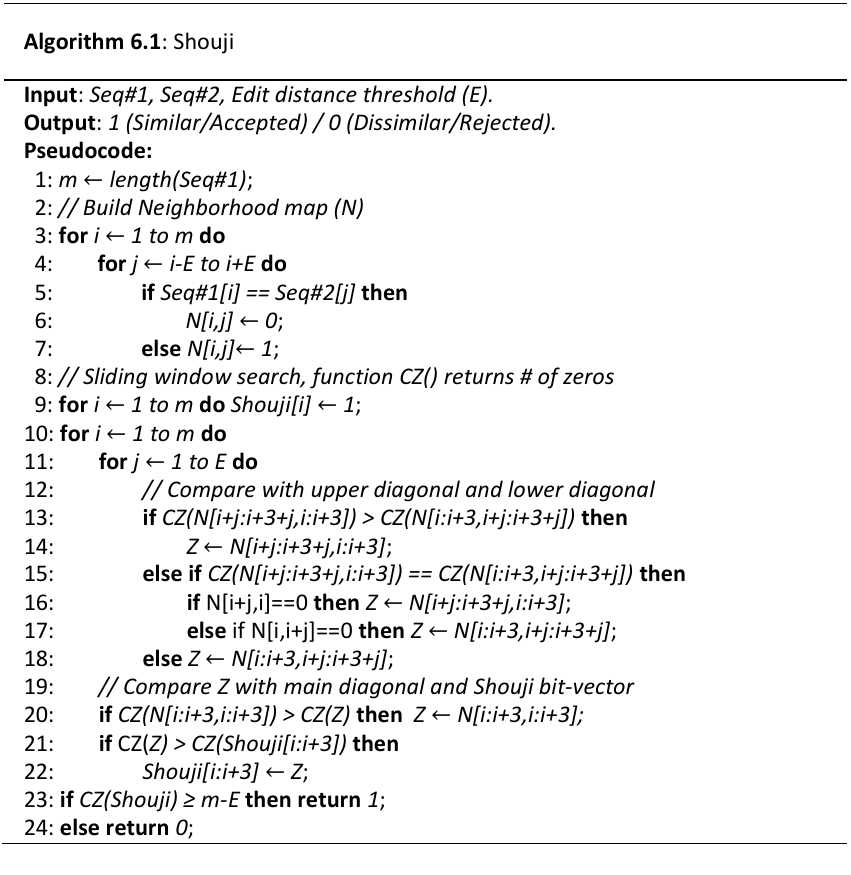}
\end{center}

\begin{figure}
\includegraphics[width=\linewidth]{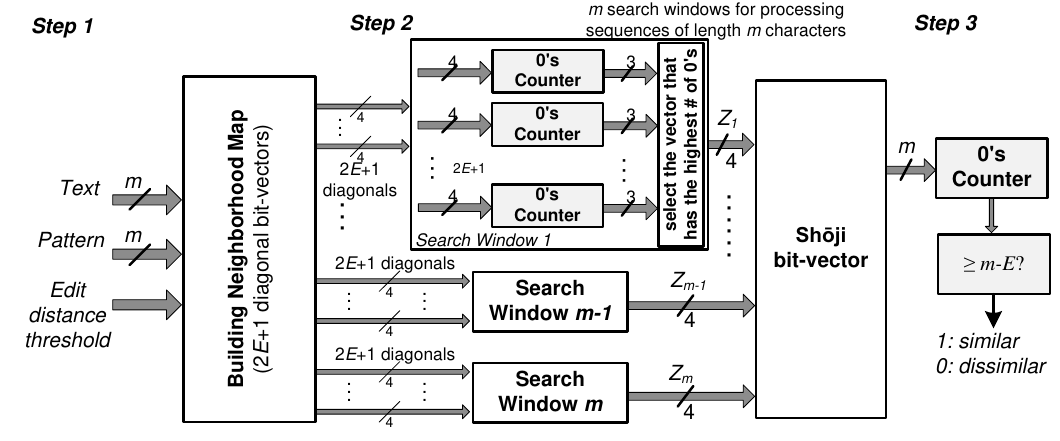}
\caption{A block diagram of the sliding window scheme implemented in FPGA for a single filtering unit.}
\label{fig:figure19SLIDER}
\end{figure}

\section{Discussion}

To make the best use of the available resources in the FPGA chip, our algorithm utilizes the operations that are easily supported on an FPGA, such as bitwise operations, bit shifts, and bit count. To build the neighborhood map on the FPGA, we use the observation that the main diagonal can be implemented using a bitwise XOR operation between the two given sequences. The vacant bits due to the shift operation are filled with ones. The upper \textit{E} diagonals can be implemented by gradually shifting the pattern (\textit{P}) to the right-hand direction and then performing bitwise XOR with the text (\textit{T}). This allows each character of \textit{P} to be compared with the right-hand neighbor characters (up to \textit{E} characters) of its corresponding character of \textit{T}. The lower \textit{E} diagonals can be implemented in a way similar to the upper \textit{E} diagonals, but here the shift operation is performed in the left-hand direction. This ensures that each character of \textit{P} is compared with the left-hand neighbor characters (up to \textit{E} characters) of its corresponding character of \textit{T}. 
We also build an efficient hardware architecture for each search window of the Shouji algorithm. It quickly finds the number of zeros in each 4-bit vector using a hardware look-up table that stores the 16 possible permutations of a 4-bit vector along with the number of zeros for each permutation. As presented in Figure \ref{fig:figure19SLIDER}, each counter counts the number of zeros in a single bit-vector. The counter takes four bits as input and generate three bits that represents the number of zeros within the window. Each counter requires three 4-input LUTs. In total, we need 6\textit{E}+6 4-input LUTs to build a single search window. All bits of the counter output are generated at the same time, as the propagation delay through an FPGA look-up table is independent of the implemented function \cite{guide7series}. The comparator is responsible for selecting the 4-bit subsequence that maximizes the number of consecutive matches based on the output of each counter and the Shouji bit-vector. Finally, the selected 4-bit subsequence is then stored in the Shouji bit-vector at the same corresponding location. Our hardware implementation of the Shouji filtering unit is independent of the specific FPGA-platform as it does not rely on any vendor-specific computing elements (e.g., intellectual property cores).

\section{Summary}
We propose Shouji, a highly parallel and accurate pre-alignment filter designed on a specialized hardware platform. The first key idea of our proposed pre-alignment filter is to provide high filtering accuracy by correctly detecting all identical subsequences shared between two given subsequences. This way leads to address the first two causes of filtering inaccuracy in SHD (i.e., random zeros and conservative counting). The second key idea is to avoid the filtering inaccuracy caused by the leading and trailing zeros (as we discuss in Chapter 3). Shouji replaces the vacant bits that result from shifting the read with ones.

%% file: chapter7.tex
\chapter{MAGNET: Accurate Hardware Pre-Alignment Filter}
In this chapter, we introduce MAGNET, a pre-alignment filtering algorithm to achieve highly accurate filtering. MAGNET is a filtering heuristic that aims at finding all non-overlapping long streaks of consecutive zeros in the neighborhood map using a divide-and-conquer approach. We discuss the details of MAGNET algorithm next.

\section{Overview}
MAGNET uses a divide-and-conquer technique to find all the \textit{E}+1 identical subsequences, if any, and summing up their length. By calculating their total length, we can estimate the total number of edits between the two given sequences. If the total length of the \textit{E}+1 identical subsequences is less than \textit{m-E}, then there exist more identical subsequences than \textit{E}+1 that are associated with more edits than allowed. If so, then MAGNET rejects the two given sequences without performing the alignment step. The filtering strategy of MAGNET makes three observations based on the pigeonhole principle to examine each mapping accurately. 

\begin{enumerate}
\item Given that the user-defined edit distance threshold, \textit{E}, is usually less than 5\% of the read length (\textit{m}) \cite{cheng2015bitmapper, xin2015shifted, hatem2013benchmarking, ahmadi2011hobbes}, the identical subsequence is usually long and ranges from a single pairwise-match to \textit{m} pairwise-matches long. 
\item The length of the longest identical subsequence is strictly not less than $\lceil ((m-E))/((E+1) ) \rceil$ and can be at most m pairwise-matches long (i.e., equivalent to the sequence length). The upper bound is trivial and holds when the alignment is free of edits. The lower bound equality occurs when all edits are equispaced and all \textit{E}+1 subsequences are of the same length. 

\item We observe that the \textit{E}+1 identical subsequences that are part of the correct alignment are always non-overlapping. We also observe that the 2\textit{E}+1 diagonal bit-vectors of the neighborhood map contain other identical subsequences that are always short. This raises a fundamental question about whether the \textit{E}+1 identical subsequences need to be strictly non-overlapping or only long and not necessarily non-overlapping. Next, we investigate both cases by providing two algorithms for finding the longest identical subsequences.
\end{enumerate}

\section{Methods}
MAGNET pre-alignment filter identifies the incorrect mappings, without calculating the optimal alignment, in three main steps. (1) It first constructs the neighborhood map (described in Chapter 6). (2) It then identifies all non-overlapping identical subsequences in the neighborhood map using a divide-and-conquer approach. (3) And finally it makes a decision (it accepts or rejects the given sequences) based on the length of the found identical subsequences.

\subsection{Method 1: Building the Neighborhood Map}
The first step of MAGNET algorithm is building our binary \textit{m} by \textit{m} neighborhood map that we describe in Chapter 6. Similarly with Shouji filter, MAGNET filter starts with building the 2\textit{E}+1 diagonal bit-vectors of the neighborhood map for the two given sequences. We use the neighborhood map to represent all the pairwise matches between the read and the reference sequences. The neighborhood map considers also the presence of substitutions, insertions, and deletions. Next step is to find the consecutive matches that are part of the correct alignment.

\subsection{Method 2: Identifying the \textit{E}+1 Identical Subsequences}
There are two methods to identify the identical subsequences. Either to find the non-overlapping subsequences of consecutive zeros or find the \textit{E}+1 top longest subsequences of consecutive zeros. We describe both methods and show which one is more effective.
\subsubsection{Identifying \textit{E}+1 non-overlapping subsequences}
Finding the \textit{E}+1 non-overlapping subsequences in the neighborhood map involves three main steps. 
\begin{enumerate}
\item \textbf{Extraction.} Each diagonal bit-vector nominates its local longest subsequence of consecutive zeros. Among all nominated subsequences, a single subsequence is selected as a global longest subsequence based on its length. (Once found, MAGNET evaluates if its length is less than  is strictly not less than $\lceil ((m-E))/((E+1) ) \rceil$, then the two sequences contains more edits than allowed, which cause the identical subsequences to be shorter (i.e., each edit results in dividing the sequence pair into more identical subsequences). If so, then the two sequences are rejected. Otherwise, MAGNET stores its length to be used towards calculating the total length of all \textit{E}+1 identical subsequences. 

\item \textbf{Encapsulation.} The next step is essential to preserve the original edit (or edits) that causes a single identical sequence to be divided into smaller subsequences. MAGNET penalizes the found subsequence by two edits (one for each side). This is achieved by excluding from the search space of all bit-vectors the indices of the found subsequence in addition to the index of the surrounding single bit from both left and right sides. 

\item \textbf{Divide-and-Conquer Recursion.} In order to locate the other \textit{E} non-overlapping subsequences, MAGNET applies a divide-and-conquer technique where we decompose the problem of finding the non-overlapping identical subsequences into two subproblems. While, the first subproblem focuses on finding the next long subsequence that is located on the right-hand side of the previously found subsequence in the first extraction step, the second subproblem focuses on the other side of the found subsequence. Each subproblem is solved by recursively repeating all the three steps mentioned above, but without evaluating again the length of the longest subsequence. MAGNET applies two early termination methods that aim at reducing the execution time of the filter. The first method is evaluating the length of the longest subsequence in the first recursion call. The second method is limiting the number of the subsequences to be found to at most \textit{E}+1, regardless their actual number for each two sequences.
\end{enumerate}

\subsubsection{Identifying top \textit{E}+1 longest subsequences}
\begin{figure}
\centering
\includegraphics[width=10cm]{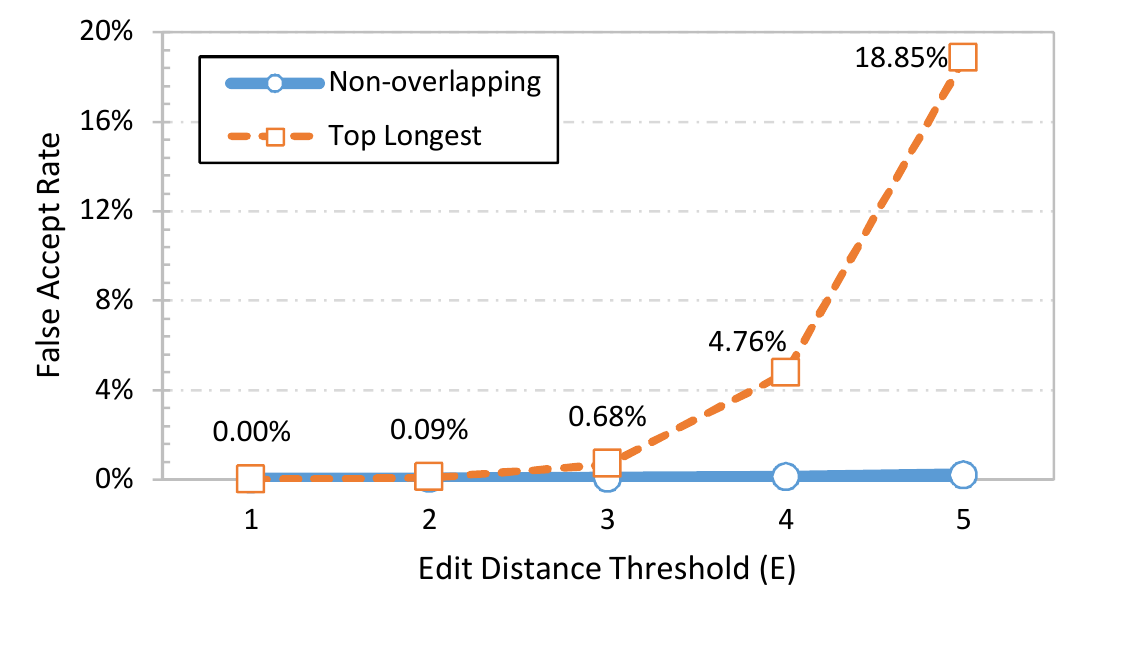}
\caption{The false accept rate of MAGNET using two different algorithms for identifying the identical subsequences. We observe that finding the \textit{E}+1 non-overlapping identical subsequences leads to a significant reduction in the incorrectly accepted sequences compared to finding the top \textit{E}+1 longest identical subsequences.}
\label{fig:figure20MAGNET}
\end{figure}

Alternatively, MAGNET can be changed to find only top \textit{E}+1 longest subsequences without the restriction of being non-overlapping ones. This can be achieved by maintaining a binary max-heap priority queue \cite{williams1964algorithm}, where it stores the length of each of the \textit{E}+1 subsequences from each mask. \newpage In total, the queue stores up to (\textit{E}+1).(2\textit{E}+1) elements. The length of the top longest subsequence is always stored at the root of the heap (the heap property). We need to extract the root of the heap structure \textit{E}+1 times in order to find the total length of the top \textit{E}+1 longest subsequences. We evaluate both algorithms for finding the longest identical subsequences in Figure \ref{fig:figure20MAGNET}. We observe that the accuracy of finding top \textit{E}+1 longest subsequences degrades exponentially as the edit distance threshold increases. We also observe that the accuracy of finding the non-overlapping subsequences remains almost linear and provides considerably more accurate filtering. Thus, we consider identifying only non-overlapping subsequences throughout the following sections. 

\subsection{Method 3: Filtering Out Dissimilar Sequences}
The last step of MAGNET is to decide if the mapping is potentially correct and needs to be examined by read alignment. Once after the termination, if the total length of all found identical subsequences is less than \textit{m-E} then the two sequences are rejected. Otherwise, they are considered to be similar and the alignment can be measured using sophisticated alignment algorithms.

\begin{figure}
\includegraphics[width=\linewidth]{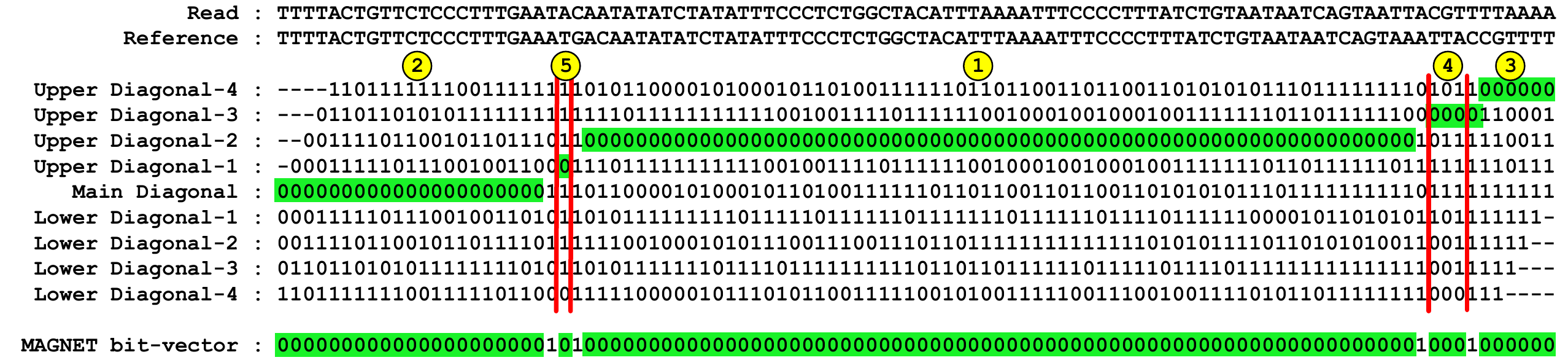}
\caption{An example of applying MAGNET filtering algorithm with an edit distance threshold of 4. MAGNET finds all the longest non-overlapping subsequences of consecutive zeros in the descending order of their length (as numbered in yellow).}
\label{fig:figure21MAGNET}
\end{figure}

\newpage We provide an example of applying MAGNET filtering algorithm in Figure \ref{fig:figure21MAGNET}.

\section{Analysis of MAGNET algorithm}
In this section, we analyze the asymptotic run time and space complexity of MAGNET algorithm. We provide the pseudocode of MAGNET in Algorithm 7.1. We also provide the divide-and-conquer algorithm that MAGNET uses in Algorithm 7.2. 
MAGNET applies a divide-and-conquer technique that divides the problem of finding the identical subsequences into two subproblems in each recursion call. In the first recursion call, the extracted identical subsequence is of length at least \textit{a}=$\lceil ((\textit{m-E}))/((\textit{E}+1)) \rceil$ bases. This reduces the problem of finding the identical subsequences from \textit{m} to at most \textit{m-a}, which is further divided into two subproblems: a left subproblem and a right subproblem. For the sake of simplicity, we assume that the size of the left and the right subproblems decreases by a factor of \textit{b} and \textit{c}, respectively, as follows:
\begin{equation} \textit{m} = a+2+m/b+m/c \end{equation}
The addition of 2 bases is for the encapsulation bits added at each recursion call. Now, let \textit{T$_{MAGNET}$(m)} be the time complexity of MAGNET algorithm, for identifying non-overlapping subsequences. 

\begin{center}
\includegraphics[width=\textwidth]{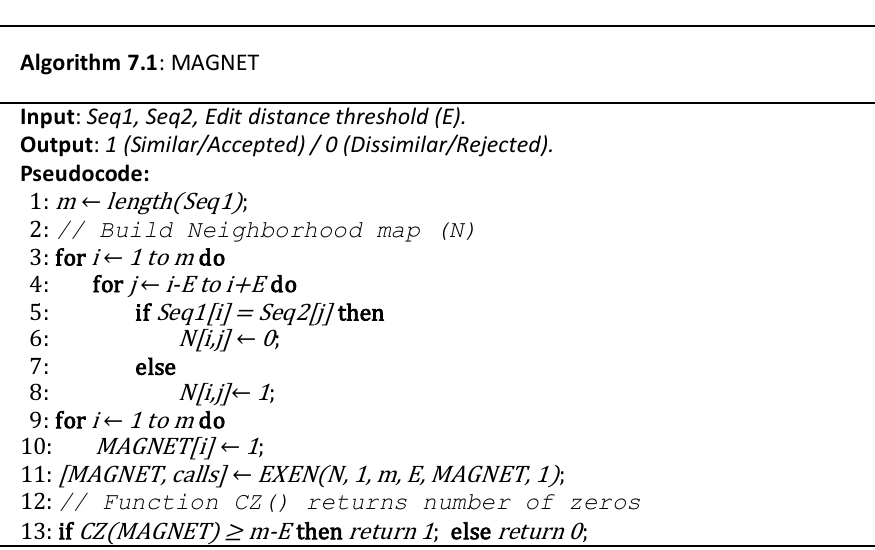}
\end{center}

If it takes O(\textit{km}) time to find the global longest subsequence and divide the problem into two subproblems, where \textit{k} = 2\textit{E}+1 is the number of bit-vectors, we get the following recurrence equation:
\begin{equation} \textit{T$_{MAGNET}$(m)} = \textit{T$_{MAGNET}$(m/b)} + \textit{T$_{MAGNET}$(m/c)} + O(km) \end{equation}
Given that the early termination condition of MAGNET algorithm restricts the recursion depth as follows:
\begin{equation} Recursion\,\,tree\,\,depth = \lceil log _{2} (E+1) \rceil -1\end{equation}
Solving the recurrence in equation (7.2) using equation (7.1) and equation (7.3) by applying the recursion-tree method provides a loose upper-bound on the time complexity as follows:
\[ \textit{T$_{MAGNET}$(m)} = O(km) \,\,. \sum_{x=0}^{ \lceil log _{2} (E+1) \rceil -1} \left ( \frac{1}{c} + \frac{1}{b}\right )^{x} \] 
\begin{equation}\textit{T$_{MAGNET}$(m)} \approx O(fkm) \end{equation}
\newpage Where \textit{f} is a fraction number satisfies the following range: 1 $\leq$ \textit{f} $<$ 2. This in turn demonstrates that the MAGNET algorithm runs in a linear time with respect to the sequence length and edit distance threshold and hence it is computationally inexpensive. The space complexity of MAGNET algorithm is as follows:
\[ \textit{D$_{MAGNET}$(m)} = \textit{D$_{MAGNET}$(m/b)} + \textit{D$_{MAGNET}$(m/c)} + O(km+m) \]
\begin{equation} \textit{D$_{MAGNET}$(m)} \approx O(fkm+fm)  \end{equation}
Hence, MAGNET algorithm requires a linear space with respect to the read length and edit distance threshold. Next, we outline the hardware implementation details of MAGNET filter.

\section{Discussion}
In this section, we outline the challenges that are encountered in implementing MAGNET filter to be used in our accelerator design. Implementing MAGNET algorithm is challenging due to the random location and variable length of each of the \textit{E}+1 identical subsequences. The Verilog-2011 imposes two challenges on our architecture as it does not support variable-size partial selection and indexing of a group of bits from a vector \cite{mcnamara2001ieee}. In particular, the first challenge lies in excluding the extracted identical subsequence along with its encapsulation bits from the search space of the next recursion call. The second challenge lies in dividing the problem into two subproblems, each of which has an unknown size at design time. To address these limitations and tackle the two design challenges, we keep the problem size fixed and at each recursion call. We exclude the found longest subsequence from the search space by amending all bits of all 2\textit{E}+1 bit-vectors that are located within the indices (locations) of the encapsulation bits to ‘1’s. This ensures to omit the found longest subsequence and all its corresponding locations in the other bit-vectors from the following recursion calls. 

\begin{center}
\includegraphics[width=\textwidth]{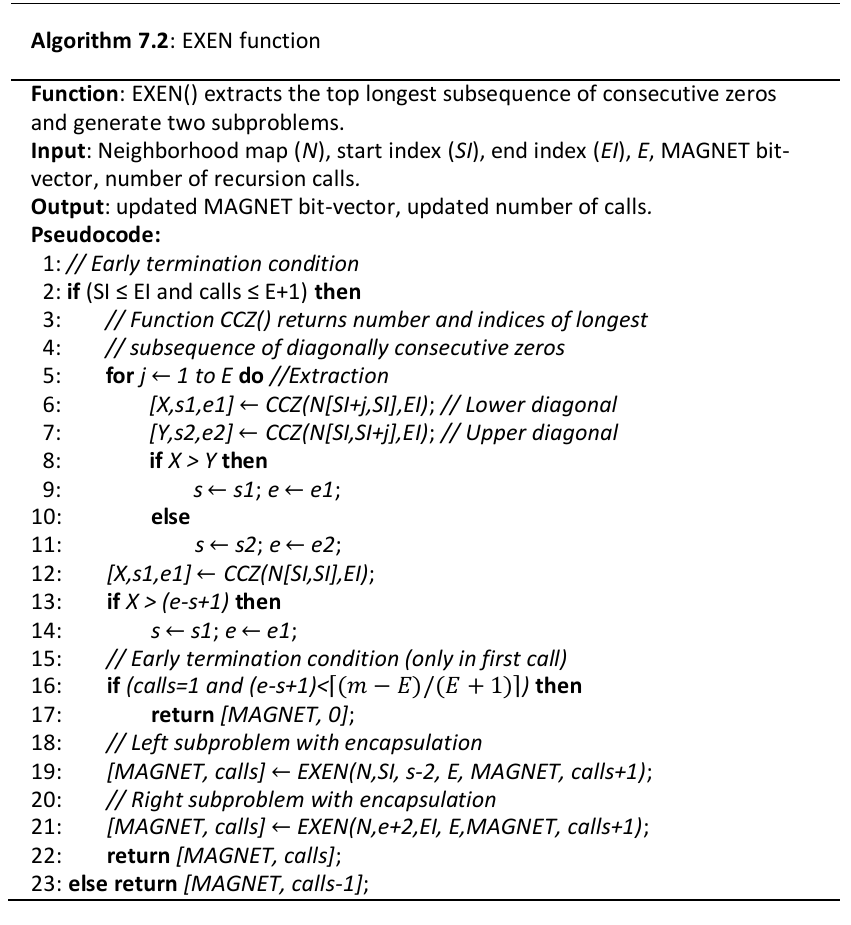}
\end{center}

\section{Summary}
We introduce MAGNET, a new filtering strategy that remarkably improves the accuracy of pre-alignment filtering and provides a minimal number of falsely accepted mappings. \newpage MAGNET gets rid of the first three causes of filtering inaccuracy that we observed in SHD \cite{xin2015shifted} (see Chapter 3). We believe that MAGNET is the most accurate pre-alignment filter in literature today but this comes at the expense of hardware implementation challenges. 

%% file: chapter8.tex
\chapter{SneakySnake: Fast, Accurate, and Cost-Effective Filter}
In this chapter, we address the issue of the long execution time of read alignment using a different approach. The use of specialized hardware chips can yield significant performance improvements. However, the main drawbacks are the higher design effort, the limited data throughput, and the necessity of algorithm redesign and tailored implementation to fully leverage the FPGA architecture. To tackle these challenges and obviate these difficulties, we introduce fast, accurate, and yet cost-effective pre-alignment filter. We call it SneakySnake. It leverages the today\rq{}s general purpose processor (i.e., CPU), which is largely available to bioinformaticians at no additional cost. Beside the CPU implementation of SneakySnake, we also provide an efficient hardware architecture to further accelerate the computations of the SneakySnake algorithm with high parallelism.

\section{Overview}
The key idea of SneakySnake filter is that it needs to know only if two sequences are dissimilar by more than the edit distance threshold or not. \newpage This fact leads to the following question. \textbf{Can one approximate the edit distance between two sequences much faster than calculating the edit distance?} Given reference sequence \textit{A}[1…\textit{m}], read sequence \textit{B}[1…\textit{m}], and an edit distance threshold \textit{E}, SneakySnake calculates the approximate edit distance between \textit{A} and \textit{B} and then checks if it is greater than the edit distance threshold then the two sequences \textit{A} and \textit{B} are rejected. Otherwise, the two sequences \textit{A} and \textit{B} are accepted by SneakySnake and the optimal alignment is calculated. The approximate edit distance calculated by SneakySnake should always be less than or equal to the edit distance threshold as long as the actual edit distance does not exceed the edit distance threshold.

\section{The Sneaky Snake Problem} 
We convert the approximate edit distance problem to as a restricted optimal path finding problem in a grid graph. We call this the Sneaky Snake Problem. It can be summarized as a traversal problem in a special grid graph, where a snake travels in the grid graph with the presence of randomly distributed obstacles. The goal is to find the path that connects the origin and the destination points with the minimal number of obstacles along the path.

We describe the grid graph of the Sneaky Snake Problem as follows:
\begin{itemize}
\item There is a two dimensional \textit{m} by \textit{m} grid graph, where \textit{m} is the read length. The grid cells are aligned in rows and columns and all cells have the same cost. The snake wants to travel through the grid from the top left corner towards the bottom right corner of the grid. There are \textit{E}+1 entrances that are next to the cells of the first column and similarly, there are \textit{E}+1 exits that are located next to the cells of the last column.
\item The snake is allowed to travel through dedicated pipes. Each pipe is represented as a stretch of diagonally consecutive pairwise matches. We define the path in this problem as a sequence of diagonal grid cells. \newpage In other words, the movement through the vertical cells is not counted towards the length of the path. The length of the path is simply the total number of diagonal cells along the path.
\item The snake is only able to switch between the pipes after skipping an obstacle (i.e., end of the pipe). The obstacle represents a pairwise mismatch. We model the obstacle as a solid black grid cell. Traveling across an obstacle requires the snake to diagonally move one step ahead, and hence the obstacle cell is counted towards the total length of the path. The snake is allowed to avoid only \textit{E} obstacles by transferring to another pipe or jumping over the obstacle within the same diagonal vector. This can be seen as the number of \textit{E} tries required to arrive the destination.
\end{itemize}

The general goal of this problem is to find a path in the grid graph with the minimum number of obstacles. The path starts at the first leftmost cell and travels out of the grid while exiting onto a destination. SneakySnake filtering algorithm can approximate the edit distance using two different approaches. While one of them relies on what we call \textit{weighted neighborhood maze}, the other approach relies on \textit{unweighted neighborhood maze}. Both approaches involve three main steps. 

\section{Weighted Maze Methods}
Next, we describe the three steps of the first approach. (1) It constructs the weighted neighborhood maze. (2) It identifies the optimal travel path that has the least number of obstacles. (3) The mapping is accepted if and only if the snake survives the grid maze.

\subsection{Method 1: Building the Weighted Neighborhood Maze}
The first step in the SneakySnake algorithm is to build the \textit{m} by \textit{m} weighted neighborhood maze (\textit{WN}). It is a modified version of our original neighborhood maze (see Chapter 6), where we change the meaning of the content of each cell. First, we change the value of the grid cell that represents a pairwise match from zero to the total number of the consecutive zeros in its lower right neighbors within the same diagonal vector. Second, we change the value of the grid cell that represents a pairwise mismatch from one to zero (modeled as a solid black cell). Given \textit{i} and \textit{j} (where 1$\leq$ \textit{i} $\leq$ \textit{m} and \textit{i-E} $\leq$ \textit{j} $\leq$ \textit{i+E}), the entry \textit{WN}[\textit{i, j}] of the weighted neighborhood maze can be calculated as follows:

\begin{equation} \textit{WN}[\textit{i,j}] =\begin{cases}WN[i+1,j+1]+1  & A[i]=B[j]\\ 0 & A[i]\neq B[j]\end{cases} \end{equation}

\begin{figure}
\centering
\includegraphics[width=8cm,keepaspectratio]{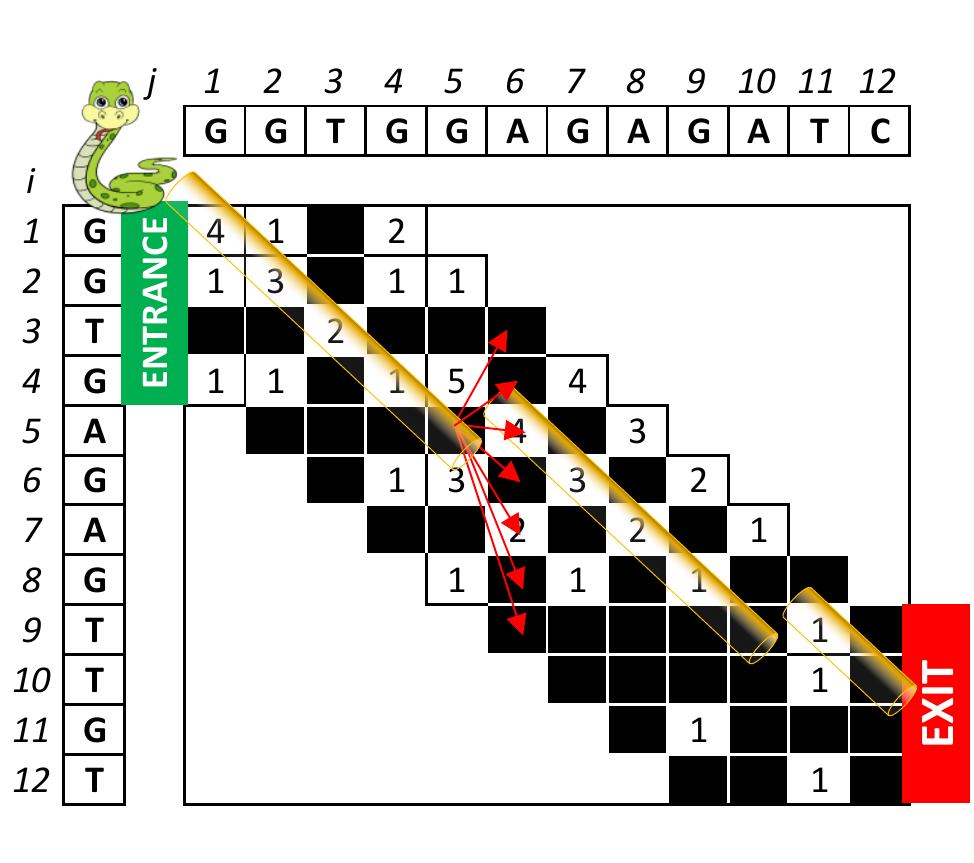}
\caption{Weighted Neighborhood maze (\textit{WN}), for reference sequence \textit{A} = GGTGGAGAGATC, and read sequence \textit{B} = GGTGAGAGTTGT for \textit{E}=3. The snake traverses the path that is highlighted in yellow which includes three obstacles and three pipes, that represents three identical subsequences: GGTG, AGAG, and T}
\label{fig:figure22SneakySnake}
\end{figure}

Computing each cell depends on its immediate lower right cell. This restricts the order of the computations to be performed starting from the lower right corner towards the upper left corner. We present in Figure \ref{fig:figure22SneakySnake} an example of a weighted neighborhood maze for two sequences, where the sequence \textit{B} differs from the sequence \textit{A} by three edits (i.e., obstacles).

\subsection{Method 2: Finding the Optimal Travel Path}
The second step of the SneakySnake algorithm is to decide on which pipe the snake should travel through, using three main steps. (1) The snake essentially uses the precomputed weight of each cell in the weighted neighborhood maze to make the decision. This is the key reason behind filling the weighted neighborhood maze from the lower right corner to the upper left corner, while the snake travels from the upper left corner towards the lower right corner. \newpage The snake examines the first cell of each diagonal vector in the first column of the grid and finds the cell with the largest weight. (2) If the snake is unable to decide on which pipe to follow (i.e., all cells of the first column are obstacles), the snake skips the current column and consumes one attempt out of the \textit{E} tries. The snake then finds the maximum weight of the cells in the next column. If it finds more than one pipe with the same length, it always chooses the first one (starting from the upper diagonal). (3) Once found, it travels through the pipe until it faces an obstacle. It repeats these three steps until it reaches its destination or consumes all the \textit{E} tries.

\subsection{Method 3: Examining the Snake Survival}
The last step is to check if the snake makes it through the grid maze or not. Given a limited number of tries (equals to \textit{E}), if the snake arrives the destination, then there exists a path that has at most \textit{E} obstacles. \newpage In the context of approximate edit distance calculation, it means that there exists an alignment that has at most \textit{E} edits. The SneakySnake algorithm considers this mapping as a correct one and accepts this mapping. Otherwise, it rejects the mapping without performing read alignment step.

\section{Unweighted Maze Methods}
Here, we explain the three steps of solving the Sneaky Snake Problem using the unweighted neighborhood maze approach. (1) It first constructs the unweighted neighborhood maze. (2) It then identifies the optimal travel path that has the least number of obstacles. (3) And finally the mapping is accepted if and only if the snake survives the grid maze without exceeding the limited number of tries.

\subsection{Method 1: Building the Unweighted Neighborhood Maze}
In the unweighted neighborhood maze, we introduce two changes to the way we build the neighborhood maze. First, each grid cell has no weight assigned to it. Second, we define a state for each grid cell, whereas a cell can be either in an available state or a blocked state. We set the grid cell to be in an available state if it represents a pairwise match. If the grid cell represents a pairwise mismatch, then we set its state to blocked. In this way, the unweighted neighborhood maze eliminates the data dependency between the grid cells that exists in the weighted neighborhood maze. Given \textit{i} and \textit{j} (where 1$\leq$ \textit{i} $\leq$ \textit{m} and \textit{i-E} $\leq$ \textit{j} $\leq$ \textit{i+E}), the state of the entry \textit{UN}[\textit{i, j}] of the unweighted neighborhood maze can be chosen as follows:

\begin{equation} \textit{UN}[\textit{i,j}] =\begin{cases}Available  & A[i]=B[j]\\ Blocked & A[i]\neq B[j]\end{cases} \end{equation}

\begin{figure}
\centering
\includegraphics[width=8cm,keepaspectratio]{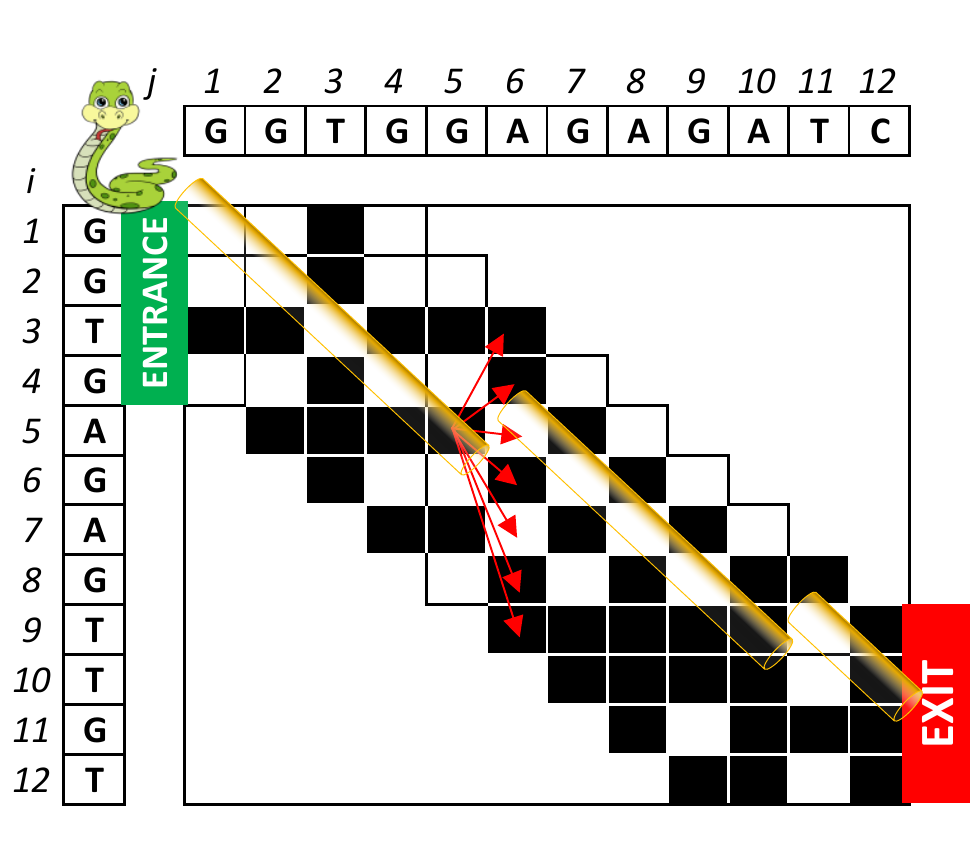}
\caption{Unweighted neighborhood maze (\textit{UN}), for reference sequence \textit{A} = GGTGGAGAGATC, and read sequence \textit{B} = GGTGAGAGTTGT for \textit{E}=3. The snake traverses the path that is highlighted in yellow which includes three obstacles and three pipes, that represents three identical subsequences: GGTG, AGAG, and T.}
\label{fig:figure23SneakySnake}
\end{figure}

We present in Figure \ref{fig:figure23SneakySnake} an example of an unweighted neighborhood maze for two sequences, where the sequence \textit{B} differs from the sequence \textit{A} by three edits. The state of the entry \textit{UN}[\textit{i,j}] is set to available if the \textit{i$^{th}$} character of the read sequence matches the \textit{j$^{th}$} character of the reference sequence. Otherwise, it is set to blocked (highlighted in a black).

\subsection{Method 2: Finding the Optimal Travel Path}
The second step of the SneakySnake algorithm is to decide on which pipe the snake should consider, in three main steps. (1) As the snake has no prior information on the length of each pipe (as in the weighted neighborhood maze), it uses its telescoping lens to perform depth-first search (DFS) \cite{tarjan1972depth} each time it faces a blocked cell (i.e., obstacle). The DFS algorithm traverses the first upper diagonal vector to reach the maximum possible depth (i.e., until it arrives a blocked cell or the end of the vector). If the DFS algorithm reaches the exit of the maze, then SneakySnake terminates the DFS search. \newpage The depth of each grid cell is equal to the value of its \textit{j} index. It stores the maximum possible depth and backtracks to the starting point (i.e., the obstacle cell). It then continues traversing the next unsearched diagonal vector and repeats the previous steps. (2) The snake selects the path with the largest depth. If it is unable to find one (i.e., all cells of the first column are obstacles), it skips the current column and consumes one path transfer out of the \textit{E} allowed transfers. (3) If it finds multiple equi-length pipes, it always chooses the first one (starting from the upper diagonal). Once found, it travels through the pipe until it faces another obstacle. It repeats these three steps until it reaches its destination or consumes all the \textit{E} tries.

\subsection{Method 3: Examining the Snake Survival}
The last step is to accept the mapping based on whether the snake arrives its destination given a limited number of path transfers (equals to \textit{E}). This means that there is an alignment for the two given sequences that has at most \textit{E} edits. Otherwise, The SneakySnake algorithm rejects the mapping without performing read alignment step.

\section{Analysis of SneakySnake Algorithm}
In this section, we analyze the asymptotic run time and space complexity of the SneakySnake algorithm. The SneakySnake algorithm builds the weighted neighborhood maze by traversing through each vector starting from the lower right corner. It compares the corresponding characters of the two given sequences and if they match each other, then it updates the current cell with the result of summing up one and the value of the previous cell. Assuming it takes O(\textit{m}) time to build one diagonal vector, where \textit{m} is the read length. Then it takes O(\textit{km}) to build the entire weighted maze, where \textit{k} = 2\textit{E}+1 is the number of the grid vectors. \newpage Upon arriving an obstacle, the SneakySnake algorithm examines the weight of each cell following the obstacle cell and picks the cell with the largest weight as a new starting point. With the existence of \textit{E} obstacles, this step is repeated at most \textit{E} times. Each time it takes O(\textit{k}) time to compare the weight of \textit{k} cells. Thus, the upper-bound on the time complexity of SneakySnake using weighted neighborhood maze is given as follows:
\begin{equation}\textit{T$_{Weighted\_SneakySnake}$(m)} = O(Em+hk) \end{equation}
Where \textit{h} is a number satisfies the following range: 0 $\leq$ \textit{h} $\leq$ \textit{E}. 

Using the unweighted neighborhood maze, the SneakSnake algorithm does not necessarily traverse all the cells of each diagonal vector (as in the weighted approach described above). Thus, its asymptotic run time is not determined. On the one hand, the lower-bound on the its time complexity is O(\textit{m}), which is achieved when the DFS algorithm reaches the exist of the maze without facing any obstacle along the path. On the other hand, the loose upper-bound run time complexity is also equal to O(\textit{km+hk}) when the DFS traverses through nearly the entire unweighted neighborhood maze. However, it is unrealistic to traverse the entire maze, as in this case the value of each and every cell of the entire maze should be equal to \lq{}0\rq{} (for example, when all characters of the two sequences are \lq{}A\rq{}s). If this is the case, then the DFS algorithm traverses only through the first upper diagonal and then get terminated. Thus, the run time complexity of SneakySnake with the unweighted neighborhood maze is given as follows:
\begin{equation} O(m) \leq \textit{T$_{Unweighted\_SneakySnake}$(m)} < O(km+hk) \end{equation}
Where \textit{h} is a number satisfies the following range: 0 $\leq$ \textit{h} $\leq$ \textit{E}. This in turn demonstrates that the unweighted neighborhood maze algorithm is asymptotically inexpensive compared to the weighted SneakySnake algorithm. Hence, we consider the unweighted SneakySnake algorithm for further analysis and evaluation. We provide the pseudocode of SneakySnake in Algorithm 8.1. While the weighted SneakySnake requires no additional auxiliary space, the weighted approach requires only storing the depth of at most 2\textit{E}+1 vectors.

\begin{center}
\includegraphics[width=\textwidth]{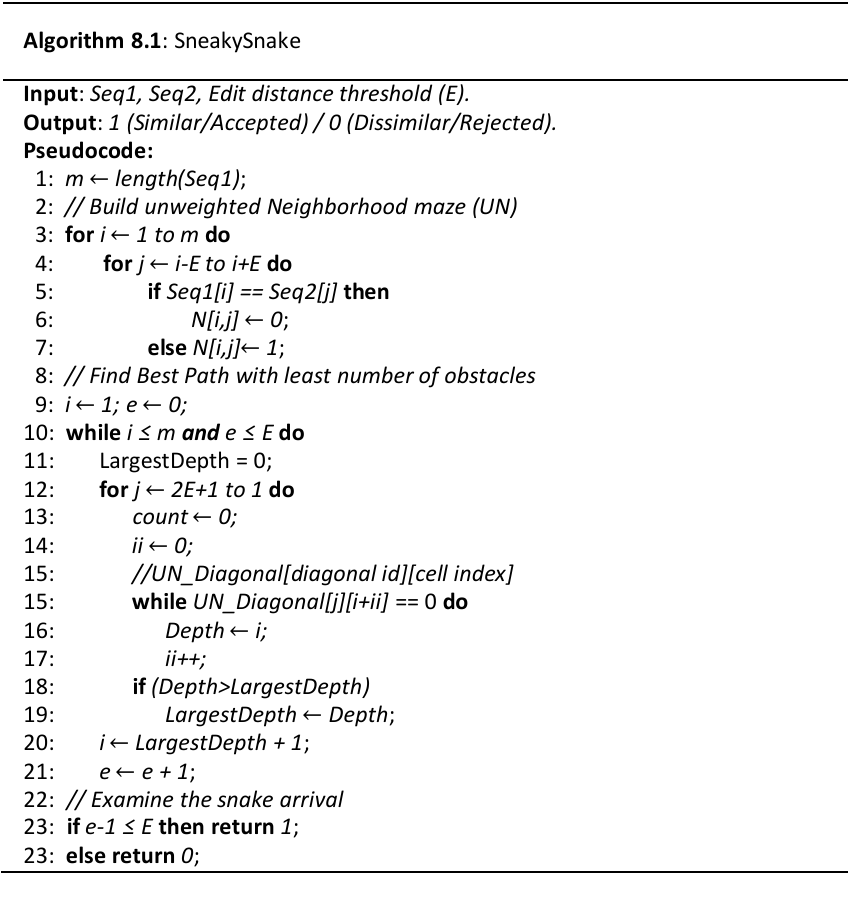}
\end{center}

\section{SneakySnake on an FPGA}
We introduce an FPGA-friendly architecture for the unweighted SneakySnake algorithm. Achieving an efficient hardware architecture raises the following question: \textbf{Can one solve many small  sub-problems of the Sneaky Snake Problem with a high parallelism by reducing the search space of the SneakySnake algorithm?} \newpage The main idea behind the hardware architecture of the SneakySnake algorithm is to partition the unweighted neighborhood maze into small non-overlapping sub-matrices with a width of \textit{t} columns (the height is always fixed to 2\textit{E}+1 rows) for some parameter \textit{t}. We then apply the SneakySnake algorithm to each sub-matrix independently from the other sub-matrices. This results in three key benefits. First, downsizing the search space into a reasonably small sub-matrix with a known dimension at the design time limits the number of all possible solutions for that sub-matrix. This reduces the size of the look-up tables (LUTs) required to build the architecture and simplifies the overall design. Second, this approach helps to maintain a modular and scalable architecture that can be implemented for any read length and edit distance threshold. Third, all the smaller sub-problems can be solved independently and rapidly with a high parallelism. This reduces the execution time of the overall algorithm as the SneakySnake algorithm does not need to evaluate the entire path. However, these benefits come at the cost of accuracy degradation. The solution for each sub-problem is not necessarily part of the solution for the main problem (with the original size of \textit{m} by \textit{m}). Though it is guaranteed to always choose the path with the least obstacles within the search space, our hardware architecture can underestimate the number of obstacles found and thereby increase the false accept rate.

\begin{figure}
\centering
\includegraphics[width=10cm,keepaspectratio]{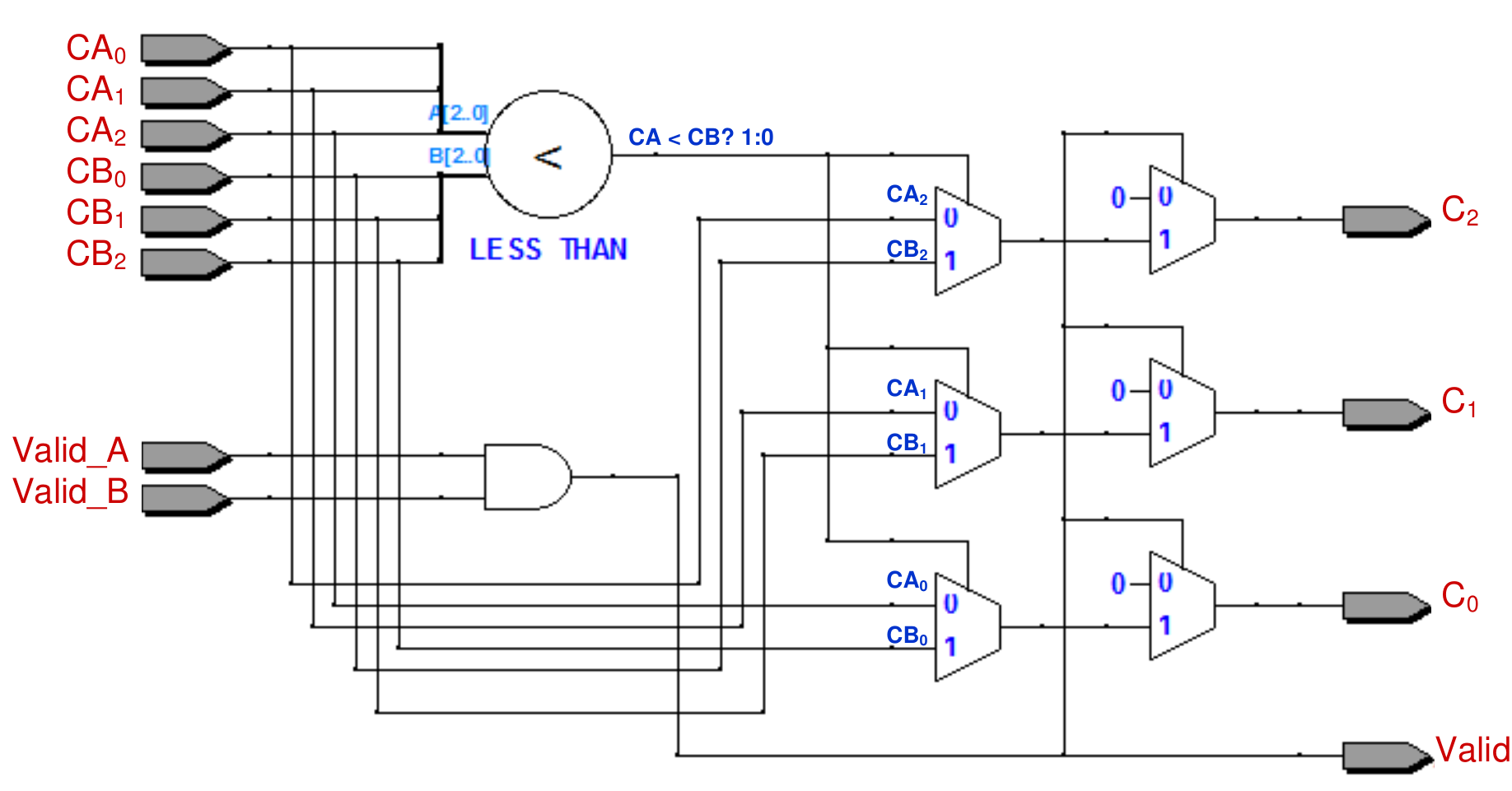}
\caption{Proposed 4-bit LZC comparator.}
\label{fig:figure391SneakySnake}
\end{figure}

\begin{figure}
\centering
\includegraphics[width=16cm,keepaspectratio]{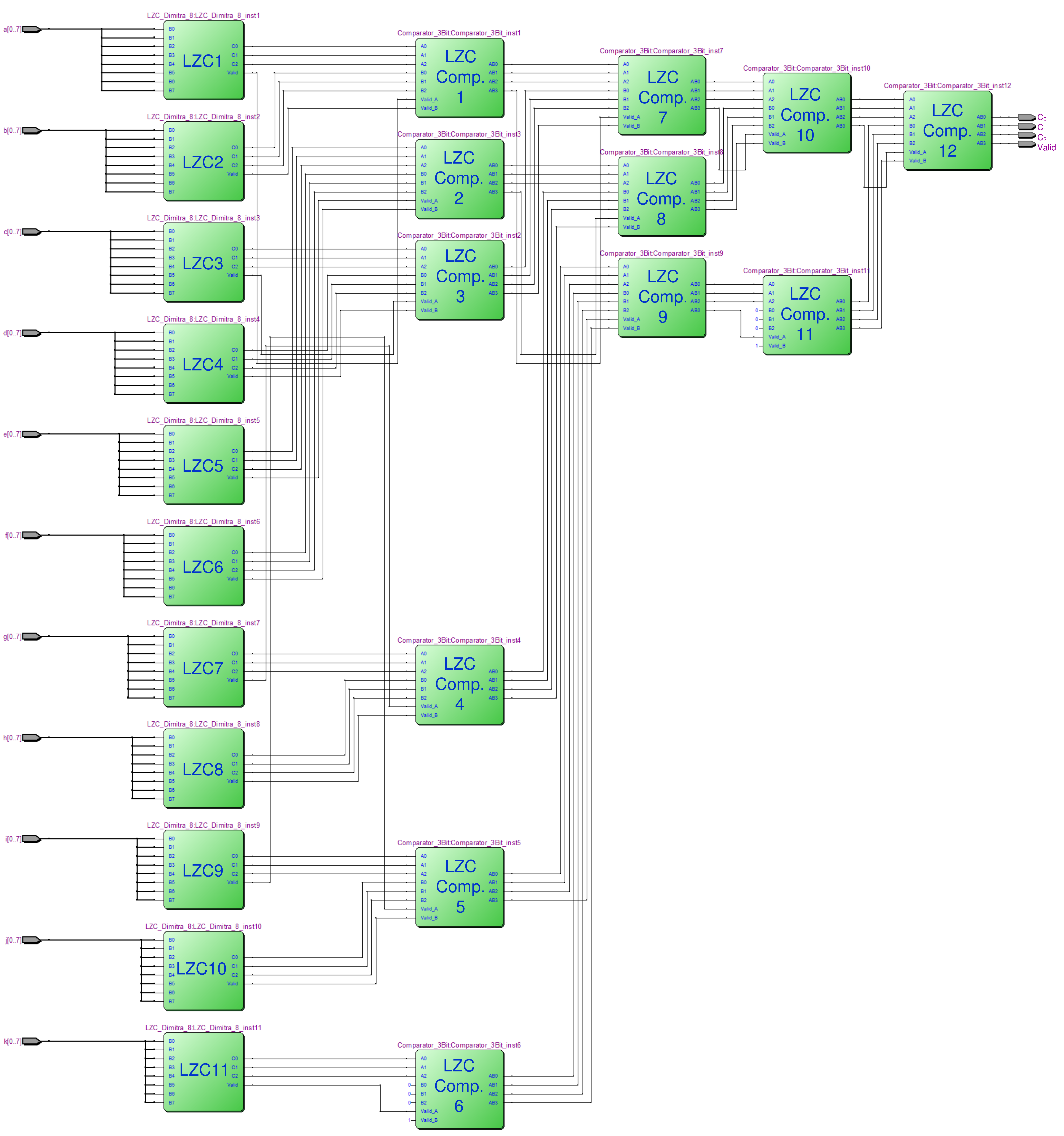}
\caption{Block diagram of the 11 LZCs and the hierarchical LZC comparator tree for computing the largest number of leading zeros in 11 rows.}
\label{fig:figure401SneakySnake}
\end{figure}

Next, we present the details of our hardware architecture. We choose the parameter \textit{t} to be 8 columns. This results in partitioning the unweighted neighborhood maze of size 100 by 100 (or 2\textit{E}+1 by 100 computed cells) into 13 sub-matrices, each of size 2\textit{E}+1 by 8. Each row in the sub-matrix is part of the diagonal vector of the unweighted neighborhood maze. Each sub-matrix represents an individual Sneaky Snake Problem. Solving each problem requires determining the optimal path (with the least number of obstacles) along each sub-matrix using five main steps. (1) The first step is to perform the DFS search for finding the optimal path along the 2\textit{E}+1 rows of the sub-matrix. We implement the DFS algorithm on FPGA as a leading-zero counter (LZC). We use the LZC design proposed in \cite{dimitrakopoulos2008low}. It counts the number of consecutive zeros that appear in a n-bit input word before the first more significant bit that is equal to one. \newpage It generates two output signals, the log$_{2}$n bits of the leading-zero count \textit{C} and a flag \textit{Valid}, for an input word \textit{A} = $A_{n-1}$, $A_{n-2}$, \dots ,$A_{0}$ , where $A_{n-1}$ is the most-significant bit. When all input bits are set to zero, the \textit{Valid} flag is set to one. For other cases of input, the value of \textit{C} represents the number of leading zeros. For the case of an 8-bit input operand, the two output signals of the LZC are given by:
\begin{equation} 
\begin{aligned}
\overline{C_2} & = A_7 + A_6 + A_5 + A_4 \\
\overline{C_1} & = A_7 + A_6 + \overline{A_5} \cdot  \overline{A_4} \cdot (A_3+A_2)  \\
\overline{C_0} & = A_7 + \overline{A_6} \cdot A_5 + \overline{A_6} \cdot \overline{A_4} \cdot A_3 + \overline{A_6} \cdot \overline{A_4} \cdot \overline{A_2} \cdot A_1 \\
Valid & = A_7 + A_6 + A_5 + A_4 + A_3 + A_2 + A_1 + A_0
\end{aligned}
\end{equation}
(2) The second step is to find the row that has the largest number of leading zeros. We build a hierarchical comparator structure with log$_{2}$(2\textit{E}+1) levels. We use a single \textit{LZC comparator} for comparing the number of leading zeros of two rows. The LZC comparator compares two numbers of leading zeros produced by two LZC circuits and passes the largest number as output signals without changing their values (maintaining the same meaning). We provide in Figure \ref{fig:figure391SneakySnake} the proposed architecture of the 4-bit LZC comparator. For an edit distance thresholds of 5 bp, we need 12 LZC comparators arranged into 4 levels (6, 3, 2, and 1 LZC comparators in each level, respectively). We provide the overall architecture of the 4-level LZC comparator tree including the 11 LZC block diagrams in Figure \ref{fig:figure401SneakySnake}.

(3) The third step is to store the largest number of leading zeros and shift the bits of all rows to the right direction. As we implement the DFS search as a LZC circuit, we always need to start the DFS search space from the least significant bit of each row. The shift operation is necessary for deleting the leading zeros and allowing the snake to continue the process of finding the optimal path. We shift each row of the sub-matrix by \textit{x}+1 bits, where \textit{x} is the largest number of leading zeros found in step 2. This guarantees to exclude the found path from the next search round along with the edit, which divides the optimal path into shorter paths.

(4) The fourth step is to repeat the previous three steps in order to find the optimal path from the least significant bit all the way to the most significant bit. The number of replications needed depends on the desired accuracy of the SneakySnake algorithm. If our target is to find at most a single edit within each sub-matrix, then we need to build two replications for the steps described above. For example, let \textit{A} be 00010000, where \textit{t} = 8. The first replication computes the value of \textit{x} as four zeros and updates the bits of \textit{A} to 11111000. The second replication computes the value of \textit{x} as three zeros and updates the bits of \textit{A} to 11111111.

(5) The last step is to calculate the total number of obstacles along the entire optimal path from the least significant bit towards the most significant bit. We first find out the number of replications that produces at least a single leading zero (i.e., \textit{x}>0). If it equals the total number of replications (\textit{y}), then the number of obstacles (edits) equals to the total number of replications included in the design. Otherwise, we compute the number of the obstacles as follows:
\begin{equation}
min(y, t - \sum_{o=1}^{y} x_{o})
\end{equation}
where \textit{y} is the total number of replications involved in the design and \textit{$x_{o}$} is the largest number of leading zeros produced by the replication of index \textit{o}.

\section{Discussion}
We introduce the SneakySnake algorithm and hardware architecture in order to address the question that we raise earlier in this chapter about weather approximating the edit distance calculation can be faster than calculating the exact edit distance. The accuracy of the edit distance approximation is another concern. Our SneakySnake filter always underestimates the total number of edits. This is mainly due to the goal of the Sneaky Snake Problem that is the search for optimal path with the least number of obstacles (or edits). This concern raises two key questions. (1) \textbf{Can one improve the accuracy of edit distance approximation such that we achieve either a new optimal read aligner or a highly-accurate pre-alignment filter?} The SneakySnake algorithm can be slightly modified such that it considers a penalty on the selected path. If the snake selects the upper diagonals, then this means that the obstacle is basically a deleted character at the \textit{i$^{th}$} index of the read sequence. Similarly, if the current path is at the main diagonal and the next selected path is at the second upper diagonal, this means that the obstacle is two deleted characters. The current implementation of SneakySnake considers this two deleted characters as a single edit. The accuracy improvement of SneakySnake is yet to be explored in this thesis. 

The second question is (2) Similarly to the way we partition the search space of the SneakySnake algorithm, \textbf{Can one reduce the search space of exact edit distance algorithms such that only the necessary cells are computed?} We observe the fact that the exact edit distance algorithms explore a large area of the dynamic programming matrix (even with the banded matrix), which is unnecessary for highly dissimilar sequences. The edit distance is considered to be a non-additive distance measure \cite{calude2002additive}. This means that its calculations can not be distributed over concatenated subsequences of the long sequence. In other words, we can not divide the read-reference pair into shorter pairs and calculate the exact edit distance for each short read-reference pair individually (aiming at concurrently computing them) and then accumulate the results. \newpage For example, take \textit{A} = \lq\lq{}CGCG\rq\rq{}, \textit{B} = \lq\lq{}GCGC\rq\rq{} and observe that calculating the edit distance for each character of \textit{A} with its corresponding character of \textit{B} yields four edits, while the edit distance between \textit{A} and \textit{B}, each as a whole, is only 2 edits. This is clear from the way that its dynamic programming matrix is computed. The computations always depend on the prefixes of both the read and the reference sequences. As a workaround, we can compute the exact edit distance in an incremental approach. One can divide the dynamic programming matrix into sub matrices, where each smaller sub-matrix is fully covered by all the larger sub-matrices. Such that each sub-matrix \textit{UN$_{s}$} strictly starts from the index \textit{i} of value 1 up to some s, where the following equation holds 1 $\leq$ \textit{i} $\leq$ \textit{s} $\leq$ \textit{m} and \textit{i-E} $\leq$ \textit{j} $\leq$ \textit{i+E}. We refer to this edit distance measure as \textbf{\textit{a prefix edit distance}}. In the next chapter, we evaluate all these scenarios in details.

\section{Summary}
In this chapter, we introduce the Sneaky Snake Problem, and we show how an approximate edit distance problem can be converted to an instance of the Sneaky Snake Problem. Subsequently, we propose a new pre-alignment filtering algorithm (we call it SneakySnake) that obviates the need for expensive specialized hardware. The solution we provide is cost-effective given a limited resources environment. Our algorithm does not exploit any SIMD-enabled CPU instructions or vendor-specific processor. This makes it superior and attractive. We also provide efficient and scalable hardware architecture along with several design optimizations for the SneakySnake algorithm. Finally, we discuss several optimizations and challenges of accelerating both approximate and exact edit distance calculations.

%% file: chapter9.tex
\chapter{Evaluation}
In this chapter, we evaluate the FPGA resource utilization, the filtering accuracy, and the memory utilization of all our proposed pre-alignment filters. We also investigate the benefits of using our hardware and CPU-based pre-alignment filtering solutions along with the state-of-the-art aligners. We compare the performance of our proposed pre-alignment filters (SneakySnake, MAGNET, Shouji, and GateKeeper) with the state-of-the-art existing pre-alignment filter, SHD \cite{xin2015shifted} and read aligners, Edlib \cite{vsovsic2017edlib}, Parasail \cite{daily2016parasail}, GSWABE \cite{liu2015gswabe}, CUDASW++ 3.0 \cite{liu2013cudasw++}, and FPGASW \cite{fei2018fpgasw}. We run all experiments using 3.6 GHz Intel i7-3820 CPU with 8 GB RAM. We use a Xilinx Virtex 7 VC709 board \cite{virtex7fpga} to implement our accelerator architecture and our hardware filters. We build the FPGA designs using Vivado 2015.4 in synthesizable Verilog. 

\section{Dataset Description}
Our experimental evaluation uses 12 different real datasets. Each dataset contains 30 million real sequence pairs. Next, we elaborate on how we obtain them. \newpage We obtain three different read sets (ERR240727\_1, SRR826460\_1, and SRR826471\_1) of whole human genome that include three different read lengths (100 bp, 150 bp, and 250 bp, respectively), as summarized in Table \ref{table:table2Data}. We download these three read sets from EMBL-ENA ( http://www.ebi.ac.uk/ena). We map each read set to the human reference genome (GRCh37) using mrFAST \cite{alkan2009personalized} mapper. We obtain the human reference genome from 1000 Genomes Project \cite{10002012integrated}. For each read set, we use four different maximum number of edits using the \textit{-e} parameter of mrFAST to generate four real datasets. We summarize the details of these 12 datasets in Table \ref{table:table3Data}. For the convenience of referring to these datasets, we number them from 1 to 12 (e.g., set\_1 represents 30 million reads from ERR240727\_1 mapped with \textit{-e} = 2 edits). The 12 real datasets enable us to measure the effectiveness of the filters in tolerating low number of edits and far more edits than the allowed edit distance threshold. We provide detailed information on the number of correct and incorrect pairs of each of the 12 datasets for different user-defined edit distance thresholds in Table \ref{table:table1Appendix}, Table \ref{table:table2Appendix}, and Table \ref{table:table3Appendix} in Appendix A.

\begin{table}
\centering
\caption{Benchmark illumina-like read sets of whole human genome, obtained from EMBL-ENA.}
\label{table:table2Data}
\includegraphics[width=11.5cm]{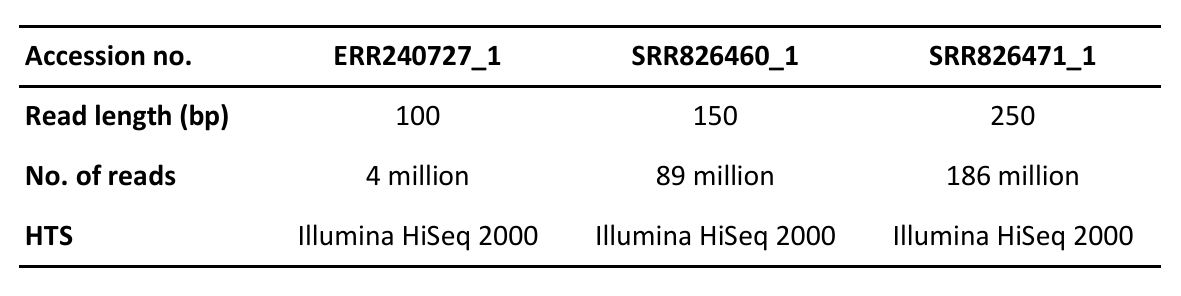}
\end{table}

\begin{table}
\centering
\caption{Benchmark illumina-like datasets (read-reference pairs). We map each read set, described in Table \ref{table:table2Data}, to the human reference genome in order to generate four datasets using different mapper\rq{}s edit distance thresholds (using \textit{-e} parameter).}
\label{table:table3Data}
\includegraphics[width=13cm]{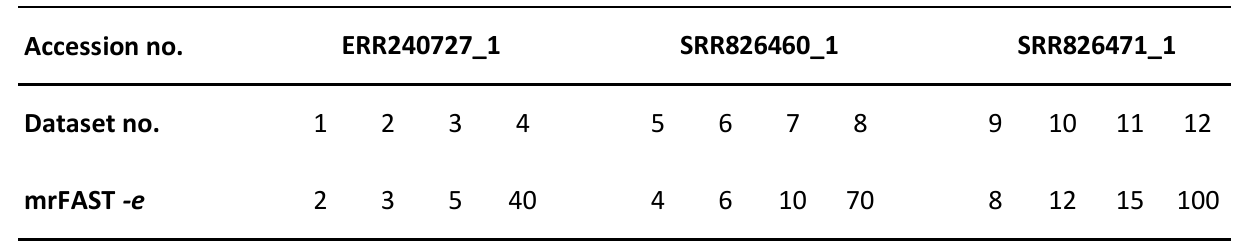}
\end{table}

\section{Resource Analysis}
We now examine the FPGA resource utilization for the hardware implementation of GateKeeper, Shouji, MAGNET, and SneakySnake pre-alignment filters. We provide several hardware designs for two commonly used edit distance thresholds, 2 bp and 5 bp as reported in \cite{cheng2015bitmapper, xin2015shifted, hatem2013benchmarking, ahmadi2011hobbes}, for a sequence length of 100 bp. The VC709 FPGA chip contains 433,200 slice LUTs (look-up tables) and 866,400 slice registers (flip-flops). Table \ref{table:table4Data} lists the FPGA resource utilization for a single filtering unit. We make five main observations. 
\begin{enumerate}
\item The design for a single MAGNET filtering unit requires about 10.5\% and 37.8\% of the available LUTs for edit distance thresholds of 2 bp and 5 bp, respectively. Hence, MAGNET can process 8 and 2 sequence pairs concurrently for edit distance thresholds of 2 bp and 5 bp, respectively, without violating the timing constraints of our hardware accelerator.
\item The design for a single Shouji filtering unit requires about 15x-21.9x less LUTs compared to MAGNET. This enables Shouji to achieve more parallelism over MAGNET design as it can have 16 filtering units within the same FPGA chip. 
\item GateKeeper requires about 26.9x-53x and 1.7x-2.4x less LUTs compared to MAGNET and Shouji, respectively. GateKeeper can also examine up to 16 sequence pairs at the same time.
\item SneakySnake requires 15.4x-26.6x less LUTs compared to MAGNET. While SneakySnake requires a slightly less LUTs compared to Shouji, it requires about 2x more LUTs compared to GateKeeper. SneakySnake can also examine up to 16 sequence pairs concurrently.
\item We observe that the hardware implementations of Shouji, MAGNET, and SneakySnake require pipelining the design (i.e., shortening the critical path delay of each processing core by dividing it into stages or smaller tasks) to enable meeting the timing constraints and achieve more parallelism. \newpage We build 8 pipeline stages for Shouji, 22 pipeline stages for MAGNET, and 5 pipeline stage for SneakySnake to satisfy the timing constraints. However, pipelining the design comes with the expense of increased register utilization. 
\end{enumerate}

We conclude that the FPGA resource usage is correlated with the filtering accuracy. For example, the least accurate filter, GateKeeper, occupies the least FPGA resource that can be integrated into the FPGA. We also conclude that the less the logic utilization of a single filtering unit, the more the number of filtering units.

\begin{table}
\centering
\caption{FPGA resource usage for a single filtering unit of GateKeeper, Shouji, MAGNET, and SneakySnake for a sequence length of 100 and under different edit distance thresholds (\textit{E}).}
\label{table:table4Data}
\includegraphics[width=13cm]{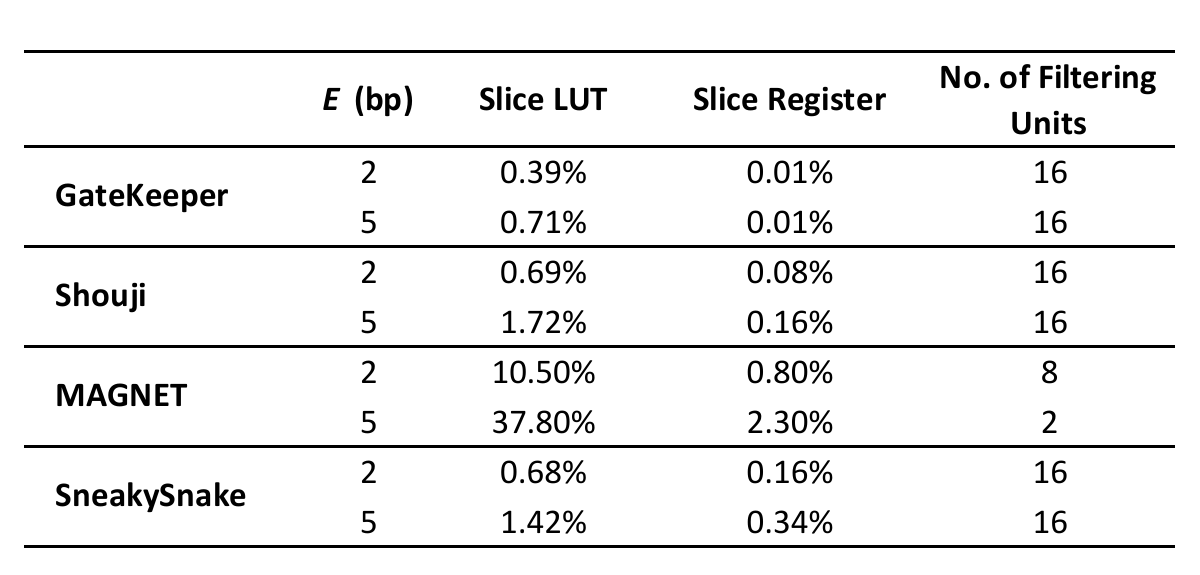}
\end{table}

\section{Filtering Accuracy}
Next, we assess the false accept rate and false reject rate of GateKeeper, Shouji, MAGNET, and SneakySnake across our 12 datasets. We also investigate and address several concerns that we raise in Chapter 8. We compare the accuracy performance of our proposed pre-alignment filters with the best performing existing pre-alignment filter, SHD \cite{xin2015shifted}. \newpage SHD supports a sequence length of up to only 128 characters (due to the SIMD register size). To ensure as fair a comparison as possible, we allow SHD to divide the long sequences into batches of 128 characters, examine each batch individually, and then sum up the results. As we describe in Chapter 3, we aim to minimize the false accept rate so that the elimination of dissimilar sequences is maximized. We also aim to maintain a 0\% false reject rate. We use Edlib \cite{vsovsic2017edlib} (set to edit distance mode) to generate the ground truth edit distance value for each sequence pair as it has a zero false accept rate and a zero false reject rate.

\subsection{Partitioning the Search Space of Approximate and Exact Edit Distance}
\begin{figure}
\centering
\includegraphics[width=\textwidth]{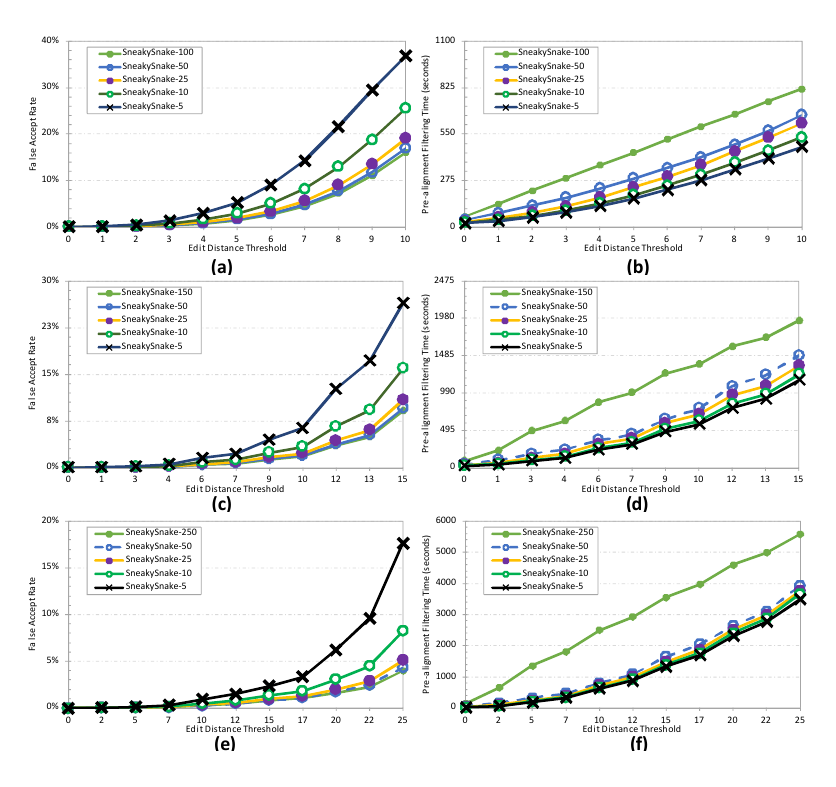}
\centering
\caption{The effects of column-wise partitioning the search space of SneakySnake algorithm on the average false accept rate ((a), (c), and (e)) and the average execution time ((b), (d), and (f)) of examining set\_1 to set\_4 in (a) and (b), set\_5 to set\_8 in (c) and (d), and set\_9 to set\_12 in (e) and (f). Besides the default size (equals the read length) of the SneakySnake\rq{}s unweighted neighborhood maze, we choose partition sizes (the number of grid\rq{}s columns that are included in each partition) of 5, 10, 25, and 50 columns.}
\label{fig:figure27Evaluation}
\end{figure}

We raise two key questions in Chapter 8. 
(1) Can one approximate the edit distance between two sequences much faster than calculating the edit distance? (2) Can one reduce the search space of approximate and exact edit distance algorithms? To answer these two questions, we first evaluate the performance of SneakySnake (approximate edit distance algorithm) and then examine the feasibility of reducing its search space without causing falsely-rejected mappings. Secondly, we examine the ability to implement the best performing existing exact edit distance algorithm, Edlib \cite{vsovsic2017edlib} such that it calculates the prefix edit distance. As we discuss in Chapter 8, we column-wise partition the unweighted neighborhood maze of SneakySnake algorithm into adjacent non-overlapping sub-matrices of the same size (2\textit{E}+1 by \textit{t}). In Figure \ref{fig:figure27Evaluation}, we illustrate the effects of this partitioning on the false accept rate and the execution time of our SneakySnake algorithm. We make two observations. 
\begin{enumerate}
\item \textbf{Partitioning the search space of SneakySnake with a partition size of 5 (\textit{t}=5) reduces its execution time by up to 5.12x (Figure \ref{fig:figure27Evaluation}a), 7.4x (Figure \ref{fig:figure27Evaluation}c), and 13.2x (Figure \ref{fig:figure27Evaluation}e)} at the expense of increased false accept rate by up to 55.4x (Figure \ref{fig:figure27Evaluation} b), 43.5x (Figure \ref{fig:figure27Evaluation}d), and 67.3x (Figure \ref{fig:figure27Evaluation}f). 
\item There is a trade-off between the speed and the accuracy of SneakySnake algorithm. For example, the least accurate filter, SneakySnake with a partition size of 5 columns (or in short SneakySnake-5) yields the fastest speed. 
\end{enumerate}

\begin{figure}
\centering
\includegraphics[width=\textwidth]{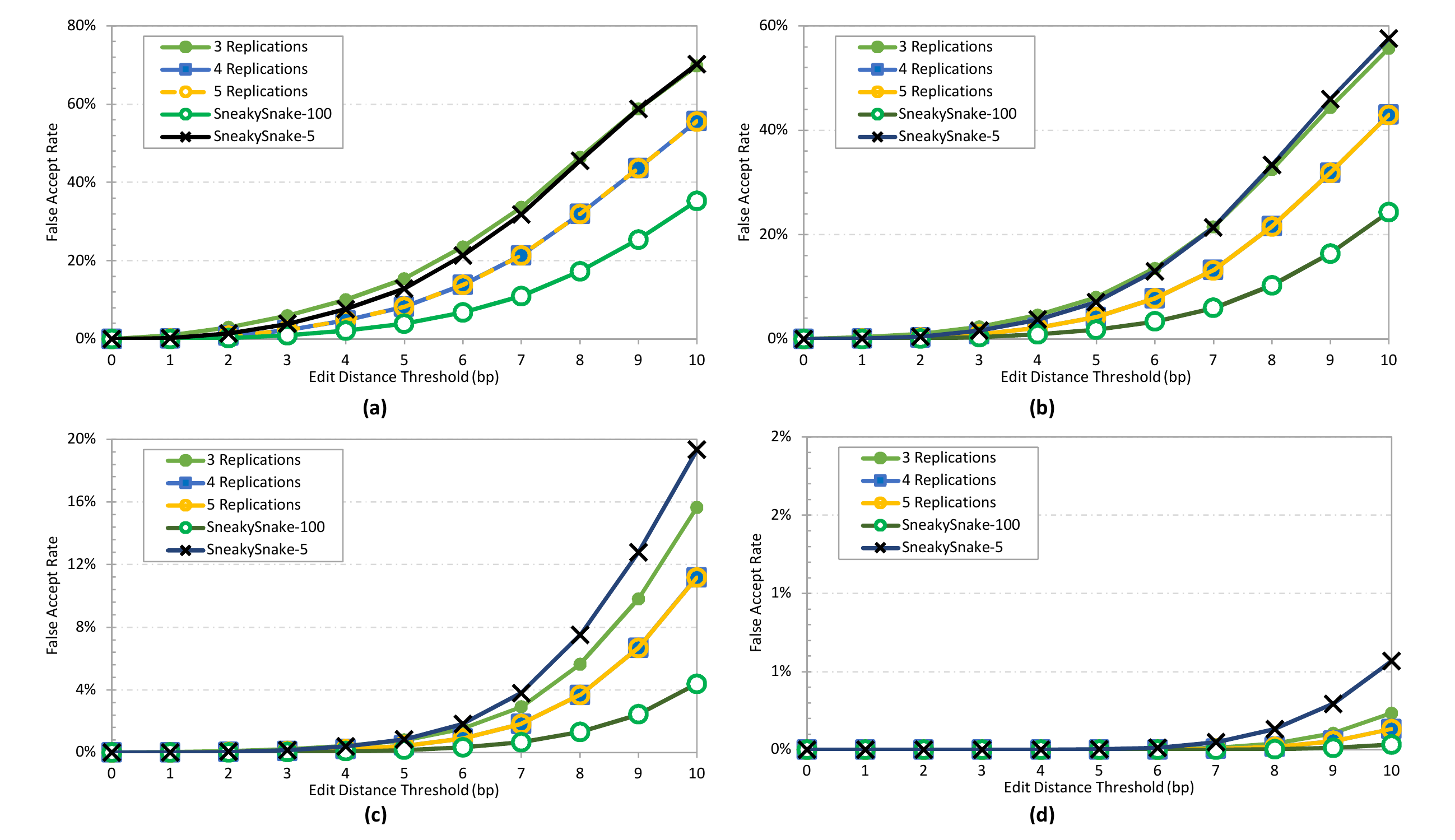}
\centering
\caption{The effects of the number of replications of the hardware implementation of SneakySnake algorithm on its filtering accuracy (false accept rate). We use a wide range of edit distance thresholds (0 bp-10 bp for a read length of 100 bp) and four datasets: (a) set\_1, (b) set\_2, (c) set\_3, and (d) set\_4.}
\label{fig:figure41Evaluation}
\end{figure}

Next, we assess the effect of the number of replications on the filtering accuracy of the hardware implementation of the SneakySnake algorithm. We use a sub-matrix's width of 8 columns (\textit{t}=8) and we vary the height of the sub-matrix from 1 row (i.e., \textit{E}=0 bp) up to 21 rows (i.e., \textit{E}=10 bp). Based on Figure \ref{fig:figure41Evaluation}, we make two observations:

\begin{enumerate}
\item We observe that increasing the number of the replications in the design improves the filtering accuracy of the SneakySnake algorithm. This observation is in accord with our expectation as each replication detects at most a single edit within each sub-matrix. 
\item We also observe that the hardware implementation of SneakySnake using 3 replications (3 iterations for finding the optimal path within each sub-matrix) achieves a similar accuracy performance (or slightly better) as that of the SneakySnake-5.  
\end{enumerate}

We conclude that partitioning the search space of the SneakySnake algorithm is also beneficial for building an efficient hardware architecture while maintaining high filtering accuracy.

\begin{figure}
\includegraphics[width=12cm,keepaspectratio]{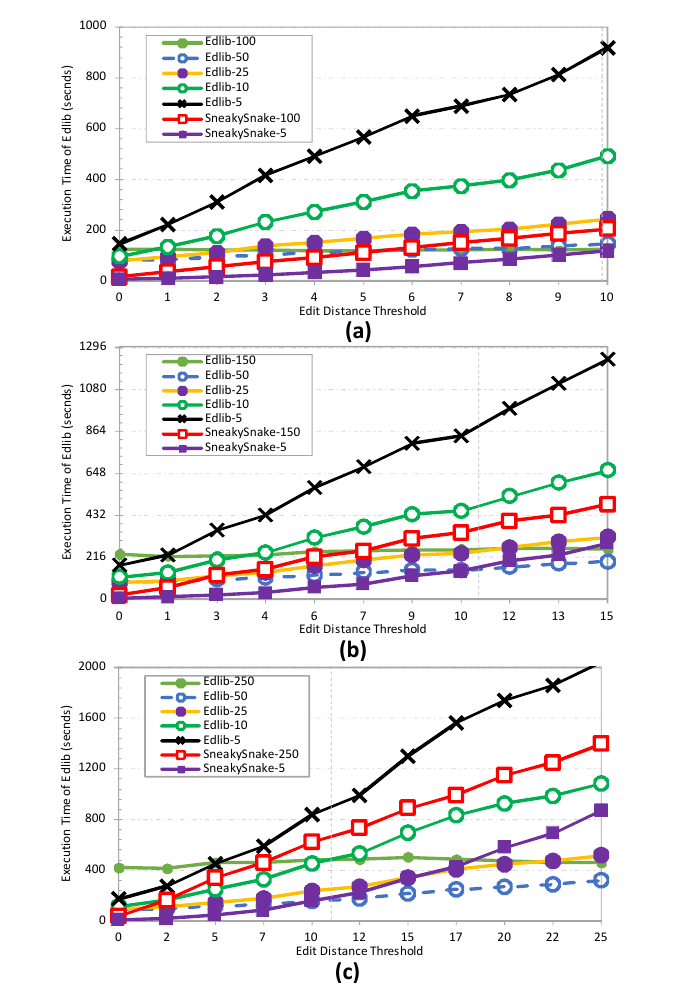}
\centering
\caption{The effects of computing the \textit{prefix edit distance} on the overall execution time of the edit distance calculations compared to the original Edlib (exact edit distance) and our partitioned implementation of SneakySnake algorithm. We present the average time spent in examining set\_1 to set\_4 in (a), set\_5 to set\_8 in (b), and set\_9 to set\_12 in (c). We choose initial sub-matrix sizes of 5, 10, 25, and 50 columns. We mark the intersection of SneakySnake-5 and Edlib-50 plots with a dashed vertical line.}
\label{fig:figure28Evaluation}
\end{figure}

Now, we modify Edlib algorithm such that it applies \textit{a prefix edit distance} (we provide the definition in Chapter 8). It starts computing only a small square sub-matrix. If the computed edit distance meets the user-defined edit distance threshold, then it extends the sub-matrix into a larger square one (which overlaps entirely with the smaller one) by increasing the number of columns and rows by a constant. For example, if the initial sub-matrix size is 5 columns by 5 rows, then we extend it into a larger one of size 10 columns by 10 rows, and next we extend it into a sub-matrix of size 15 columns by 15 rows. \newpage We keep extending the size of the sub-matrix until we cover the entire dynamic programming matrix or the prefix edit distance exceeds the edit distance threshold. In Figure \ref{fig:figure28Evaluation}, we present the effects of computing the prefix edit distance (partitioning the edit distance matrix) on the overall execution time. We also provide the performance of the original and the partitioned SneakySnake algorithm for a better comparison. We provide more detailed results in Table \ref{table:table4Appendix}, Table \ref{table:table5Appendix}, and Table \ref{table:table6Appendix} in Appendix A. We make two key observations. 
\begin{enumerate}
\item \textbf{Our SneakySnake algorithm with a partition size of 5 (SneakySnake-5) is up to 25.5x (Figure 
\ref{fig:figure28Evaluation}a), 52.5x (Figure 
\ref{fig:figure28Evaluation}b), and 94.5x (Figure 
\ref{fig:figure28Evaluation}c) faster than the best performing edit distance algorithm, Edlib \cite{vsovsic2017edlib}}. 

\item Prefix edit distance with large enough initial sub-matrix size provides a slight reduction in the execution time of Edlib. \newpage We observe that setting the initial sub-matrix size to 50 bp provides the highest reduction in the Edlib\rq{}s execution time over a wide range of edit distance thresholds. Edlib with an initial sub-matrix size of 50 bp (or in short Edlib-50) is up to 1.9x (Figure 
\ref{fig:figure28Evaluation}a), 4.4x (Figure 
\ref{fig:figure28Evaluation}b), and 7.8x (Figure 
\ref{fig:figure28Evaluation}c) faster than the original Edlib. However, SneakySnake-5  is still up to an order of magnitude (19.8x) faster than Edlib-50 when the edit distance threshold is set to 9 (for \textit{m}=100) or 10 (for \textit{m}=150 or 250) and below, as highlighted with a dashed vertical line in Figure \ref{fig:figure28Evaluation}. Note that \textit{E} is typically less than or equal 5\% of the read length \cite{cheng2015bitmapper, xin2015shifted, hatem2013benchmarking, ahmadi2011hobbes}.
\end{enumerate}

We conclude that reducing the search space of both approximate and exact edit distance algorithms is beneficial. Combining SneakySnake-5 and Edlib-50 can provide a fast examination for incorrect mappings across a wide range of edit distance thresholds (0\% - 10\% of the read length).

\subsection{False Accept Rate}
\begin{figure}
\centering
\includegraphics[width=\textwidth]{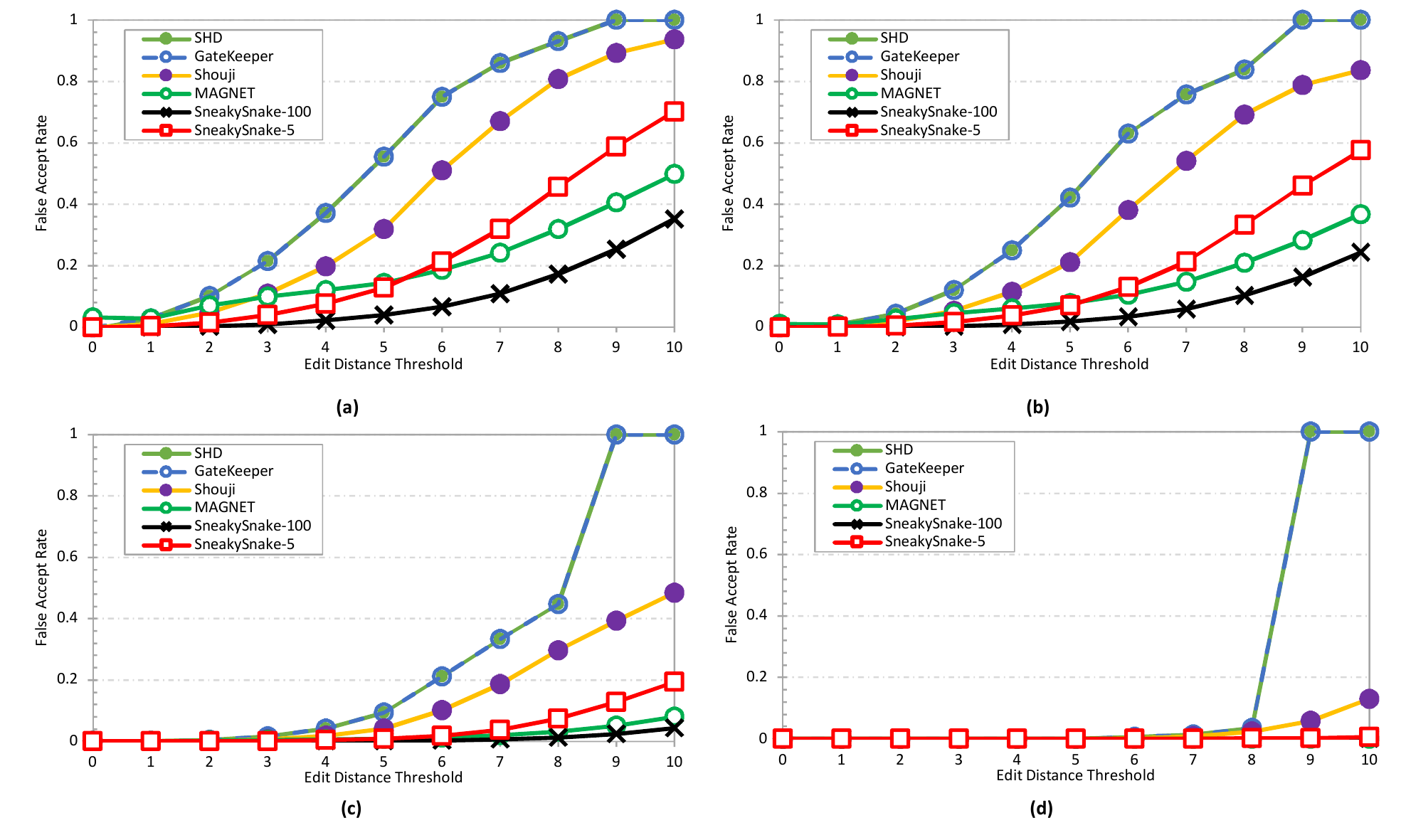}
\caption{The false accept rate produced by our pre-alignment filters, GateKeeper, Shouji, MAGNET, and SneakySnake, compared to the best performing filter, SHD \cite{xin2015shifted}. We use a wide range of edit distance thresholds (0-10 edits for a read length of 100 bp) and four datasets: (a) set\_1, (b) set\_2, (c) set\_3, and (d) set\_4.}
\label{fig:figure26Evaluation}
\end{figure}

\begin{figure}
\centering
\includegraphics[width=\textwidth]{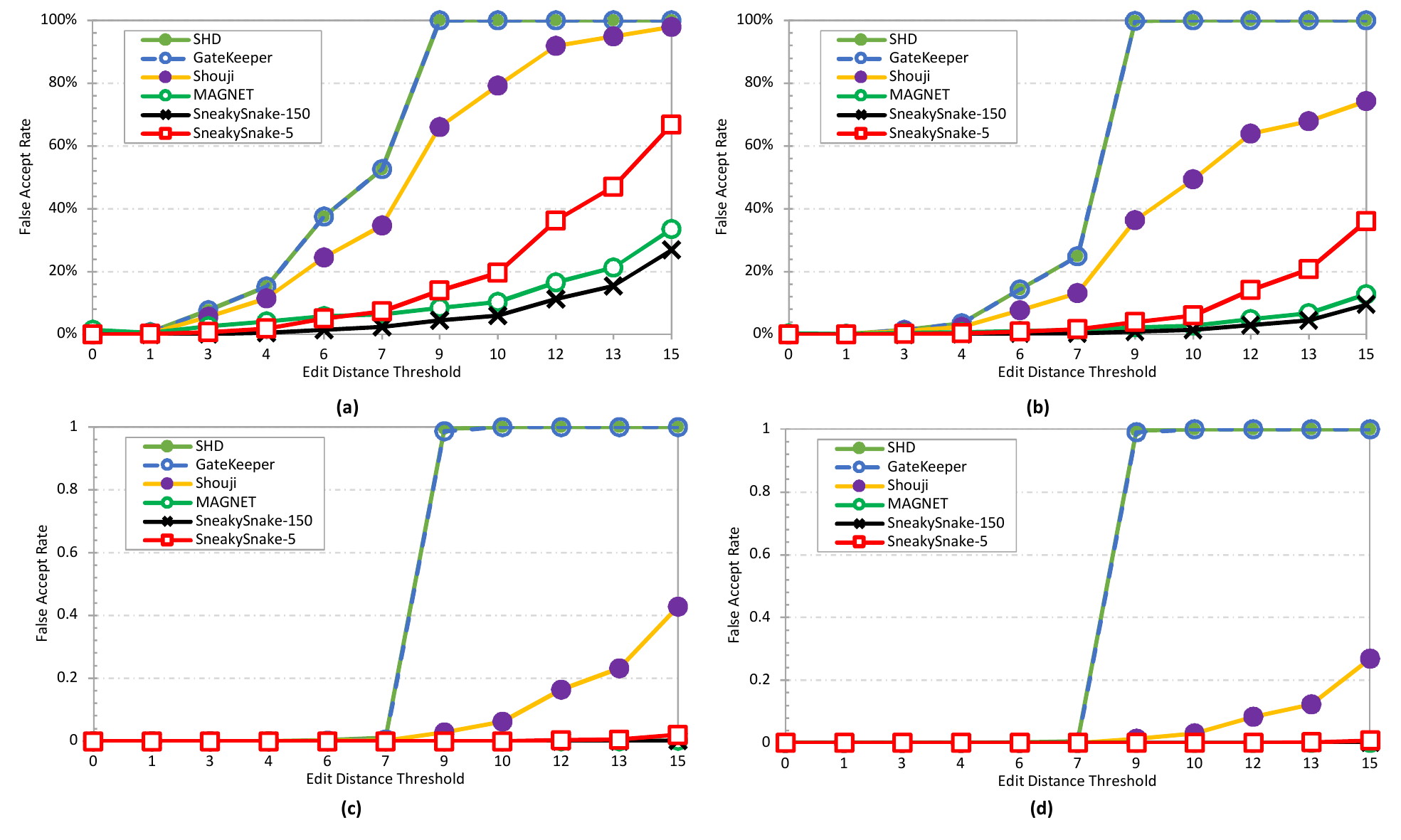}
\caption{The false accept rate produced by our pre-alignment filters, GateKeeper, Shouji, MAGNET, and SneakySnake, compared to the best performing filter, SHD \cite{xin2015shifted}. We use a wide range of edit distance thresholds (0-15 edits for a read length of 150 bp) and four datasets: (a) set\_5, (b) set\_6, (c) set\_7, and (d) set\_8.}
\label{fig:figure29Evaluation}
\end{figure}

\begin{figure}
\centering
\includegraphics[width=\textwidth]{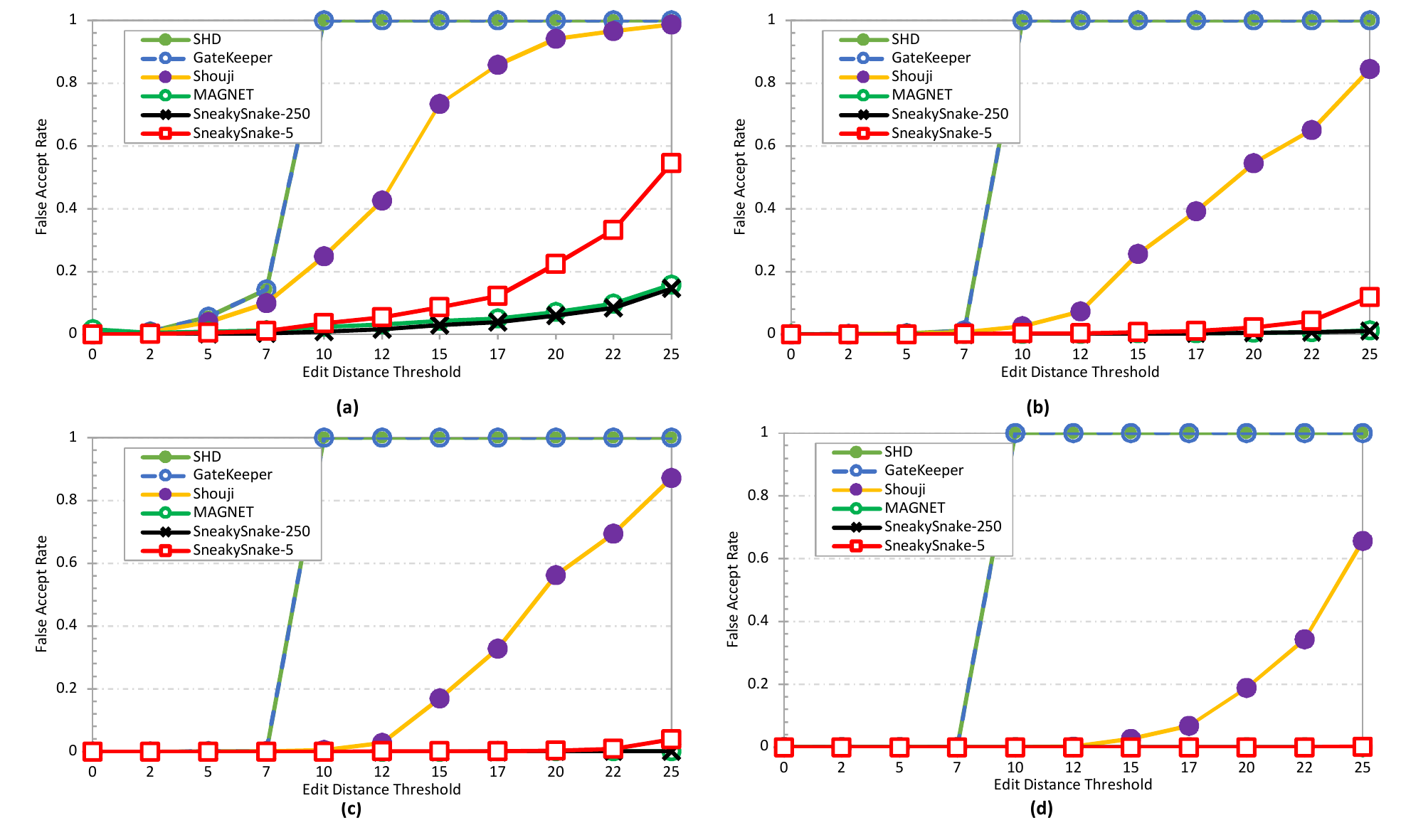}
\caption{The false accept rate produced by our pre-alignment filters, GateKeeper, Shouji, MAGNET, and SneakySnake, compared to the best performing filter, SHD \cite{xin2015shifted}. We use a wide range of edit distance thresholds (0-25 edits for a read length of 250 bp) and four datasets: (a) set\_9, (b) set\_10, (c) set\_11, and (d) set\_12.}
\label{fig:figure30Evaluation}
\end{figure}

We present in Figure \ref{fig:figure26Evaluation}, Figure \ref{fig:figure29Evaluation}, and Figure \ref{fig:figure30Evaluation} the false accept rate of our pre-alignment filters compared to SHD for read lengths of 100 bp, 150 bp, and 250 bp, respectively. We make six key observations.
\begin{enumerate}
\item We observe that Shouji, MAGNET, GateKeeper, and SneakySnake are less accurate in examining the low-edit sequences (Figure \ref{fig:figure26Evaluation}a,b, Figure \ref{fig:figure29Evaluation}a,b, and Figure \ref{fig:figure30Evaluation}a,b) than the edit-rich sequences (Figure \ref{fig:figure26Evaluation}c,d, Figure \ref{fig:figure29Evaluation}c,d, and Figure \ref{fig:figure30Evaluation}c,d). While SneakySnake pre-alignment filter yields the highest accuracy, SHD \cite{xin2015shifted} and GateKeeper provide the least accuracy compared to all other pre-alignment filters. The slope of MAGNET plot is almost comparable to that of the SneakySnake (with the default maze size).
\newpage 
\item GateKeeper and SHD \cite{xin2015shifted} become ineffective for edit distance thresholds of greater than 8\% for \textit{m} = 100 bp (Figure \ref{fig:figure26Evaluation}), 5\% for \textit{m} = 150 bp (Figure \ref{fig:figure29Evaluation}), and 3\% for \textit{m} = 250 bp (Figure \ref{fig:figure30Evaluation}) for both low-edit and edit-rich sequences, where \textit{m} is the read length. This leads to examining each sequence pair twice unnecessarily, by both GateKeeper or SHD and the full alignment step. 

\item Shouji provides up to 17.2x, 73x, and 467x less false accept rate compared to GateKeeper and SHD for read lengths of 100 bp, 150 bp, and 250 bp, respectively. 

\item MAGNET, SneakySnake (with the default maze size), and SneakySnake-5 show a slow exponential degradation in their filtering inaccuracy for low-edit sequences with a false accept rate of up to 50\%, 30\%, and 70\%, respectively, as we show in Figure \ref{fig:figure26Evaluation}a,b, Figure \ref{fig:figure29Evaluation}a,b, and Figure \ref{fig:figure30Evaluation}a,b. They also show almost a linear growth in their false accept rate of less than 4\% for edit-rich sequences of different read lengths, as we show in Figure \ref{fig:figure26Evaluation}c,d, Figure \ref{fig:figure29Evaluation}c,d, and Figure \ref{fig:figure30Evaluation}c,d.

\item MAGNET shows up to 1577x, 3550x, and 25552x less false accept rate compared to GateKeeper and SHD for read lengths of 100 bp, 150 bp, and 250 bp, respectively. MAGNET also provides up to 205x, 951x, and 16760x less false accept rate compared to Shouji for read lengths of 100 bp, 150 bp, and 250 bp, respectively.

\item SneakySnake (with the default maze size) produces up to four and five orders of magnitude less false accept rate compared to Shouji and GateKeeper, respectively. SneakySnake shows up to 55.4x, 46.7x, and 67.1x less false accept rate compared to SneakySnake-5 for read lengths of 100 bp, 150 bp, and 250 bp, respectively. SneakySnake also shows up to 64.1x, 16x, and 22x less false accept rate compared to MAGNET for read lengths of 100 bp, 150 bp, and 250 bp, respectively.

\end{enumerate}

We conclude that SneakySnake, MAGNET, and Shouji are very effective and superior to the state-of-the-art pre-alignment filter, SHD \cite{xin2015shifted} in both situations (low-edit and edit-rich mappings). They maintain a very low rate of falsely-accepted incorrect mappings and significantly improves the accuracy of pre-alignment filtering by up to five orders of magnitude compared to GateKeeper and SHD.

\subsection{False Reject Rate}
\begin{figure}
\centering
\includegraphics[width=\textwidth,keepaspectratio]{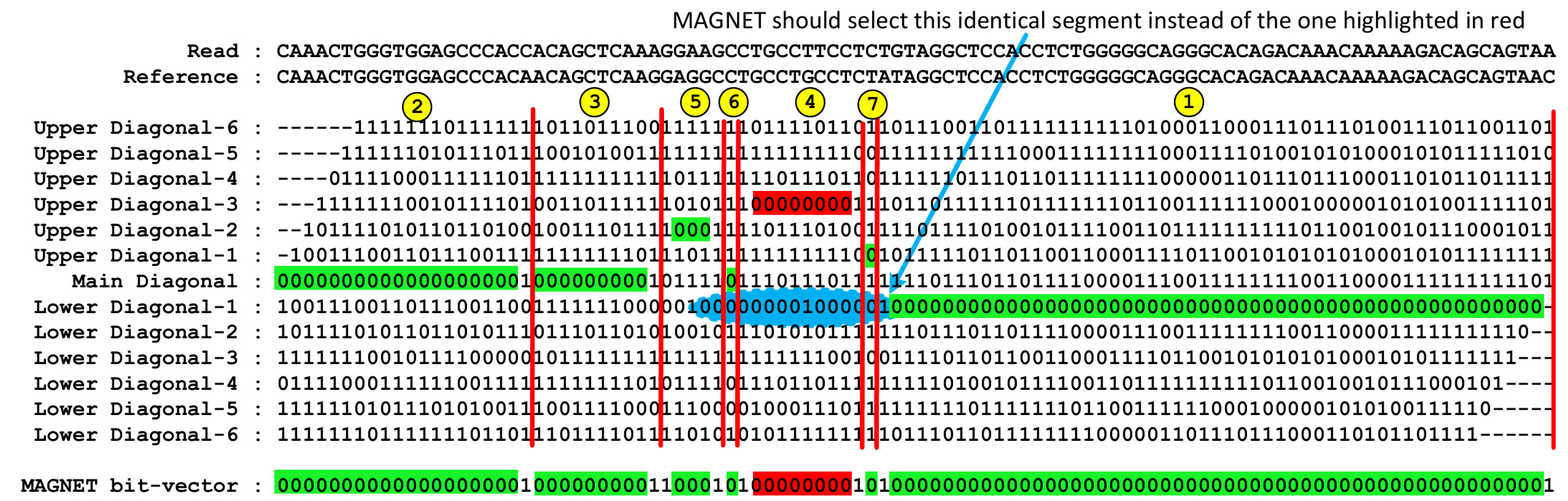}
\caption{An example of a falsely-rejected mapping using MAGNET algorithm for an edit distance threshold of 6. The random zeros (highlighted in red) confuse MAGNET filter causing it to select shorter segments of random zeros instead of a longer identical subsequence (highlighted in blue).}
\label{fig:figure31Evaluation}
\end{figure}

Using our 12 low-edit and edit-rich datasets for three different read lengths, we observe that SneakySnake (for all partition sizes), Shouji, GateKeeper, and SHD do not filter out correct mappings; hence, they provide a 0\% false reject rate. The reason behind that is the way we find the identical subsequences. \newpage We aim to find the subsequences that has the largest number of zeros, such that we maximize the number of matches and minimize the number of edits that cause the division of one long identical sequence into shorter subsequences. However, this is not the case for MAGNET. We observe that MAGNET shows a very low false reject rate of less than 0.00045\% for an edit distance threshold of at least 4 bp. This is due in large part to the greedy choice of always selecting the longest identical subsequence based on their descending length and regardless of their source (i.e. the vectors that originate them). We provide in Figure \ref{fig:figure31Evaluation} an example of where MAGNET fails to maintain the correct mappings. A potential solution is to relate the number of encapsulated bits to the source of each segment. For instance, if any extracted segment in the MAGNET mask has a different source than its neighboring segments, then we need to penalize that segment by adding more encapsulation bits. This can help us to produce accurate \textit{CIGAR} string for each mapping. With the help of an auxiliary data structure, we can keep track of the source of each extracted segment at each position. While the matches coming from the \lq\lq{}upper diagonal-1\rq\rq{} mean there is a single deletion, the matches coming from the \lq\lq{}upper diagonal-2\rq\rq{} mean that there are two deletions. A detailed algorithm of this topic is beyond the scope of this thesis and is part of our future work. 

\section{Effects of Hardware Pre-Alignment Filtering on Read Alignment}
We first analyze the execution time of our hardware pre-alignment filters, GateKeeper, MAGNET, Shouji, and SneakySnake. We build the FPGA implementation of SneakySnake using a sub-matrix\rq{}s width of 8 columns (\textit{t}=8) and we include 3 replications in the design. We use GateKeeper as an optimized and efficient hardware implementation of SHD \cite{xin2015shifted}. We evaluate our four pre-alignment filters using a single FPGA chip. We use 120 million sequence pairs, each of which is 100 bp long, from set\_1, set\_2, set\_3, and set\_4. We summarize the execution time of the CPU implementations along with that of their hardware accelerators in Table \ref{table:table6HWACC}. We make two key observations based on Table \ref{table:table6HWACC}.

\begin{enumerate}
\item Our hardware accelerators provide two to three orders of magnitude (322x to 7,250x) speedup over their CPU implementations. GateKeeper provides up to two orders of magnitude of acceleration over SHD.

\item The execution time of the hardware implementation of SneakySnake and Shouji are as low as that of GateKeeper and 2x-8x lower than that of MAGNET pre-alignment filter. This observation is in accord with our expectation and can be explained by the fact that MAGNET has more computational overhead that limits the number of filtering units. Yet SneakySnake is four and five orders of magnitude more accurate than both Shouji and GateKeeper (as we show earlier). 

\end{enumerate}

We conclude that our hardware accelerator provides two and three orders of magnitude of speedup over their CPU implementations. Additionally, the execution time of the hardware accelerator is proportional to the FPGA resource utilization (the less the resource utilization the lower the execution time).

\begin{table}
\centering
\caption{The execution time (in seconds) of GateKeeper, MAGNET, Shouji, and SneakySnake under different edit distance thresholds. We use set\_1 to set\_4 with a read length of 100. We provide the performance results for the CPU implementations and the hardware accelerators with the maximum number of filtering units.}
\label{table:table6HWACC}
\includegraphics[width=14cm]{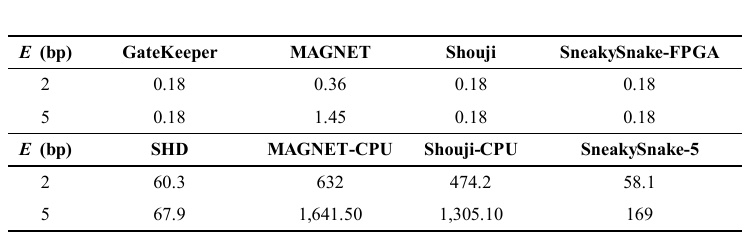}
\end{table}

Next we analyze the benefits of integrating our hardware pre-alignment filters with the state-of-the-art aligners. GateKeeper, MAGNET, Shouji, and SneakySnake are standalone pre-alignment filters and can be integrated with any existing alignment algorithm. In Table \ref{table:table7HWACC}, we present the effects of our four hardware pre-alignment filters on the overall alignment\rq{}s execution time. We use a sub-matrix\rq{}s width of 8 columns (\textit{t}=8) and we include 3 replications in the design of the hardware architecture of the SneakySnake algorithm. We also compare the effect of our pre-alignment filters with that of SHD \cite{xin2015shifted}. We select five best performing aligners, each of which is designed for different type of computing platforms. While Edlib \cite{vsovsic2017edlib} algorithm is implemented in C for standard CPUs, Parasail \cite{daily2016parasail} exploits SIMD capable CPUs. GSWABE \cite{liu2015gswabe} is designed for GPUs. CUDASW++ 3.0 \cite{liu2013cudasw++} exploits SIMD capability of both CPUs and GPUs. FPGASW \cite{fei2018fpgasw} exploit the very large number of hardware execution units offered by the same FPGA chip (i.e., VC709) as our accelerator. We evaluate the execution time of Edlib \cite{vsovsic2017edlib} and Parasail \cite{daily2016parasail} on our machine. \newpage However, FPGASW \cite{fei2018fpgasw}, CUDASW++ 3.0 \cite{liu2013cudasw++}, and GSWABE \cite{liu2015gswabe} are not open-source and not available to us. Therefore, we scale the reported number of computed entries of the dynamic programming matrix in a second. We use a total of 120 million real sequence pairs from our previously described four datasets (set\_1, set\_2, set\_3, and set\_4) in this analysis. We make three key observations. 

\begin{enumerate}
\item The execution time of Edlib \cite{vsovsic2017edlib} reduces by up to 21.4x, 18.8x, 16.5x, 13.9x, and 5.2x after the addition of SneakySnake, Shouji, MAGNET, GateKeeper, and SHD, respectively, as a pre-alignment filtering step. We also observe nearly a similar trend for Parasail \cite{daily2016parasail} combined with each of the four pre-alignment filters. 

\item Aligners designed for FPGAs and GPUs follow a different trend than that we observe in the CPU aligners. We observe that the ability of SHD \cite{xin2015shifted} to reduce the alignment time of GSWABE \cite{liu2015gswabe}, CUDASW++ 3.0 \cite{liu2013cudasw++}, and FPGASW \cite{fei2018fpgasw} diminishes. SHD even provides unsatisfactory performance as it increases the execution time of the aligner instead of reducing it. This is due to the fact that SHD is 6x slower than CUDASW++ 3.0 \cite{liu2013cudasw++} and FPGASW \cite{fei2018fpgasw} and it is lightly slower than GSWABE \cite{liu2015gswabe}.

\item SneakySnake, Shouji, MAGNET, and GateKeeper still contribute significantly towards reducing the overall execution time of FPGA and GPU based aligners. SneakySnake reduces the execution time of FPGASW \cite{fei2018fpgasw}, CUDASW++ 3.0 \cite{liu2013cudasw++} and GSWABE \cite{liu2015gswabe} by factors of up to 16x, 15.5x, and 20.3x, respectively. This is slightly higher (up to 1.3x) than the effect of Shouji on the execution time of these aligners. Shouji reduces the overall alignment time of FPGASW \cite{fei2018fpgasw}, CUDASW++ 3.0 \cite{liu2013cudasw++} and GSWABE \cite{liu2015gswabe} by factors of up to 14.5x, 14.2x, and 17.9x, respectively. This is up to 1.5x, 1.4x, and 85x more than the effect of MAGNET, GateKeeper, and SHD on the end-to-end alignment time.

\end{enumerate}

We conclude that among the four hardware pre-alignment filters, SneakySnake (3-replication design and with \textit{t}=8) is the best performing filter in terms of both speed and accuracy. \newpage Integrating SneakySnake with aligner does not lead to negative effects. We also conclude that the concept of pre-alignment filtering is still effective in boosting the overall performance of the alignment step, even the dynamic programming algorithm is accelerated by the state-of-the-art hardware accelerators such as SIMD-capable CPUs, FPGAs, and GPUs.

\begin{table}
\centering
\caption{
End-to-end execution time (in seconds) for several state-of-the-art sequence alignment algorithms, with and without pre-alignment filters (SneakySnake, Shouji, MAGNET, GateKeeper, and SHD) and across different edit distance thresholds. We use four datasets (set\_1, set\_2, set\_3, and set\_4) across different edit distance thresholds.}
\label{table:table7HWACC}
\includegraphics[width=14.5cm]{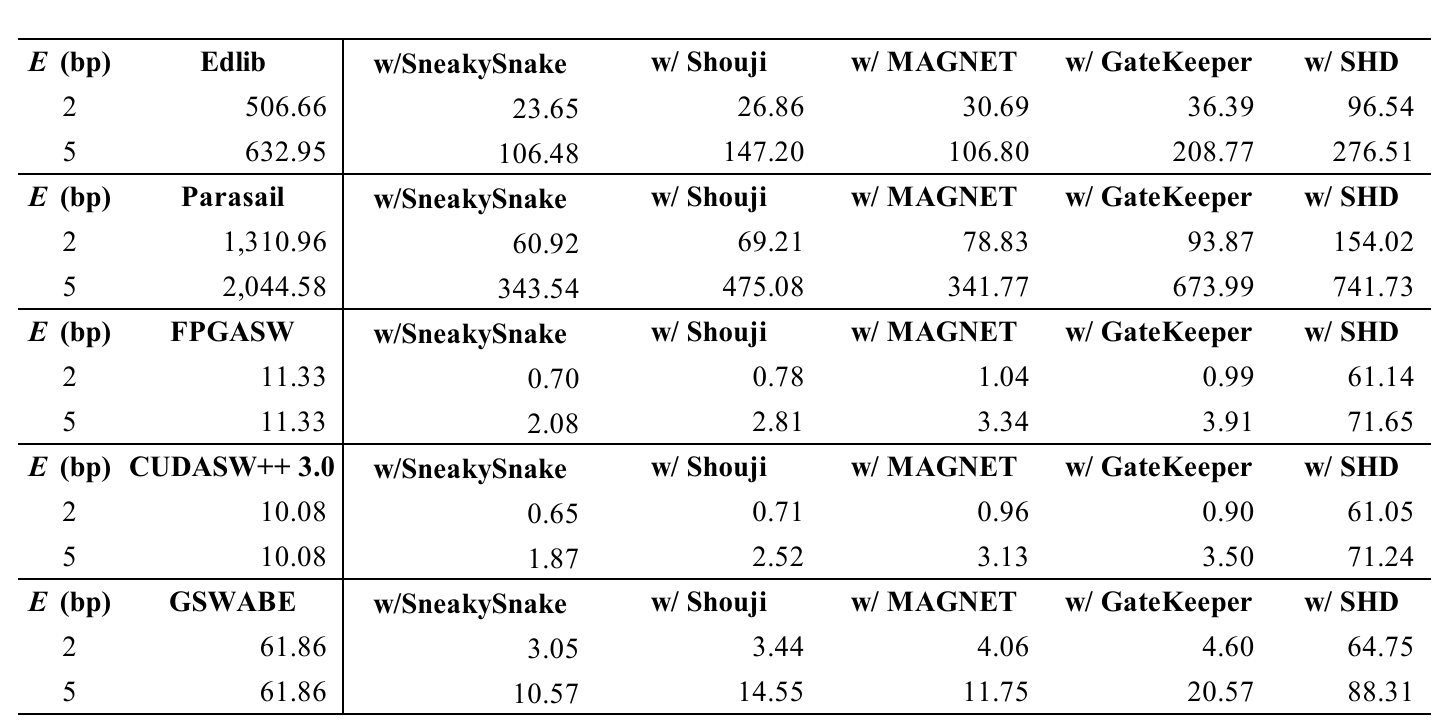}
\end{table}

\section{Effects of CPU Pre-Alignment Filtering on Read Alignment}
\begin{figure}
\centering
\includegraphics[width=\textwidth,keepaspectratio]{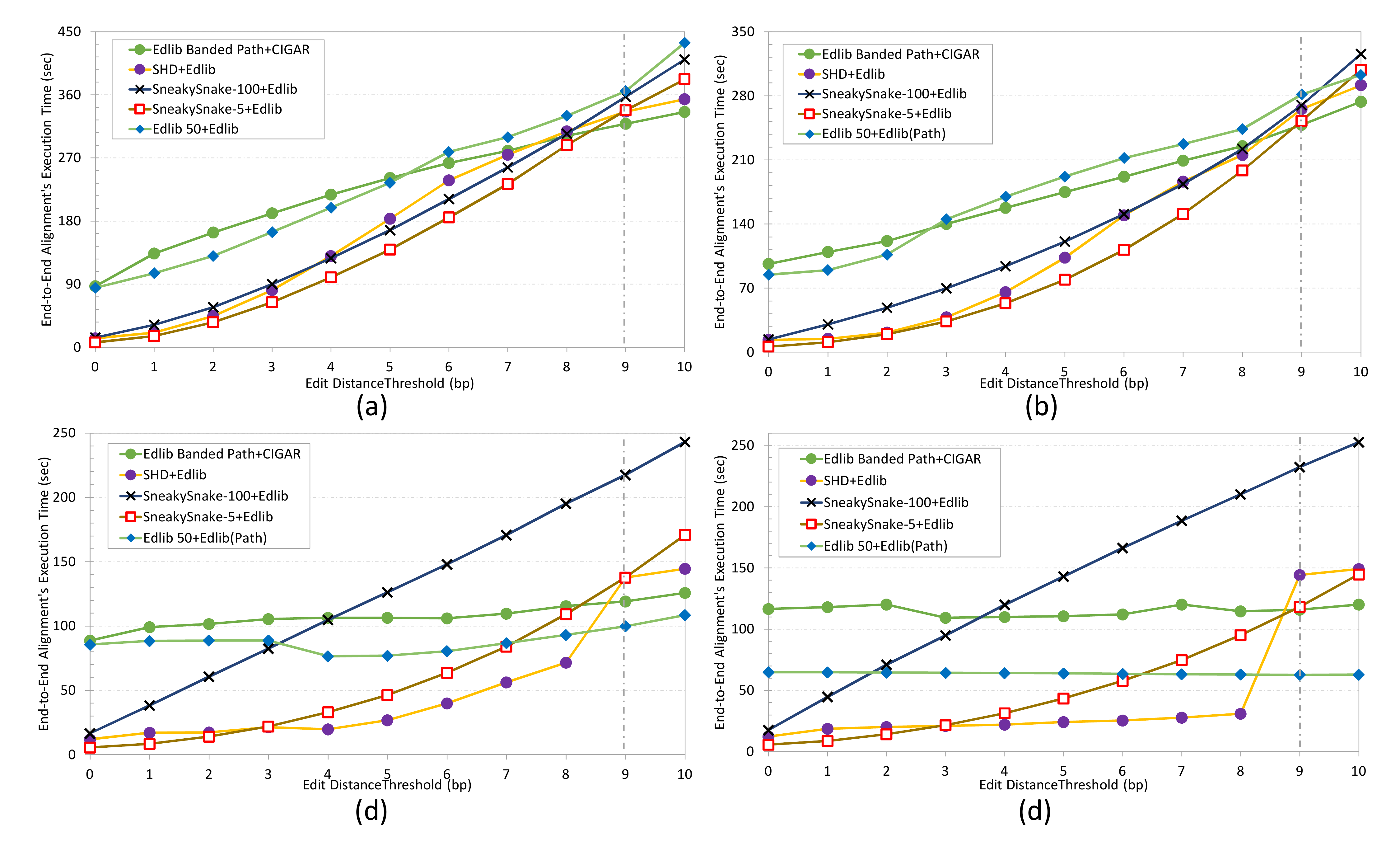}
\caption{End-to-end execution time (in seconds) for Edlib \cite{vsovsic2017edlib} (full read aligner), with and without pre-alignment filters. We use four datasets ((a) set\_1, (b) set\_2, (c) set\_3, and (d) set\_4) across different edit distance thresholds. We highlight in a dashed vertical line the edit distance threshold where Edlib starts to outperform our SneakySnake-5 algorithm.}
\label{fig:figure33Appendix}
\end{figure}

\begin{figure}
\centering
\includegraphics[width=\textwidth,keepaspectratio]{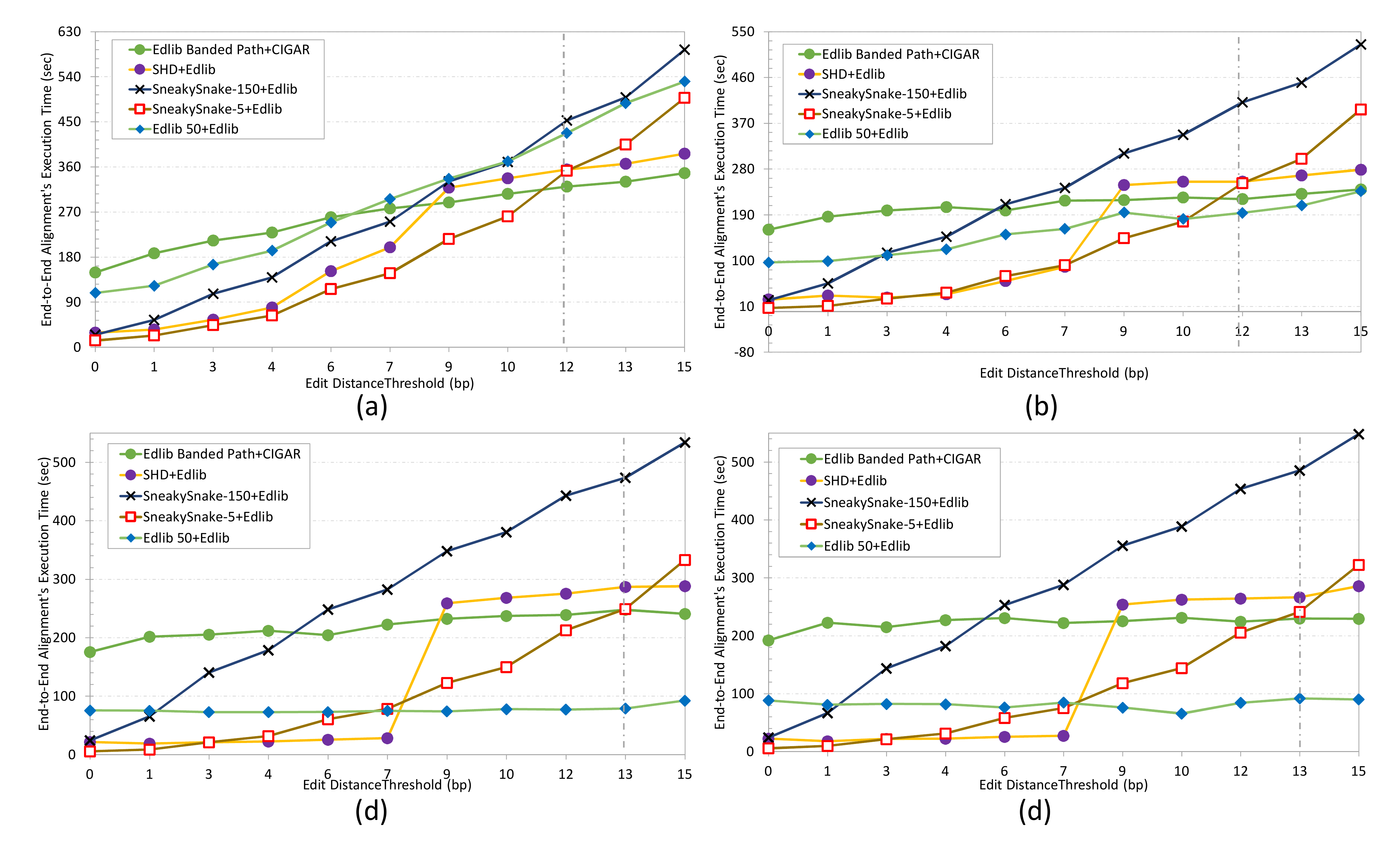}
\caption{End-to-end execution time (in seconds) for Edlib \cite{vsovsic2017edlib} (full read aligner), with and without pre-alignment filters. We use four datasets ((a) set\_5, (b) set\_6, (c) set\_7, and (d) set\_8) across different edit distance thresholds. We highlight in a dashed vertical line the edit distance threshold where Edlib starts to outperform our SneakySnake-5 algorithm.}
\label{fig:figure34Appendix}
\end{figure}

\begin{figure}
\centering
\includegraphics[width=\textwidth,keepaspectratio]{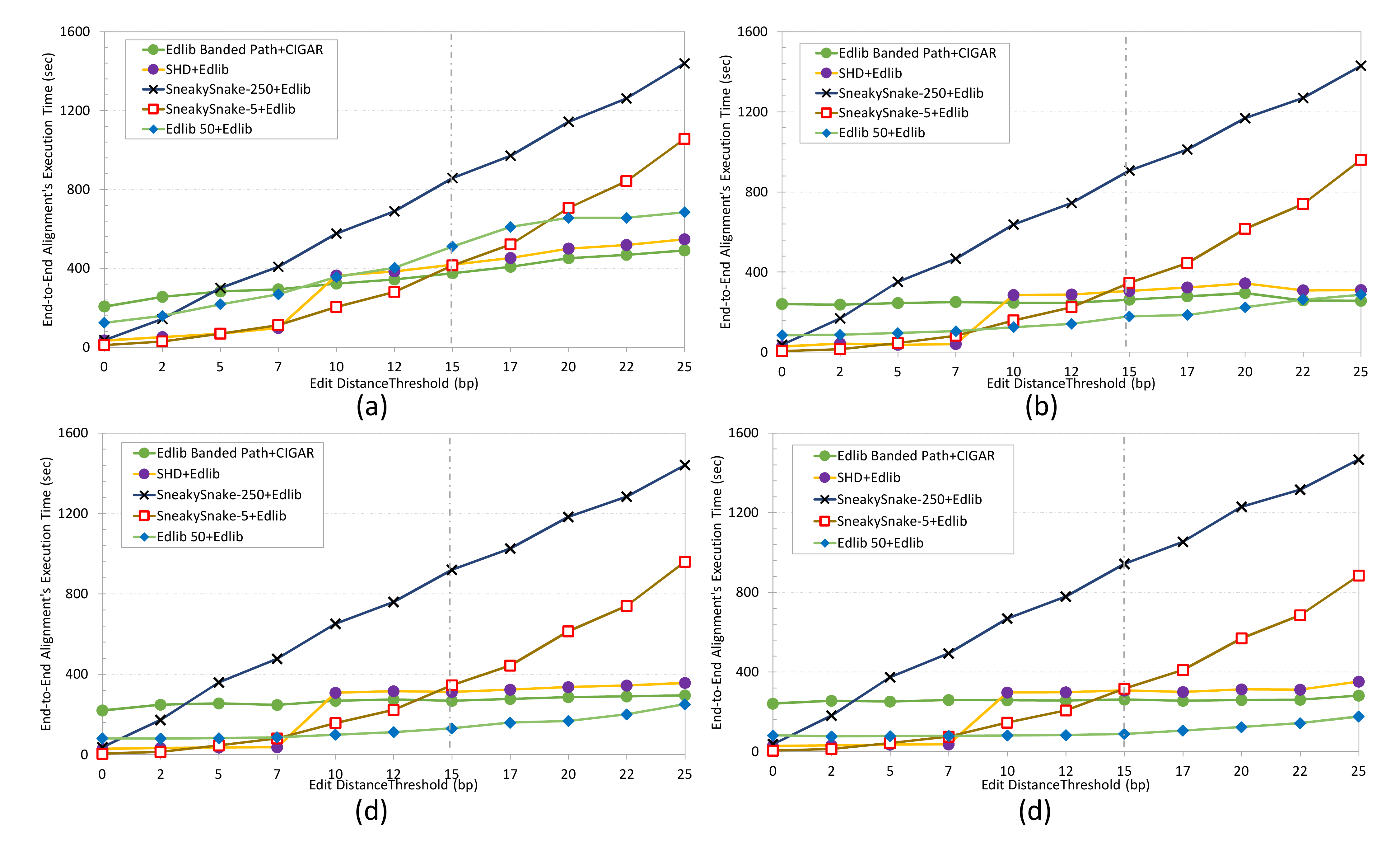}
\caption{End-to-end execution time (in seconds) for Edlib \cite{vsovsic2017edlib} (full read aligner), with and without pre-alignment filters. We use four datasets ((a) set\_9, (b) set\_10, (c) set\_11, and (d) set\_12) across different edit distance thresholds. We highlight in a dashed vertical line the edit distance threshold where Edlib starts to outperform our SneakySnake-5 algorithm.}
\label{fig:figure35Appendix}
\end{figure}

Now we analyze the benefits of integrating our CPU implementations of SneakySnake and the prefix edit distance with the state-of-the-art CPU aligners, Edlib \cite{vsovsic2017edlib} and Parasail \cite{daily2016parasail}. \newpage We evaluate the end-to-end execution time of Edlib (referred to as path in \cite{vsovsic2017edlib} and configured as a banded global Levenshtein distance with CIGAR-enabled output) with and without the pre-alignment filtering step for read lengths of 100 bp (Figure \ref{fig:figure33Appendix}), 150 bp (Figure \ref{fig:figure34Appendix}), and 250 bp (Figure \ref{fig:figure35Appendix}). We make three key observations.

\begin{enumerate}
\item SneakySnake-5 combined with the banded Edlib is up to 20.4x, 33x, and 43x faster than Edlib without pre-alignment filter for \textit{E}$<$9 bp (\textit{m}=100 bp), \textit{E}$<$12 bp (\textit{m}=150 bp), and \textit{E}$<$15 bp (\textit{m}=250 bp), where \textit{E} is the edit distance threshold and \textit{m} is the read length.

\item Edlib-50 combined with the banded Edlib is slower than SneakySnake-5 combined with the same aligner for low-edit datasets (Figure \ref{fig:figure33Appendix}a-b, Figure \ref{fig:figure34Appendix}a-b, and Figure \ref{fig:figure35Appendix}a-b). Edlib-50 becomes more effective over SneakySnake-5 in reducing the execution time of Edlib across edit-rich datasets for \textit{E}$>$6 bp (\textit{m}=100 bp and 150 bp) and \textit{E}$>$7 bp (\textit{m}=250 bp).

\item SHD leads to a negative effect as it slows down the alignment speed of Edlib for \textit{E}$>$8 (\textit{m}=100 bp), \textit{E}$>$7 (\textit{m}=150 bp and 250 bp).
\end{enumerate}

\begin{figure}
\centering
\includegraphics[width=\textwidth,keepaspectratio]{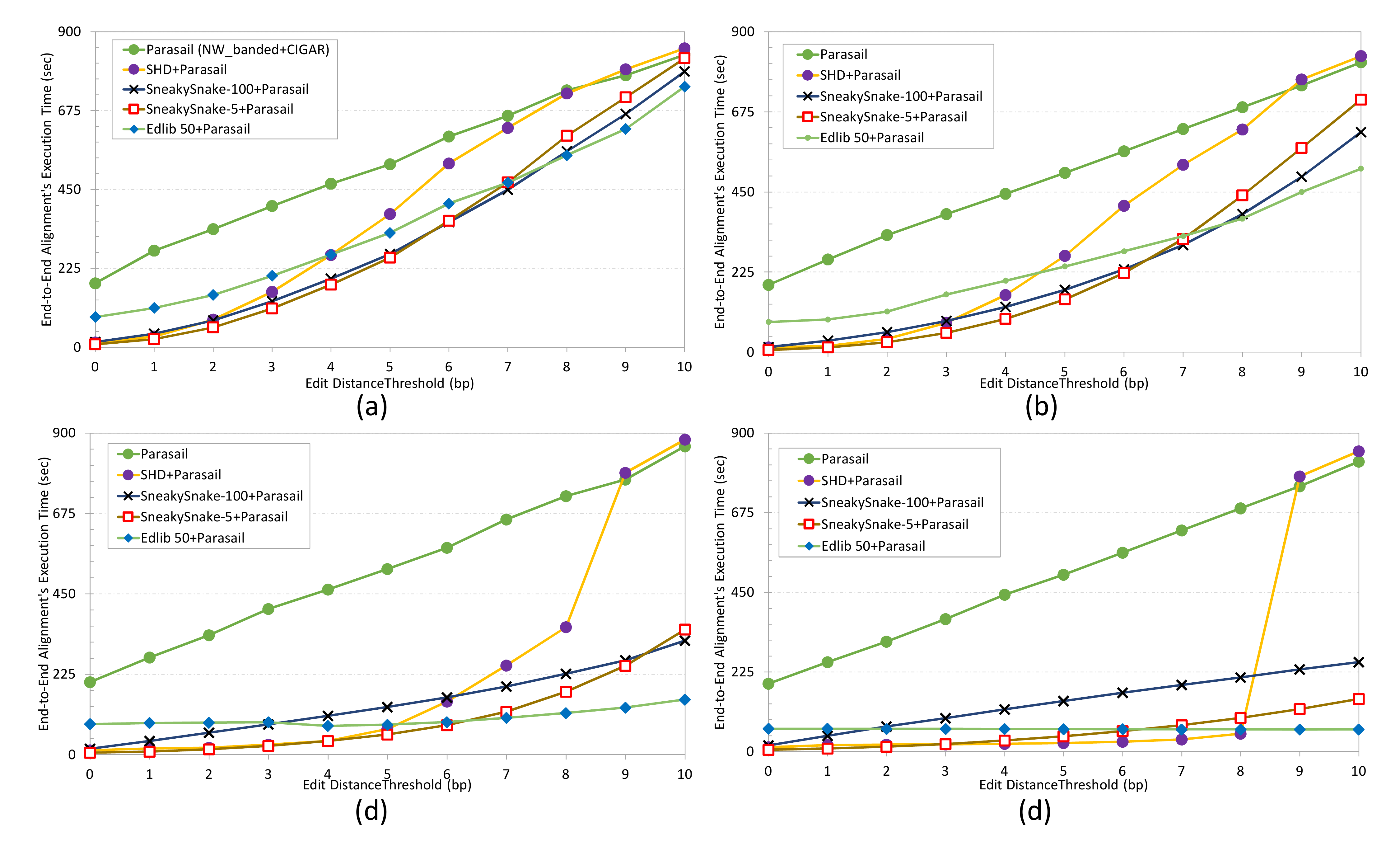}
\caption{End-to-end execution time (in seconds) for Parasail \cite{daily2016parasail} (full read aligner), with and without pre-alignment filters. We use four datasets ((a) set\_1, (b) set\_2, (c) set\_3, and (d) set\_4) across different edit distance thresholds.}
\label{fig:figure36Appendix}
\end{figure}

\begin{figure}
\centering
\includegraphics[width=\textwidth,keepaspectratio]{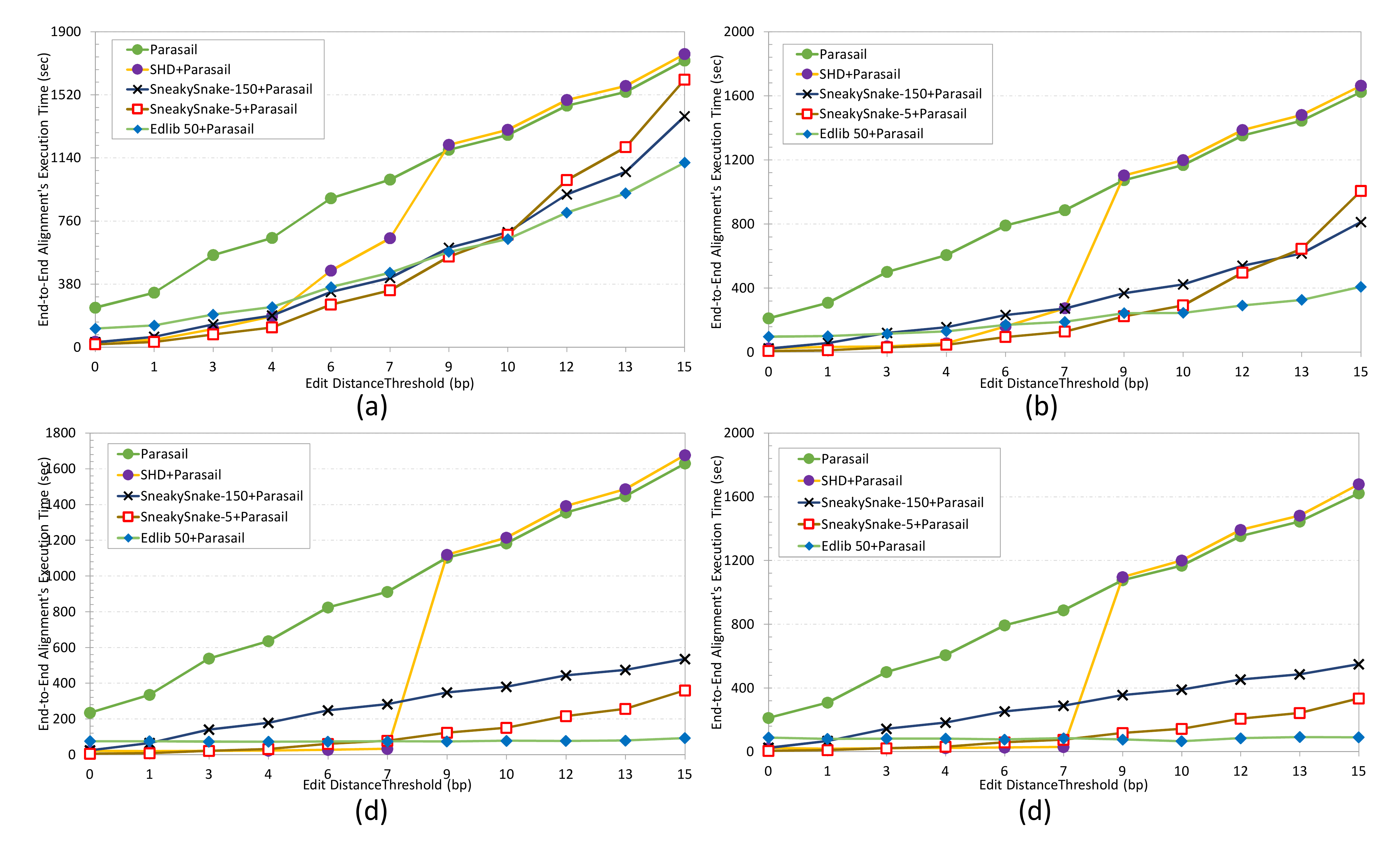}
\caption{End-to-end execution time (in seconds) for Parasail \cite{daily2016parasail} (full read aligner), with and without pre-alignment filters. We use four datasets ((a) set\_5, (b) set\_6, (c) set\_7, and (d) set\_8) across different edit distance thresholds.}
\label{fig:figure37Appendix}
\end{figure}

\begin{figure}
\centering
\includegraphics[width=\textwidth,keepaspectratio]{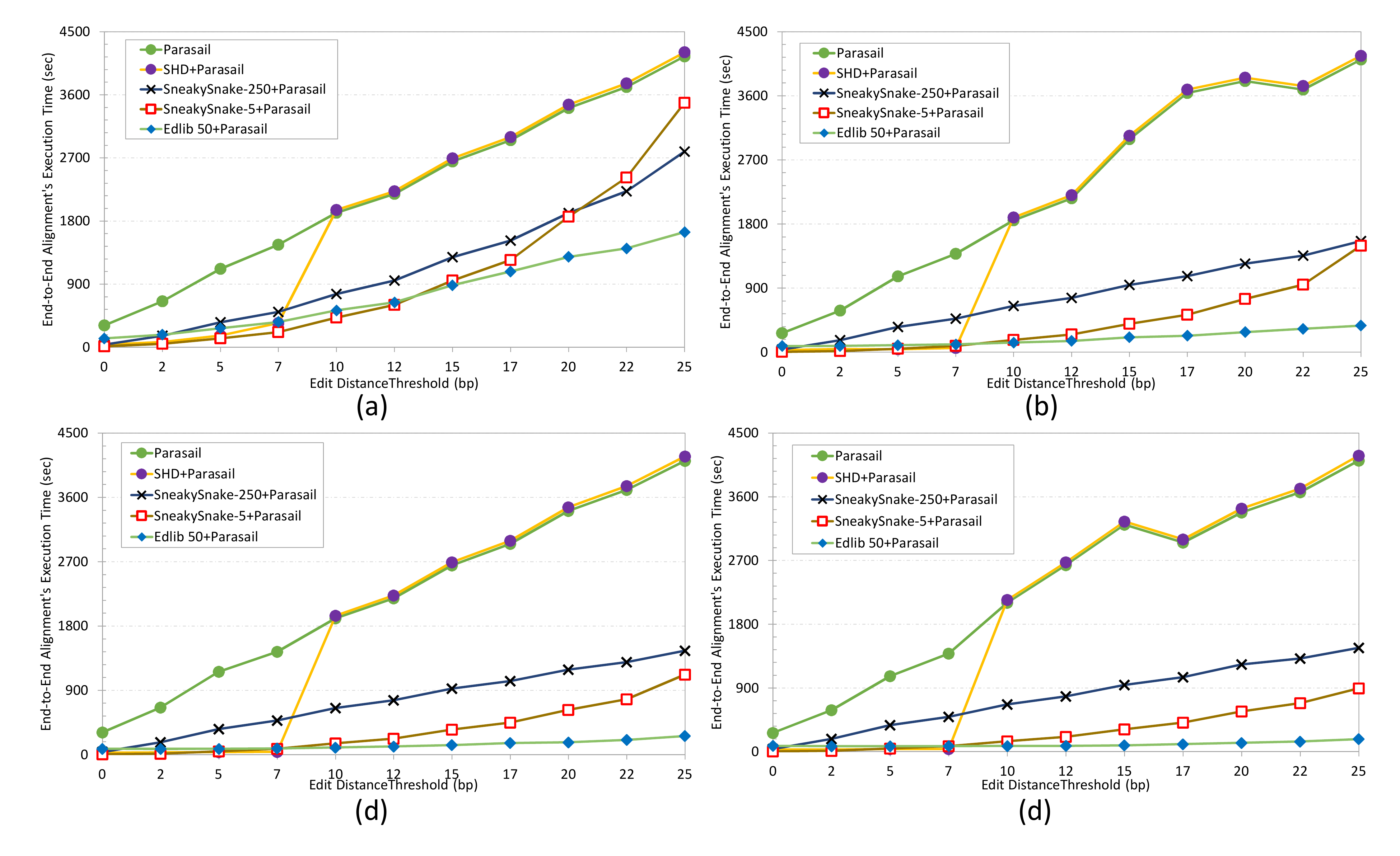}
\caption{End-to-end execution time (in seconds) for Parasail \cite{daily2016parasail} (full read aligner), with and without pre-alignment filters. We use four datasets ((a) set\_9, (b) set\_10, (c) set\_11, and (d) set\_12) across different edit distance thresholds.}
\label{fig:figure38Appendix}
\end{figure}

Secondly, we evaluate the effects of adding these pre-alignment filters to Parasail \cite{daily2016parasail} for read lengths of 100 bp (Figure \ref{fig:figure36Appendix}), 150 bp (Figure \ref{fig:figure37Appendix}), and 250 bp (Figure \ref{fig:figure38Appendix}). We configure Parasail as NW\_banded and with CIGAR-enabled output. We make three key observations.

\begin{enumerate}
\item SneakySnake-5 still provides significant benefits to the the highest end-to-end speedup over all other pre-alignment filters when combined with Parasail for \textit{E}$<$7 bp (\textit{m}=100 bp), \textit{E}$<$9 bp (\textit{m}=150 bp), and \textit{E}$<$10 bp (\textit{m}=250 bp). SneakySnake-5 combined with Parasail yields up to 36.3x, 42x, and 57.9x speedup over Parasail without a pre-alignment filter for read lengths of 100 bp, 150 bp, and 250 bp, respectively.

\item Edlib-50 combined with Parasail becomes more effective than SneakySnake-5 combined with Parasail for \textit{E}$>$6 bp (\textit{m}=100 bp), \textit{E}$>$7 bp (\textit{m}=150 bp), and \textit{E}$>$7 bp (\textit{m}=250 bp).

\item SHD leads to slowing down the alignment speed of Parasail for \textit{E}$>$8 (\textit{m}=100 bp), \textit{E}$>$7 (\textit{m}=150 bp and 250 bp).

\end{enumerate}

We conclude that our SneakySnake algorithm is the best performing CPU pre-alignment filter in terms of both speed and accuracy. It accelerates the state-of-the-art read alignment algorithms by up to an order of magnitude of acceleration. We demonstrate that SneakySnake algorithm does not lead to negative effects for edit distance thresholds of 0\% to 10\% of the read length. We also want to emphasize that combining our SneakySnake pre-alignment filter (i.e., SneakySnake-5) with a prefix edit distance algorithm (i.e., Edlib-50) provides the fastest and the most accurate pre-alignment filtering across wide range of edit distance thresholds and read lengths.

\section{Memory Utilization}
\begin{figure}
\centering
\includegraphics[width=13cm,keepaspectratio]{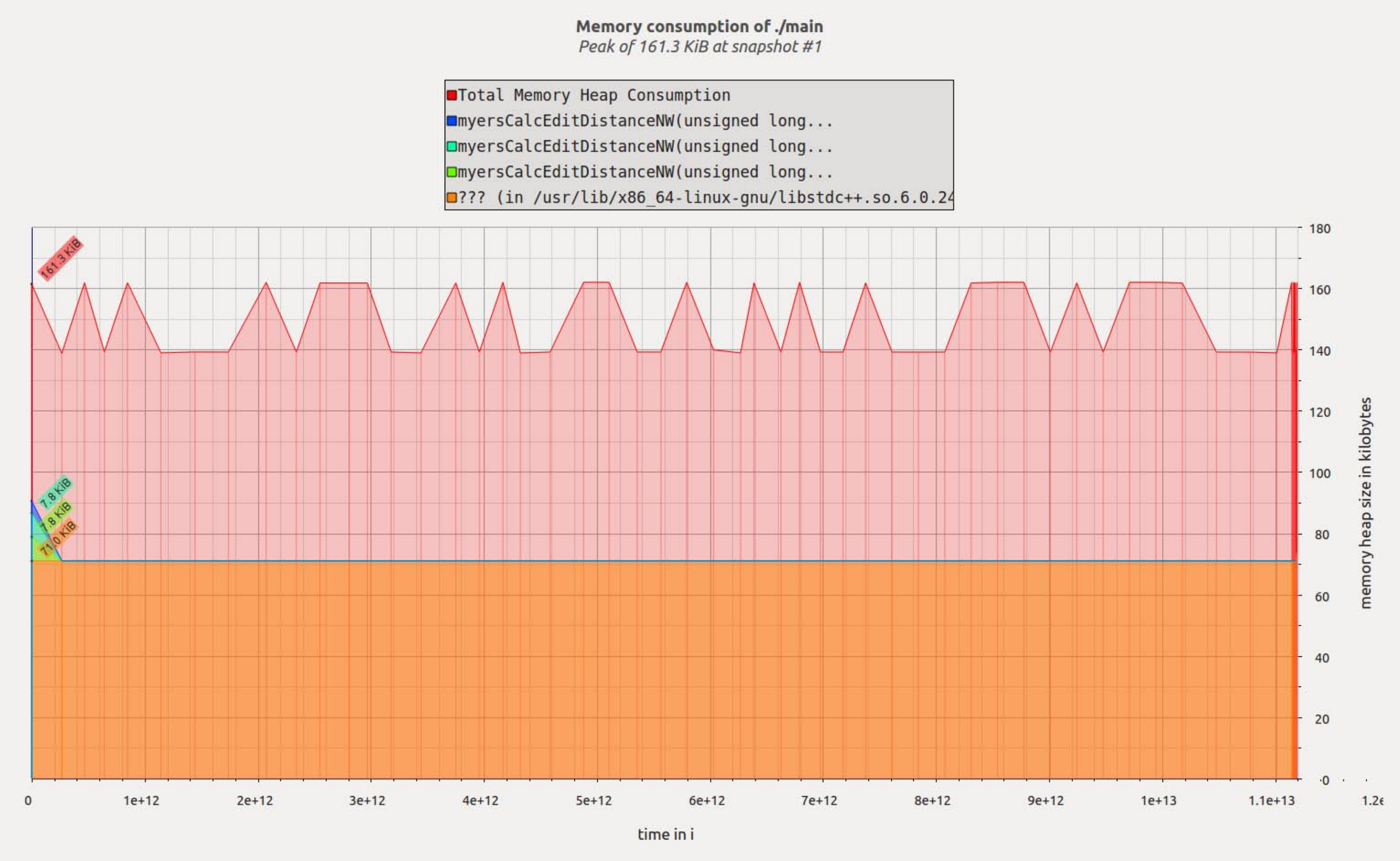}
\caption{Memory utilization of Edlib (path) read aligner while evaluating set\_12 for an edit distance threshold of 25.}
\label{fig:figureE100_EdlibEvaluation}
\end{figure}

\begin{figure}
\centering
\includegraphics[width=13cm,keepaspectratio]{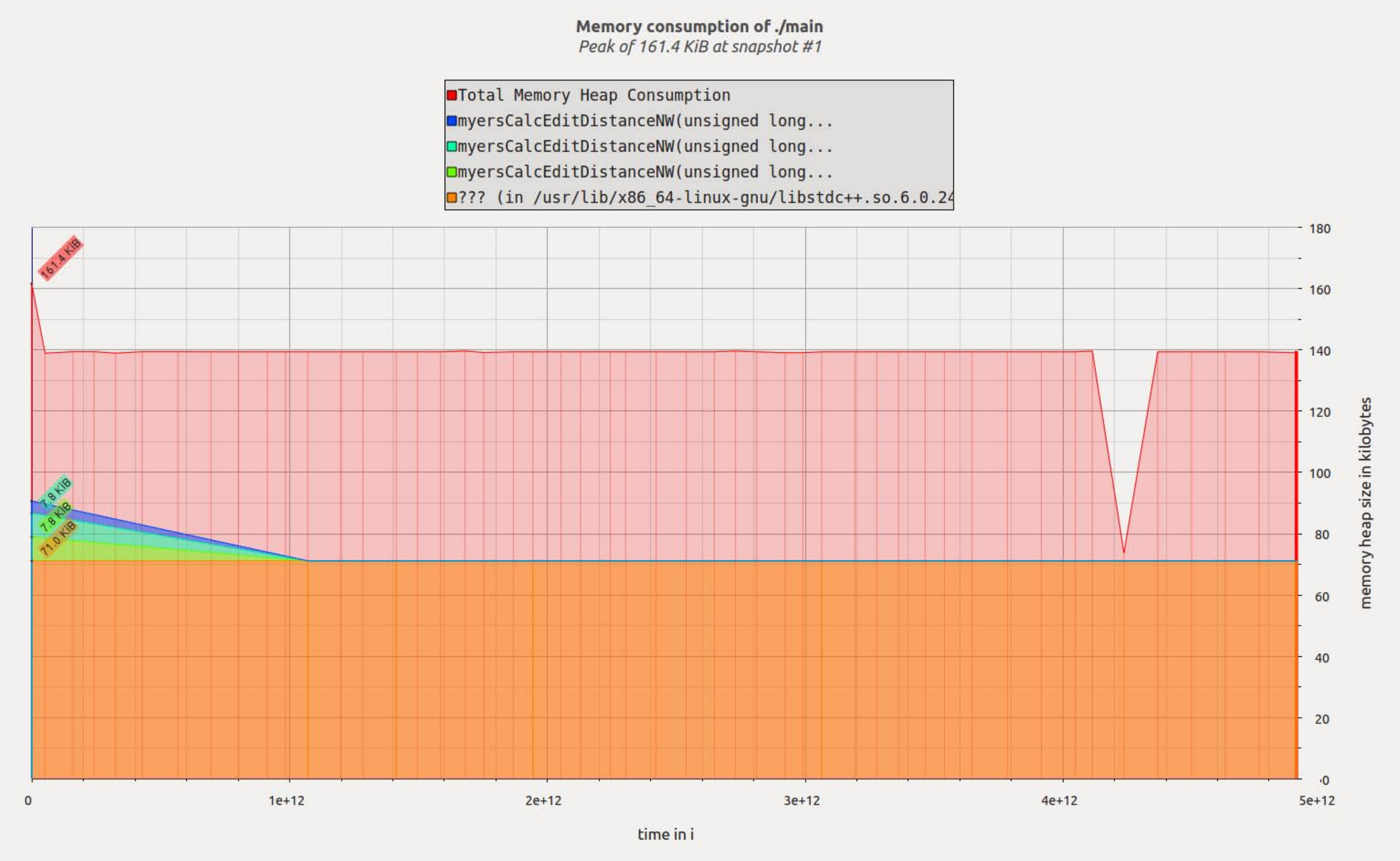}
\caption{Memory utilization of exact edit distance algorithm (Edlib ED) combined with Edlib (path) read aligner while evaluating set\_12 for an edit distance threshold of 25.}
\label{fig:figureE100_EdlibED_EdlibEvaluation}
\end{figure}

\begin{figure}
\centering
\includegraphics[width=13cm,keepaspectratio]{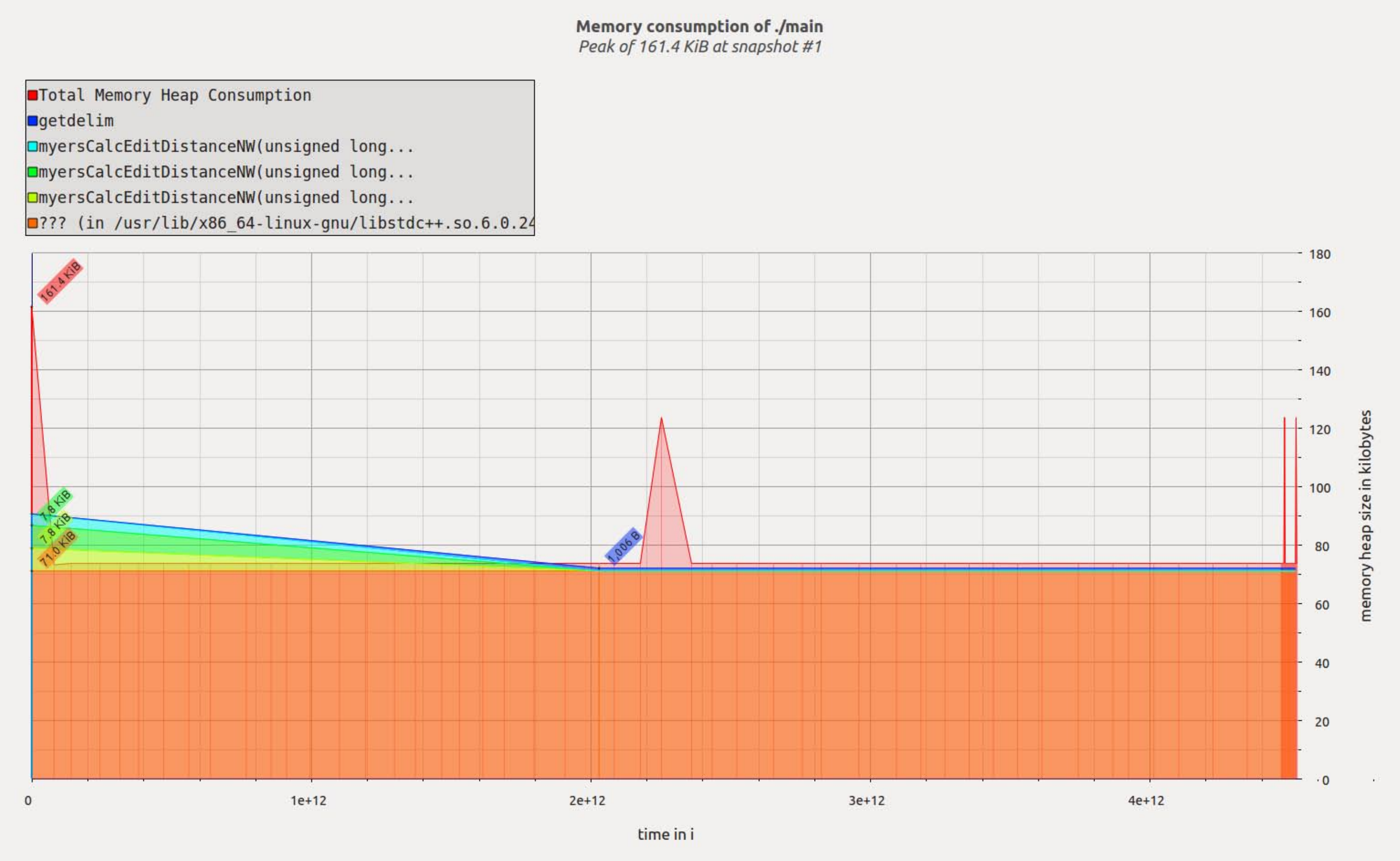}
\caption{Memory utilization of SneakySnake-5 combined with Edlib (path) read aligner while evaluating set\_12 for an edit distance threshold of 25.}
\label{fig:figureE100_SneakySnake_EdlibEvaluation}
\end{figure}

In this section, we evaluate the space-efficiency benefits of integrating our SneakySnake algorithm with the state-of-the-art full read aligner algorithm, Edlib \cite{vsovsic2017edlib}. We use Valgrind massif tool to examine the memory utilization. We provide the memory footprint of Edlib in Figure \ref{fig:figureE100_EdlibEvaluation}. On average, Edlib shows a memory footprint of 150 KB. We then evaluate the memory utilization of integrating Edlib (in edit distance mode and without backtracking) with Edlib (path) in Figure \ref{fig:figureE100_EdlibED_EdlibEvaluation}. The addition of exact edit distance algorithm to the read alignment shows a slight reduction ($\sim$ 5\%) in the memory utilization. With the addition of our SneakySnake-5 as a pre-alignment step before performing Edlib\rq{}s full alignment, we observe that the memory footprint drops significantly by at least 50\%, as we demonstrate in Figure \ref{fig:figureE100_SneakySnake_EdlibEvaluation}. \newpage This observation is in accord with our expectation and can be explained by the fact that SneakySnake-5 requires a spaces of as small as a sub-matrix of size 5 x 2\textit{E}+1 and at most \textit{m} x 2\textit{E}+1, whereas Edlib always requires a space of \textit{m} x \textit{m}. We conclude that SneakySnake algorithm is fast, accurate, and yet memory-efficient.

%% file: chapter10.tex
\chapter{Conclusions and Future Directions}
Our goal in this thesis is to considerably minimize the time spent on calculating the optimal alignment in genome analysis, given limited computational resources (i.e., personal computer or small hardware). To this end, we first provide a comprehensive accuracy analysis of the pre-alignment filtering. Understanding the causes for the filtering inaccuracy helps us to design new fast and accurate pre-alignment filters. Second, we propose the first hardware accelerator architecture for pre-alignment in genome read mapping. We leverage the large number of filtering units that our hardware accelerator offers for accelerating our proposed hardware-aware algorithms. We propose four hardware pre-alignment filters, GateKeeper, Shouji, MAGNET, and SneakySnake. In our experimental evaluation, our hardware pre-alignment filters show, on average, three orders of magnitude speedup over their equivalent CPU implementations. We demonstrate that GateKeeper occupies the least percentage of the FPGA resource and it is the least accurate filter. We show that MAGNET provides a low false accept rate but incurs a very low rate of falsely rejected mappings. We also demonstrate that Shouji is more accurate than GateKeeper and faster than MAGNET. However, SneakySnake is our best performing pre-alignment filter in terms of both speed and accuracy. \newpage SneakySnake has a very low false accept rate and 0\% false reject rate. We demonstrate that SneakySnake reduces the execution time of existing read aligners by up to an order of magnitude.

Third, we introduce a fast and cost-effective CPU implementation of our best performing pre-alignment filter, SneakySnake. In our comprehensive evaluation, we demonstrate that the CPU implementation of SneakySnake reduces the execution time of the best performing read aligner, Edlib and Parasail, by up to 43x and 57.9x, respectively. We also experimentally demonstrate that SneakySnake has 50\% less memory footprint compared with that of Edlib. The CPU implementation of SneakySnake obviates the need for costly hardware and high hardware design efforts by providing a fast and cost-effective implementation. 

We demonstrate that the concept of pre-alignment filtering provides substantial benefits to the existing and future read alignment algorithms. New accelerated sequence aligners are frequently introduced that offer different strengths and features. Our proposed pre-alignment filters offer the ability to accelerate existing aligners by an order of magnitude without sacrificing any of their capabilities and features. As such, we hope that it catalyzes the adoption of our proposed pre-alignment filters in genome sequence analysis, which are becoming increasingly necessary to cope with the processing requirements of greatly increasing amounts of genomic data.

\section{Future Research Directions}
This thesis opens up several avenues of future research directions. In this section,
we describe five directions based on the ideas and approaches proposed in this thesis. These ideas can lead to a new read mapper or improve existing ones, which we will explore for various mappers in our future research. \newpage

\begin{enumerate}
\item The first potential target of our research is to influence the design of more intelligent and attractive sequencing machines by integrating SneakySnake or Shouji inside them, to perform a real-time pre-alignment filtering. Sequencing machines (e.g., Illumina HiSeq 2500, HiSeq 4000, and MiSeq) are equipped with FPGA chips for accelerating their internal computations. Integrating our pre-alignment filters with the sequencing machine has two benefits. First, it allows a significant reduction in the total execution time of genome analysis by starting read mapping while still sequencing \cite{lindner2016hilive}. Second, it can hide the complexity and details of the underlying hardware from users who are not necessarily fluent in FPGAs. 

\item Cloud-enabled pre-alignment filtering in a pay-per-use fashion is also a promising solution to encourage the use of such pre-alignment filters and address the concern of lack of experience in dealing with hardware designs. Cloud computing offers access to a large number of advanced FPGA chips that can be used concurrently via a simple user-friendly interface. However, such scenario requires the development of privacy-preserving pre-alignment filters due to privacy and legal concerns related to the share of sensitive genome data to a third party \cite{salinas2017secure, alser2015can}. Our next efforts will focus on exploring privacy-preserving real-time pre-alignment filtering.

\item Since a single-core SneakySnake/GateKeeper/Shouji has only a small footprint on the FPGA, we can combine our architecture with any of the FPGA-based accelerators for BWT-FM or hash-based mapping techniques on a single FPGA chip. With such a combination, the end result would be an efficient and fast multi-layer mapping system: alignments that pass our pre-alignment filter can be further verified using a dynamic programing based alignment algorithm within the same chip.

\item To further improve the performance of our SneakySnake pre-alignment filter, we can take full advantage of the redundancy across both reference genome and reads present in large sequencing data sets and store the optimal path for small subsequences of the read and the reference sequences. We later use the pre-computed path towards calculating the approximate edit distance quickly. \newpage This requires the understanding of the trade-offs between the memory footprint of seed filtering and the speed of pre-alignment filtering. Another approach for accelerating our SneakySnake algorithm is to exploit GPUs or multi-threaded processors to achieve a highly parallel implementation. 

\item We also aim to explore the possibility of achieving full alignment step in linear time complexity in term of read length. We believe that we can achieve this challenging goal by improving the accuracy of DFS algorithm in SneakySnake. If we penalize each potential path based on its distance from the main diagonal, we can eventually infer the the exact number of edit distance along the entire path. This leads to not only calculate the edit distance accurately, but also to on-the-fly provide the optimal alignment without performing \lq\lq{}backtracking\rq\rq{} by printing the snake\rq{}s path. The path contains all needed information such as the location, the number, and the type of the edits involved.
\end{enumerate}

%% file: chapter11.tex
\chapter{Other Works of This Author}
During the course of my doctoral studies, I had the opportunity to work closely and collaborate with my fellow graduate students and researchers from many other institutions. These projects help me to 1) learn more about genome analysis, metagenomics, and bioinformatics in general, 2) develop my own ideas and skills to conduct a successful research, and 3) widen my network of good mentors and researchers. In this chapter, I would like to acknowledge my collaborators and these projects related to this dissertation.

In a collaboration with Eleazar Eskin, Serghei Mangul, David Koslicki, and Nathan LaPierre, we worked on developing new methods for accurate and comprehensive microbial community profiling. I also worked on developing an interactive visualization tool for facilitating the analysis of the profiling results, which leaves the exploration and interpretation to the user. This gave me the opportunity to extend my knowledge about metagenomics and gain more insights about the need for accelerating metagenomics profiling tools. This work is called MiCoP and published in BMC Genomics \cite{lapierre2019micop} and in bioRxiv \cite{lapierre2018micop}.

I am very happy to start this collaboration and continue working together on other related projects. During the last year of my doctoral studies, we started another two new projects: 1) improving the speed and accuracy of our metagenomics profiler using a new seed filtering algorithm and a new read mapper and 2) providing a survey of algorithmic foundations and methodologies across read mapping methods for both short and long reads. We discuss how longer read lengths produce unique advantages and limitations to read mapping techniques. We also discuss how general mapping algorithms have been tailored to the specific needs of various domains in biology, including whole transcriptome, adaptive immune repertoire, and human microbiome studies. This work is the first of its kind to provide a clear roadmap of how technology dictates read mapping algorithms.

In a collaboration with Onur Mutlu, Jeremie Kim, and Damla Senol Cali, we developed GRIM-Filter, a pre-alignment filter for reducing the number of invalid mapping in genome analysis. It exploits the high memory bandwidth and the logic layer of 3D-stacked memory to perform highly-parallel filtering inside the 3D-stacked memory itself. This gave me the opportunity to learn more about 3D-stacked memory architecture, which has a lot in common with memory architectures. This work is published in BMC Genomics \cite{kim2018grim} and available on arXiv \cite{kim2017grim}.

In a collaboration with Erman Ayday, Nour Almadhoun, and Azita Nouri, we survey a wide spectrum of cross-layer privacy breaching strategies to human genomic data (using both public genomic databases and other public non-genomic data). We outline the principles and outcomes of each technique, and assess its technological complexity and maturation. We then review potential privacy-preserving countermeasure mechanisms for each threat. This work introduces me into the importance of developing privacy-preserving tools for genome analysis applications. This work is presented in DPM \cite{alser2015can} and in PRIVAGEN \cite{alseridentifying}. 

All of these works are closely related to this dissertation. During these collaborations, we all gained valuable expertise on improving genome analysis by learning from each other.

%% file: appendix.tex
\appendix
\chapter{Data}
In this chapter, we use Edlib \cite{vsovsic2017edlib} (edit distance mode) to assess the number of accepted (i.e., having edits less or equal to the edit distance threshold) and rejected (i.e., having more edits than the edit distance threshold) pairs for each of the 12 datasets. We also provide the time, in seconds, that is needed for Edlib to complete clustering the pairs. We provide these details for set\_1, set\_2, set\_3, and set\_4 in Table \ref{table:table1Appendix}. We also provide the same details for set\_5, set\_6, set\_7, and set\_8 in Table \ref{table:table2Appendix} and for set\_9, set\_10, set\_11, and set\_12 in Table \ref{table:table3Appendix}. 

\begin{table}
\centering
\caption{Details of our first four datasets (set\_1, set\_2, set\_3, and set\_4). We use Edlib \cite{vsovsic2017edlib} (edit distance mode) to benchmark the accepted and the rejected pairs for edit distance thresholds of 0 up to 10 edits.}
\label{table:table1Appendix}
\includegraphics[width=\textwidth]{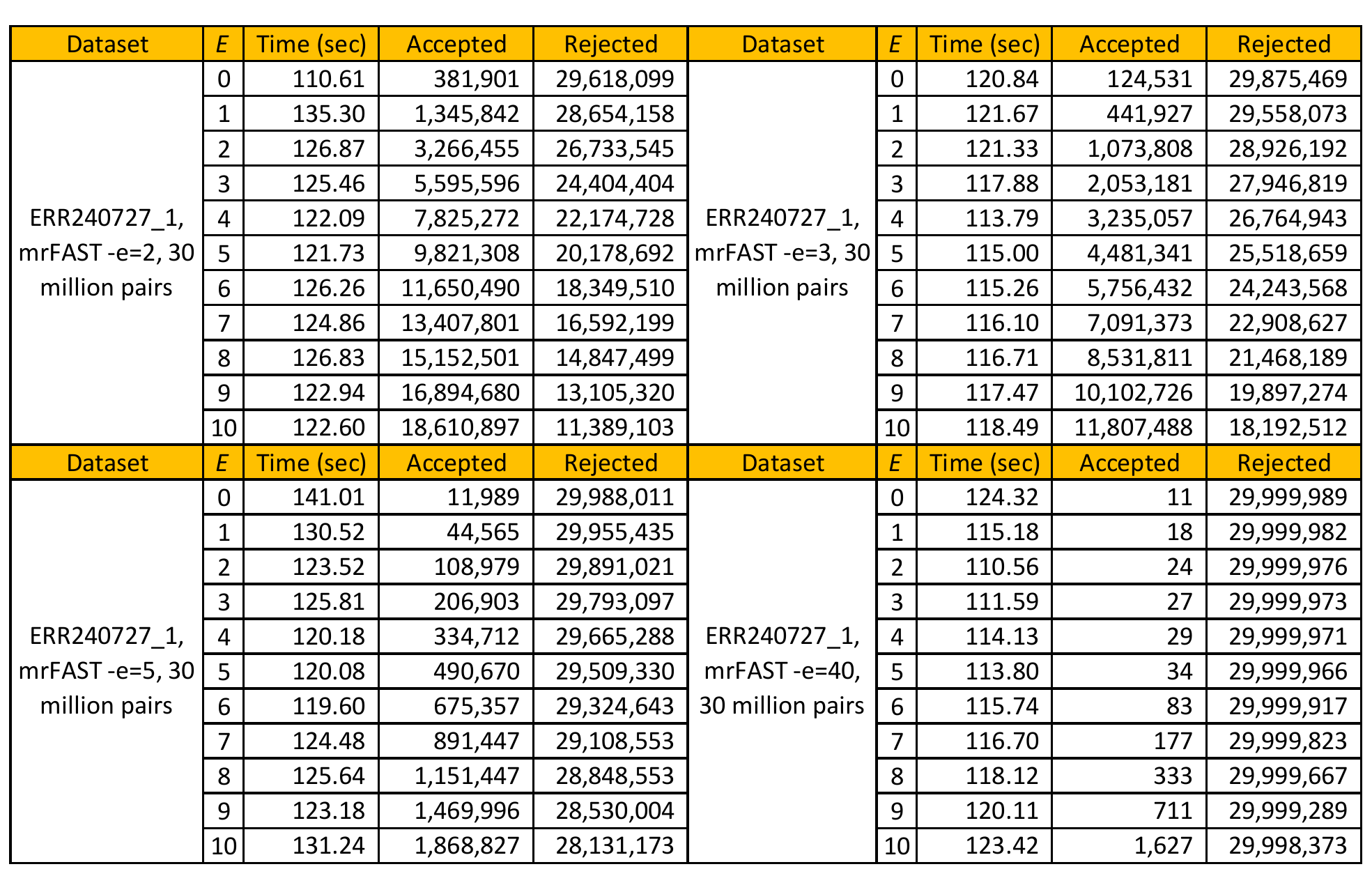}
\end{table}

\begin{table}
\centering
\caption{Details of our second four datasets (set\_5, set\_6, set\_7, and set\_8). We use Edlib \cite{vsovsic2017edlib} (edit distance mode) to benchmark the accepted and the rejected pairs for edit distance thresholds of 0 up to 15 edits.}
\label{table:table2Appendix}
\includegraphics[width=\textwidth]{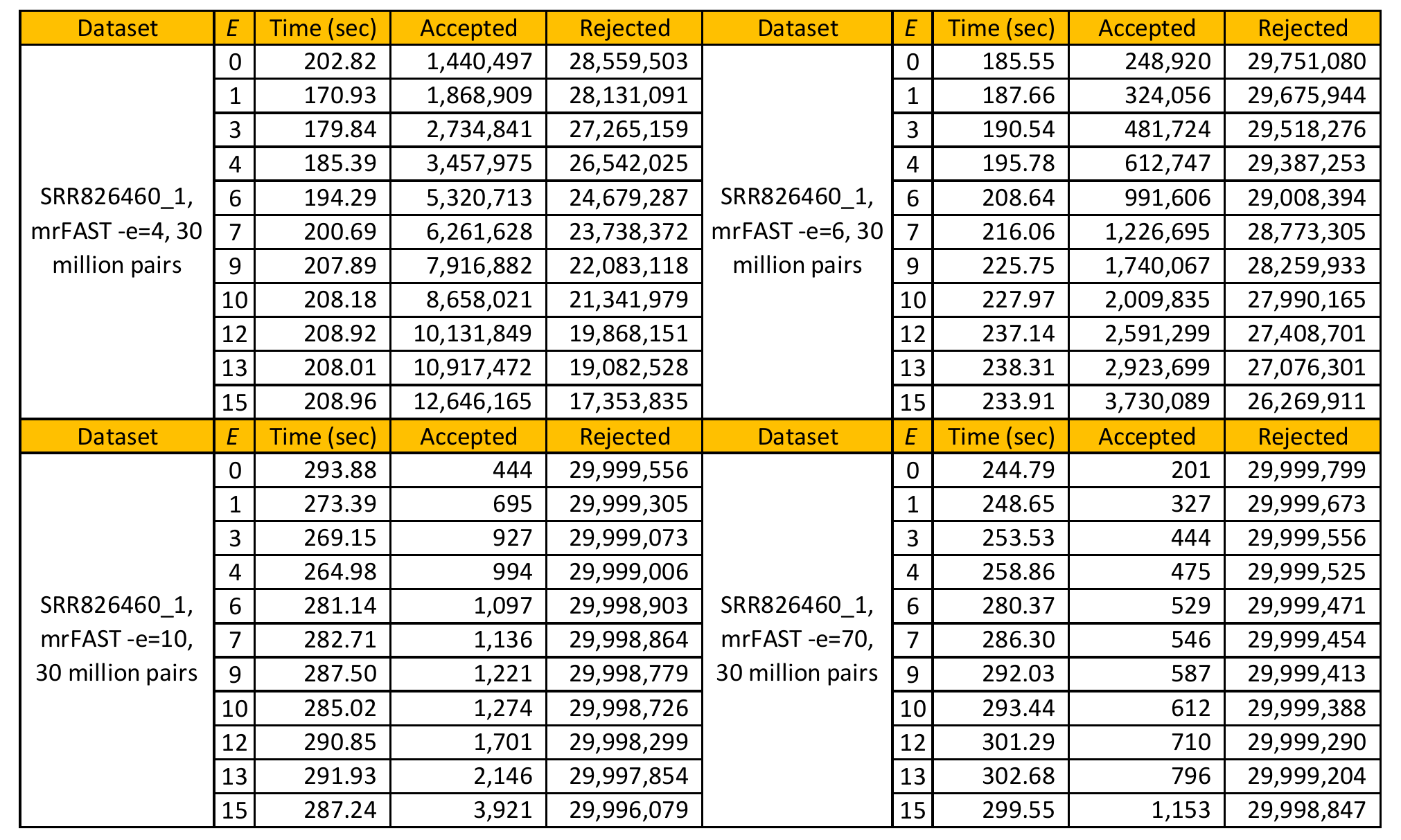}
\end{table}

\begin{table}
\centering
\caption{Details of our last four datasets (set\_9, set\_10, set\_11, and set\_12). We use Edlib \cite{vsovsic2017edlib} (edit distance mode) to benchmark the accepted and the rejected pairs for edit distance thresholds of 0 up to 25 edits.}
\label{table:table3Appendix}
\includegraphics[width=\textwidth]{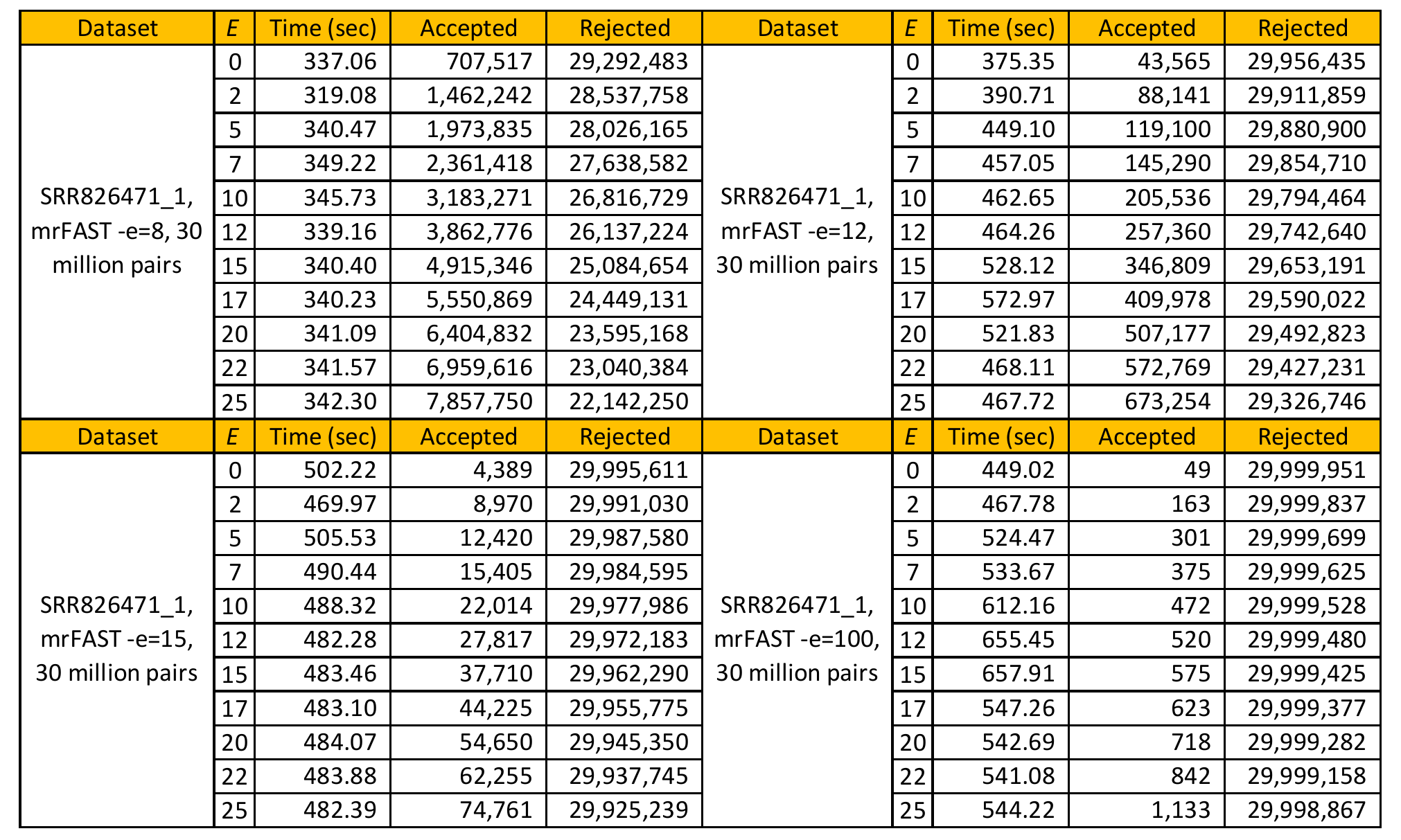}
\end{table}

Next, we provide the effect of reducing the search space of our SneakySnake algorithm and the best exact edit distance algorithm, Edlib \cite{vsovsic2017edlib}. We use the following datasets set\_1, set\_2, set\_3, and set\_4 in Figure \ref{table:table4Appendix}. We also use these datasets set\_5, set\_6, set\_7, and set\_8 in Figure \ref{table:table5Appendix}. Finally, in Figure \ref{table:table6Appendix}, we use the following datasets set\_9, set\_10, set\_11, and set\_12.

\begin{table}
\centering
\caption{Details of evaluating the feasibility of reducing the search space for SneakySnake and Edlib, evaluated using set\_1, set\_2, set\_3, and set\_4 datasets.}
\label{table:table4Appendix}
\includegraphics[width=\textwidth]{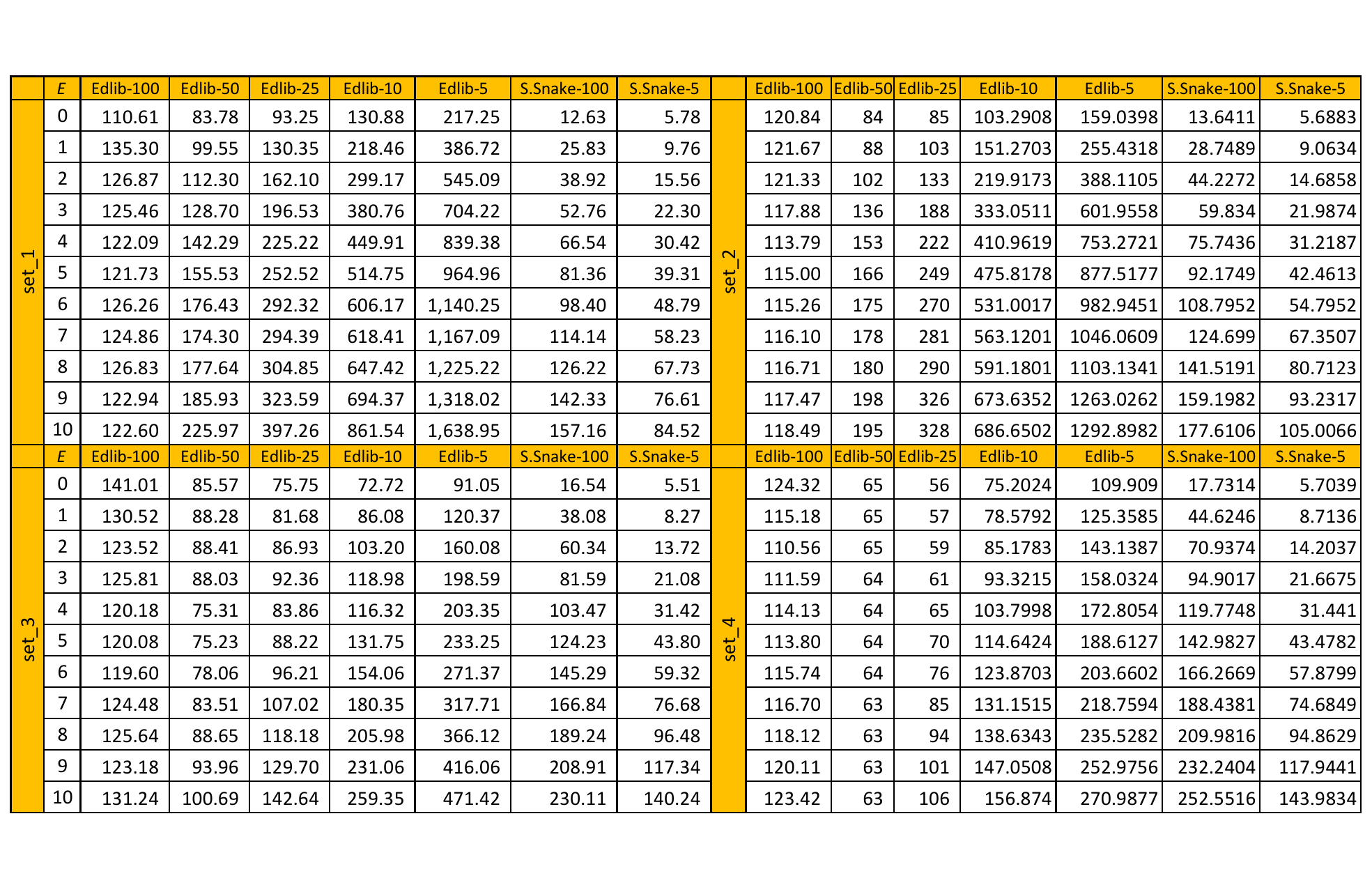}
\end{table}

\begin{table}
\centering
\caption{Details of evaluating the feasibility of reducing the search space for SneakySnake and Edlib, evaluated using set\_5, set\_6, set\_7, and set\_8 datasets.}
\label{table:table5Appendix}
\includegraphics[width=\textwidth]{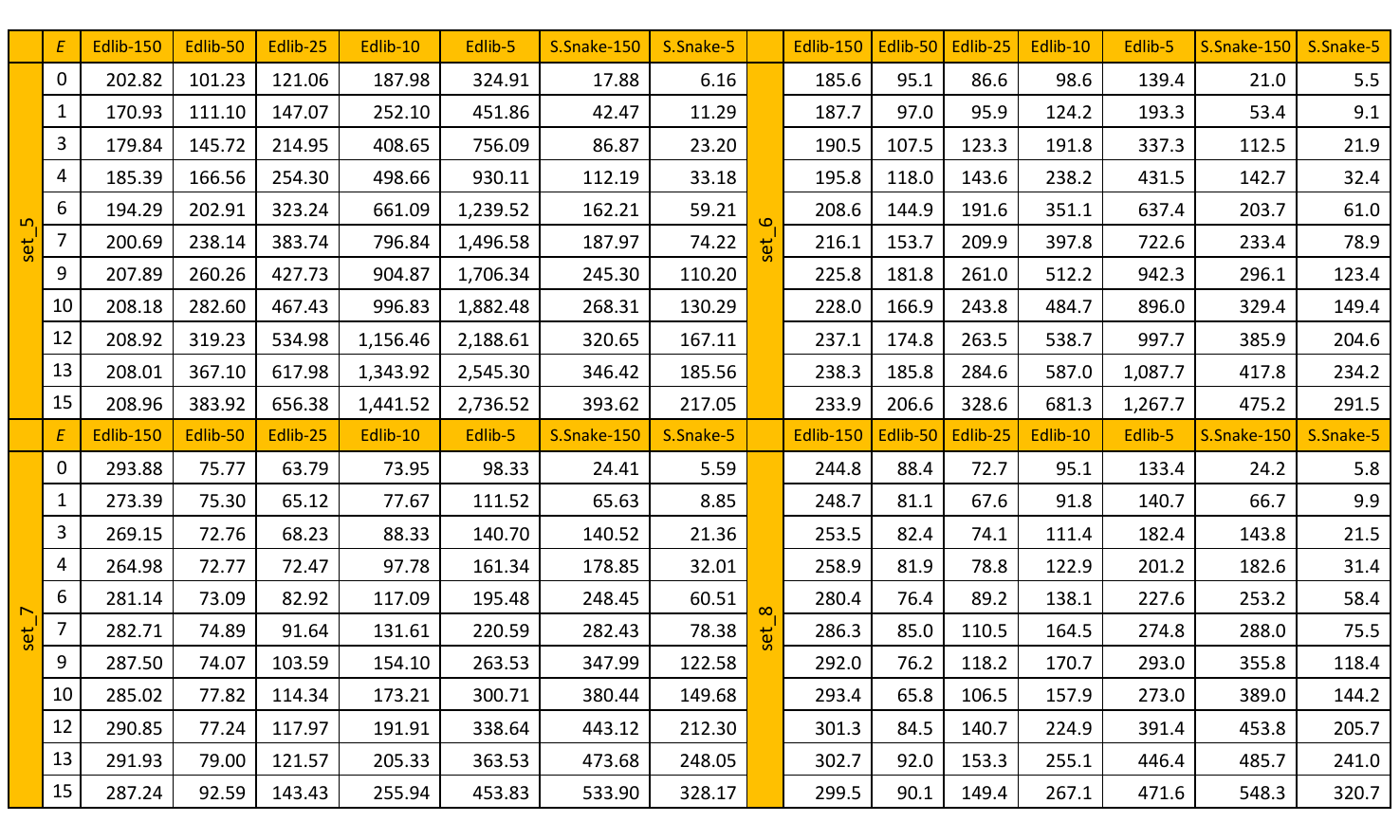}
\end{table}

\begin{table}
\centering
\caption{Details of evaluating the feasibility of reducing the search space for SneakySnake and Edlib, evaluated using set\_9, set\_10, set\_11, and set\_12 datasets.}
\label{table:table6Appendix}
\includegraphics[width=\textwidth]{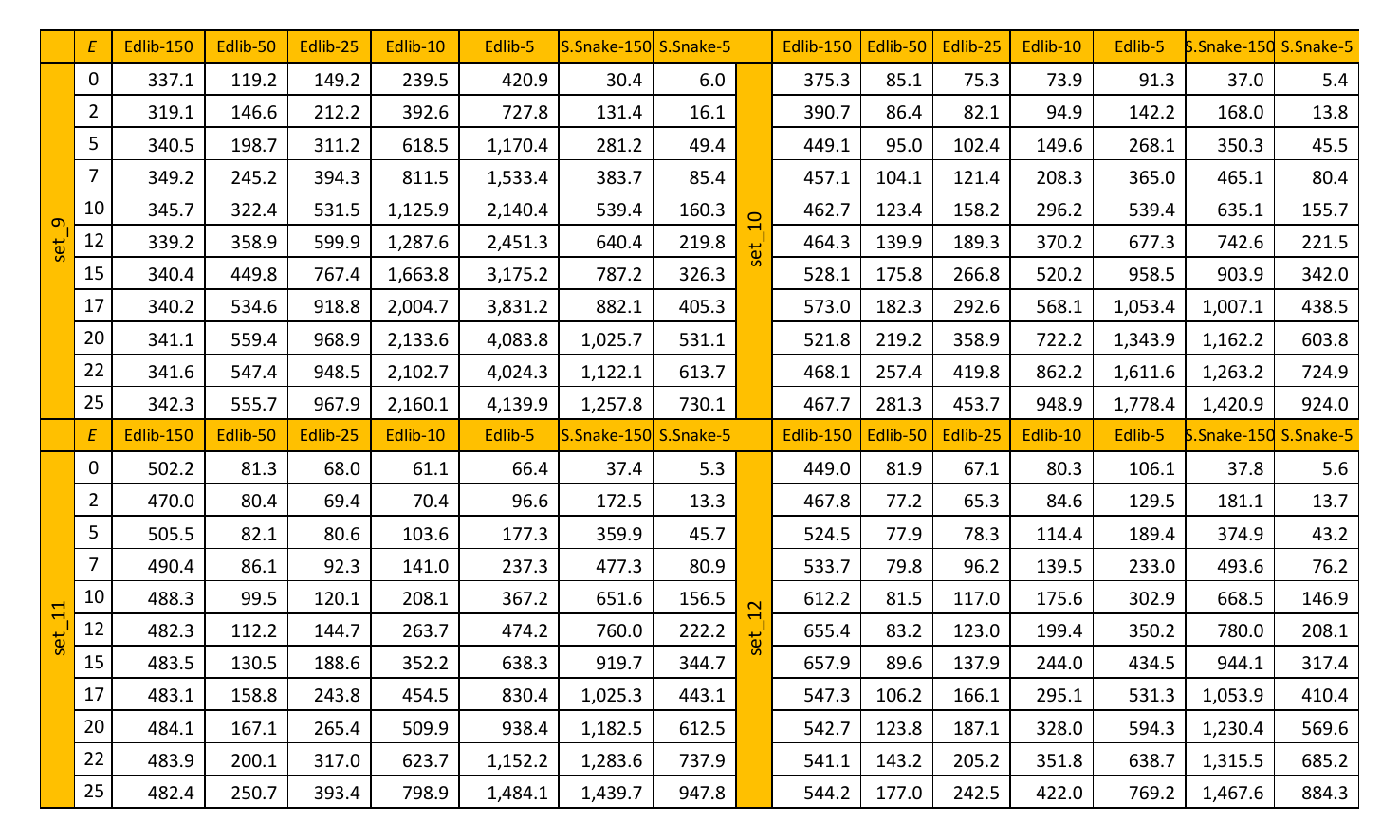}
\end{table}

%\chapter{Code}